\documentclass[a4paper,11pt]{solarphysics}

\usepackage{amsmath}
\usepackage{graphicx}
\usepackage{textcomp}
\usepackage{txfonts}
\usepackage{epstopdf}
\usepackage{lineno}
\usepackage[hyperref,optionalrh]{spr-sola-addons}
\usepackage{url}
\usepackage{array}
\usepackage{longtable}
\usepackage{multirow}
\usepackage{enumerate}
\usepackage{float}
\usepackage{pdflscape}
\usepackage{footnote}
\usepackage{marginnote}

\begin{document}

\begin{article}

\begin{opening}

\title{Catalogue of $>$55 MeV Wide-longitude Solar Proton Events Observed by SOHO, ACE, and the STEREOs at $\approx$1 AU during 2009\textendash 2016}

\author[addressref=add1,corref,email={mimapa@utu.fi}]{\inits{M.}\fnm{Miikka }\lnm{Paassilta}\orcid{0000-0002-7982-6213}}
\author[addressref=add2]{\inits{A.}\fnm{Athanasios }\lnm{Papaioannou}}
\author[addressref=add3]{\inits{N.}\fnm{Nina }\lnm{Dresing}}
\author[addressref=add1]{\inits{R.}\fnm{Rami }\lnm{Vainio}}
\author[addressref=add1]{\inits{E.}\fnm{Eino }\lnm{Valtonen}}
\author[addressref=add3]{\inits{B.}\fnm{Bernd }\lnm{Heber}}

\runningtitle{Catalogue of Wide Proton Events 2009\textendash 2016}
\runningauthor{M. Paassilta \textit{et al.}}

\address[id={add1}]{Department of Physics and Astronomy, University of Turku, 20014 Finland}
\address[id={add2}]{Institute for Astronomy, Astrophysics, Space Applications and Remote Sensing (IAASARS), National Observatory of Athens, I. Metaxa \& Vas. Pavlou St. GR-15236, Penteli, Greece}
\address[id={add3}]{Institut f\"ur Experimentelle und Angewandte Physik, Christian-Albrechts-Universit\"at zu Kiel, 24118 Kiel, Germany}

\begin{abstract}  
Based on energetic particle observations made at $\approx$1 AU, we present a catalogue of 46 wide-longitude ($>$45\textdegree) solar energetic particle (SEP) events detected at multiple locations during 2009\textendash 2016. The particle kinetic energies of interest were chosen as $>$55 MeV for protons and 0.18\textendash 0.31 MeV for electrons. We make use of proton data from the \textit{Solar and Heliospheric Observatory/Energetic and Relativistic Nuclei and Electron experiment} (SOHO/ERNE) and the \textit{Solar Terrestrial Relations Observatory/High Energy Telescopes} (STEREO/HET), together with electron data from the \textit{Advanced Composition Explorer/Electron, Proton, and Alpha Monitor} (ACE/EPAM) and the STEREO/\textit{Solar Electron and Proton Telescopes} (SEPT). We consider soft X-ray data from the \textit{Geostationary Operational Environmental Satellites} (GOES) and coronal mass ejection (CME) observations made with the SOHO/\textit{Large Angle and Spectrometric Coronagraph} (LASCO) and STEREO/\textit{Coronagraphs 1} and \textit{2} (COR1, COR2) to establish the probable associations between SEP events and the related solar phenomena. Event onset times and peak intensities are determined; velocity dispersion analysis (VDA) and time-shifting analysis (TSA) are performed for protons; TSA is performed for electrons. In our event sample, there is a tendency for the highest peak intensities to occur when the observer is magnetically connected to solar regions west of the flare. Our estimates for the mean event width, derived as the standard deviation of a Gaussian curve modelling the SEP intensities (protons $\approx$44\textdegree, electrons $\approx$50\textdegree), largely agree with previous results for lower-energy SEPs. SEP release times with respect to event flares, as well as the event rise times, show no simple dependence on the observer's connection angle, suggesting that the source region extent and dominant particle acceleration and transport mechanisms are important in defining these characteristics of an event. There is no marked difference between the speed distributions of the CMEs related to wide events and the CMEs related to all near-Earth SEP events of similar energy range from the same time period.
\end{abstract}


\end{opening}

\section{Introduction}

Solar energetic particle (SEP) events are considerable increases in the \textit{in situ} measured fluxes of charged particles, primarily protons and electrons, ejected from the Sun (\citealp{Reames1999, Reames2013}). They are accompanied by coronal mass ejections (CMEs) and also by X-ray flares; all these three phenomena are understood to be ultimately the result of massive discharges of energy stored in magnetic field structures at and near the surface of the Sun. The relationships between flares, CMEs, and SEP events, as well as the transport of particles from the Sun to observing spacecraft at various locations in the solar system, are being actively studied.

Solar Cycle 24, which commenced in late 2008, stands in contrast to its two immediate predecessors because of its lower level of SEP event activity (\citealp{Gopalswamy2015b}; \citealp{Mewaldt2015}; \citealp{Vainio2017}). Due to the Sun being relatively quiescent, the total heliospheric plasma pressure is lower than before (see \textit{e.g.} \citealp{Gopalswamy2015}). It is possible that this, perhaps together with other circumstances characteristic of Cycle 24, has facilitated the occurrence of very wide SEP events, which offer an interesting object of study with regard to particle acceleration and transport processes. On the other hand, the overall number of detected events is about 2/3 of that of the previous solar cycle, leading to less favourable statistics.

\citet{Paassilta2017} reported that high-energy SEP events observed near the Earth with a far-eastern origin appeared to be more frequent in Solar Cycle 24 than in Solar Cycle 23, raising the possibility that very wide SEP events in the 55\textendash 80 MeV proton energy range could have been more common in general during the later cycle than during the earlier one. It appears that a more detailed look into these events, and the recent SEP events in general, would add to our understanding of the SEP acceleration and transport. Since suitable data are available for Solar Cycle 24, it is desirable to view the same events from multiple points in the heliosphere and look for possible correlations between their properties, which could provide new insight into the physical processes involved in SEP events and particle transport in interplanetary space.

The benefits of multi-spacecraft observations of SEP events have been recognized for quite some time. Some notable studies, mainly concentrating on the topic of the dependence of the SEP intensity time profiles on the observer location with respect to the solar source region, include \citet{McKibben1972}, \citet{McGuire1983}, \citet{Kallenrode1993}, and \citet{Reames1996}; more recent articles are mentioned in the following. A number of interplanetary missions\textemdash such as \textit{Helios} in the 1970s and 1980s\textemdash made multi-spacecraft studies of SEP events possible, and the 2006 launch of the twin \textit{Solar Terrestrial Relations Observatory} (STEREO) spacecraft into almost circular solar orbits at 1 AU from the Sun, together with a number of near-Earth platforms in operation at the same time, brought considerable new opportunities within reach in this field.

This paper aims to identify wide high-energy (55\textendash 80 MeV) proton events detected by the \textit{Energetic and Relativistic Nuclei and Electron experiment} (ERNE; \citealp{Torsti1995}) on board the \textit{Solar and Heliospheric Observatory} (SOHO) spacecraft, \textit{i.e.} events that had sufficient width to be detected also by the comparable instrumentation on board the two STEREO probes, and investigate them further. It is additionally intended to supplement other similar studies previously published on multi-spacecraft SEP event observations, such as \citet{Lario2013} (15\textendash40 MeV and 25\textendash 53 MeV protons, 71\textendash 112 keV and 0.7\textendash 3 MeV electrons), \citet{Dresing2014} (55\textendash 105 keV electrons), \citet{Papaioannou2014}(6\textendash 10 MeV protons, 55\textendash 85 keV electrons), and \citet{Richardson2014} ($>$25 MeV protons) by concentrating on protons of somewhat higher energy and extending the timeframe of interest to the end of 2016. We attempt to associate the near-Earth observations with those of the STEREO probes so as to gain a multi-point spatial coverage for each SEP event. While this approach limits our study to the events that have occurred during Solar Cycle 24 and so unfortunately precludes a direct comparison between Cycles 23 and 24, it does offer\textemdash in ideal cases\textemdash the possibility of comparing such quantities of interest as onset times, maximum particle intensities, and proton fluences at three widely separated locations within the inner solar system. In addition, we also investigate the particle injection times near the Sun, using both time-shifting analysis (TSA) and velocity dispersion analysis (VDA), as well as the SEP/solar flare and SEP/CME associations. For the near-Earth proton and electron observations, we partially rely on the results given in \citet{Paassilta2017} but revisit electron event onset times and maximum intensities together with the SEP/flare/CME associations determined for Solar Cycle 24 SEP events in light of additional data available.

This paper is structured in the following manner. In Section \ref{Sec2}, we introduce our data sources and basic methods together with the event catalogue; a selection of example events is included as Section \ref{Sec3}; Section \ref{Sec4} contains a statistical analysis of our results and discussion; and finally, our conclusions and outlook are presented in Section \ref{Sec5}.

\section{The Multi-spacecraft Proton Event Catalogue for 2009\textendash 2016}
\label{Sec2}

\subsection{Proton Data and Event Selection}
\label{Sec2.1}

As regards to the near-Earth proton observations, our study relies primarily on the ERNE experiment (\citealp{Torsti1995}) carried by the SOHO spacecraft. The two particle telescopes of ERNE, the \textit{Low-Energy Detector} (LED) and the \textit{High-Energy Detector} (HED), jointly cover a nominal ion 
energy range that extends from $\approx$1 MeV/nucleon to a few hundred MeV/nucleon. The upper limit lies at $\approx$140 MeV/nucleon for protons. In keeping with several previous ERNE-related articles (\textit{e.g.} \citealp{Vainio2013}; \citealp{Paassilta2017}), the 54.8\textendash 80.3 MeV proton channel (average 67.7 MeV) was chosen as the reference energy range; this is higher than that considered in many other similar studies (\textit{e.g.} \citealp{Papaioannou2014}). Aside from the practical advantages offered by the comparatively easy identification of small, relatively closely spaced SEP events due to the fact that the intensities tend to decrease after the event more quickly than at low particle energies, as well as the possibility of comparing our results with those of earlier works pertaining to ERNE, the significance of high-energy particles to space weather (\citealp{Reames2013}) continues to serve as a primary motivator in our work. On the other hand, some relatively small SEP events, while detectable at lower energies (for instance, $<$ 10 MeV), cannot be readily discerned in this energy range.

Launched on 25 October 2006, the two STEREO spacecraft orbit the Sun at about 1 AU, with STEREO-A (Ahead) moving in front and STEREO-B (Behind) behind the Earth. The angular separation between the two is increasing gradually (by some 22\textdegree \,\textit{per} year relative to the Earth), allowing observations to be made from two widely separated locations, in addition to the Earth and its vicinity. The \textit{In situ Measurements of Particles and CME Transients} (IMPACT) experiment on board both STEREO-A and -B is designed to carry out measurements of solar wind, interplanetary magnetic fields, and SEPs. Among other instrumentation, IMPACT comprises four particle detectors, including the \textit{High Energy Telescope} (HET; \citealp{Rosenvinge2008}) and the \textit{Low Energy Telescope} (LET; \citealp{Mewaldt2008}). In addition to electrons, HET is capable of measuring protons from 13 MeV to 100 MeV, while LET covers the range between 1.8 MeV/nucleon and 15 MeV/nucleon in the case of protons and helium.

As the division of the proton energy channels is not identical for ERNE and HET, we decided to combine the 40\textendash 60 MeV and 60\textendash 100 MeV proton channels of HET so as to obtain a reference energy range matching that of ERNE as closely as possible, as well as to improve particle statistics. While the resulting HET reference channel is wider than its ERNE counterpart, its average (geometric mean) proton energy of 63.2 MeV is reasonably similar. Additionally, combining the channels offers the benefit of mitigating the somewhat high background intensity usually present in the HET 60\textendash 100 MeV proton channel, since the background in the 40\textendash 60 MeV channel is lower by a factor of a few and detected particle intensities during SEP events greater than in the 60\textendash 100 MeV channel. During our study, a brief investigation into the effect of the HET reference energy channel width on the observed event onset times was carried out. The results (not shown) indicated that for most events, this effect is probably of the order of a few minutes, and therefore it was not considered significant.

Our period of interest includes the years 2009\textendash 2016, starting immediately after the commencement of the current solar cycle in December 2008. At the start of 2009, the STEREO spacecraft had in their orbits gained a considerable ($>$40\textdegree) angular separation from the Earth, enabling wide-longitude SEP events to be studied. During 2009\textendash 2016, significant data gaps of several days occurred for STEREO-A/HET in early July 2014 and in late August 2014, following which an extensive (albeit non-continuous) loss of coverage took place from September 2014 to mid-November 2015, as several scientific instruments on board were switched off or set to limited operation due to the spacecraft undergoing superior solar conjunction. In addition to a data gap of a few days in early January 2011, all radio contact with STEREO-B was lost on 1 October 2014, so no data are available beyond that date. ERNE, for its part, experienced data gaps longer than 10 consecutive days between 30 November 2011 and 5 January 2012, 28 January and 10 February 2012, 9 December 2012 and 31 January 2013, and 29 October and 4 December 2013; for a comprehensive listing of major ERNE data gaps in the 2009\textendash 2016 period, the reader is referred to Table 1 in \citet{Paassilta2017}. Any events that have taken place during the loss of data coverage (aside from two special cases discussed further in Section \ref{Sec2.8}) are necessarily missing from our catalogue.

Our main criterion in selecting the candidate events to be investigated was that an increase of a factor of $\gtrsim$3.0 over the quiet-time background\footnote{This limit was intentionally somewhat flexible, so as to allow some small events, better detectable in data averaged over several minutes, to be taken into account.} in 1-minute average proton intensity had to have occurred either at SOHO and at least one of the STEREO spacecraft or at both STEREO spacecraft nearly simultaneously; it was additionally required that the observers be separated by more than $\approx$45 degrees of longitude at the time of proton event onset. The candidate events detected by instrumentation carried by SOHO and STEREO-A and -B were identified by scanning visually through the one-minute averaged intensity data of the respective reference proton channels (explained above). Cases where coverage loss or possible masking by a previous event had occurred at any one of the spacecraft were included in the analysis as candidate events. In addition, the 60\textendash 100 MeV HET proton intensity data were visually scanned independently from those of the combined 40\textendash 100 MeV energy channel. An intensity enhancement was required to be detectable in both of these channels according to the selection criterion explained above before it could be considered a potential event. This was done to ensure that the candidate events indeed reached proton energies of $>$60 MeV at both SOHO and the STEREOs, making them easier to compare.

The quiet-time background intensity values of the proton energy channels of respective instruments were estimated during the years of interest.\footnote{Here and everywhere else in this work, 1 pfu = 1 particle cm$^{-2}$ s$^{-1}$ sr$^{-1}$.} In the case of STEREO/HET data, the background in the 40\textendash 100 MeV channel was $\approx$5.0$\times 10^{-4}$ pfu MeV$^{-1}$ during the early part of the cycle (2009) and then decreased slowly to $\approx$2.0$\times 10^{-4}$ pfu MeV$^{-1}$ near the solar maximum (late 2013 and early 2014). For the 55\textendash 80 MeV proton channel of ERNE the background intensity was estimated to be $\approx$7.0$\times 10^{-4}$ pfu MeV$^{-1}$ in the early phase of the cycle, decreasing to $\approx$3.5$\times 10^{-4}$ pfu MeV$^{-1}$ by the end of 2013, after which it again began to rise, reaching $\approx$7$\times 10^{-4}$ pfu MeV$^{-1}$ by the end of 2016. This variation is mainly due to solar modulation of galactic cosmic rays.

\subsection{Electron Data}
\label{Sec2.2}

Besides proton data, we examined energetic electron intensity data recorded by the \textit{Electron, Proton, and Alpha Monitor} (EPAM; \citealp{Gold1998}) on board the \textit{Advanced Composition Explorer} (ACE) and the \textit{Solar Electron and Proton Telescope} (SEPT; \citealp{Mueller-Mellin2008}) on board both STEREO spacecraft for electron events co-occurring with the proton events. This enabled us to study each event in more detail than proton observations alone would have allowed. Aside from maximum intensities, the event onset times were an item of major interest, since they are necessary for electron solar release time analysis (see the next subsection).

With regard to near-Earth electron observations, our study was partially based on the results reported in \citet{Paassilta2017}, which also includes a somewhat more involved explanation of the ACE/EPAM data and its handling than what will be given here. In keeping with that paper, the reference energy channel chosen for ACE/EPAM was 0.18\textendash 0.31 MeV, and the default data type used for onset determination was direction-averaged (\textit{i.e.} omnidirectional) one-minute electron intensity recorded by the LEFS60 (\textit{Low-Energy Foil Spectrometer}) particle telescope. In cases where ion contamination of the electron data was evident or suspected, direction-averaged intensity data from the DE30 (\textit{Deflected Electrons}) telescope was substituted.
 
To obtain comparable results for both ACE/EPAM and STEREO/SEPT electron observations (SEPT data is available at \url{http://www2.physik.uni-kiel.de/stereo}), five SEPT energy channels spanning the range of 0.165\textendash 0.335 MeV were combined in our study. While the resulting reference channel is wider than the corresponding ACE/EPAM energy channel, their average energies are almost identical (both $\approx$0.24 MeV). The time resolution was again one minute. Since the pitch-angle distributions or intensity anisotropies of electrons were not among the immediate topics of interest for our study as such, directional intensity data were not utilized. As with the LEFS60 data recorded by ACE/EPAM, ion contamination presented an occasional issue with STEREO/SEPT electron intensity data; unfortunately, because no intrinsically contamination-free alternative is available for the latter, any onset times or intensity values derived during periods of possible ion contamination had to be discarded from our analysis.

\subsection{Maximum SEP Intensities of Events}
\label{Sec2.3}

The particle intensity data recorded during the events were smoothed with a 5-minute sliding average for all instruments and spacecraft, except for SOHO/ERNE, and the maximum of the sliding average was taken as an estimate for the event maximum intensity. The same procedure was performed for proton and electron data. Aside from two proton and seven electron events with an exceptionally prolonged rise, we generally disregarded any rapid intensity enhancement that followed more than $\approx$36 hours after event onset so as to exclude shock peaks and possible new events. Identifiable shock peaks occurring within 36 hours of the event onset were also excluded. Intercalibration of measured proton and electron intensities between different instruments was considered and implemented in the case of electrons; for discussion, see Appendix \ref{App1}.

In the case of ERNE, apparent large and rapid intensity fluctuations due to temporary instrument saturation in the reference energy channel near the peaks of some near-Earth proton events frustrated attempts to determine the event maximum intensity automatically. In these events, methods such as sliding average or median smoothing yielded values that were judged to be unreliable. We therefore decided to resort to visual estimation for ERNE proton maximum intensities in a bid to maintain consistency. Due to the approximative nature of this method, it was generally not possible to determine the time of the event peak intensity with satisfactory accuracy, and so the event rise times, listed in this work for the other instruments, were omitted for ERNE. Furthermore, gaps in ERNE data obscured the peaks of seven near-Earth proton events, and only visually estimated lower limits for maximum intensity are given for them.

It should be noted that the method used here has resulted in some differences in the maximum intensities for ACE/EPAM electron events when compared to those reported for the same events in \citet{Paassilta2017}, who relied solely on visual estimation using data with a time resolution of one minute; as no smoothing is involved, this tends to produce slightly larger values (usually by a factor of not more than $\approx$1.2). In a few events, the conservative exclusion of a late intensity enhancement possibly originating from another particle injection has caused a somewhat greater disagreement, most prominently for ACE/EPAM electrons in event 10 (23 January 2012), where our peak intensity estimate is an order of magnitude lower than the one given in the previous paper.

\subsection{Onset and Solar Release Time Determination}
\label{Sec2.4}

After identifying candidate events for each spacecraft by a visual scan of the intensity data, we used the so-called Poisson-cumulative sum (CUSUM) method to determine the observed proton event onset times for SOHO/ERNE and STEREO-A and -B/HET and LET. Applying this scheme to the SEP event onset determination, the pre-event background intensity is considered in terms of a process under control, and the increasing intensity at event onset constitutes a loss of control. The Poisson-CUSUM method is used to monitor the statistical quality of the process and it alerts when the specific conditions that signal a failure (event onset) occur. A detailed description of the algorithm as well as the onset criteria are presented in \citet{Huttunen-Heikinmaa2005}. Aside from different energy ranges of the proton channels for ERNE on one hand and for HET and LET on the other, onset determination parameters were kept identical for all data sets to obtain results that are directly comparable and internally consistent as far as possible.

In the context of the velocity dispersion analysis (VDA), the observed time of event onset at 1 AU for particles with kinetic energy $E$ can be written as

\begin{equation}
\label{VDA_eq}
t_{\rm onset} (E) = t_0 + 8.33 \, {\rm[min/AU]} \, s \, \beta^{-1}(E),
\end{equation}
where $t_0$ is the time of release of the particles (in minutes), $s$ is their apparent path length (here expressed in AU) from the source to the observer, and $\beta^\textrm{-1}(E)$ is their reciprocal speed in units of 1/$c$ where $c$ is the speed of light. Thus, if the observed onset times for several energy channels are known, a linear fit to these values as a function of the reciprocal speed results in estimates for both $t_0$ and $s$. It is assumed here that both the release time and the apparent path length are independent of the kinetic energy of the particles, \textit{i.e.} they are all released at the same time and travel an equally long distance. With the STEREO observations, the distance of each spacecraft from the equatorial surface of the Sun was used to scale the calculated apparent path length given by Equation \ref{VDA_eq} to 1 AU. No such correction was applied to ERNE VDA results, as SOHO has stayed at nearly 1 AU from the Sun at all times since it became operational.

The energy channels used in VDA are listed in Table \ref{VDA_channels} for ERNE and HET and LET. For ERNE, 20 channels spanning from 1.58 MeV to 131 MeV were applied; both LED and HED provide ten individual channels, with those of LED ranging from 1.58 MeV to 12.7 MeV and those of HED from 13.8 MeV to 131 MeV. In the case of HET and LET, the number of channels was 14: of these, LET contributed three (1.8 MeV to 10.0 MeV) and HET 11 (13.6 MeV to 100 MeV). Overall, they cover a similar energy range. However, the considerable width of the LET channels, compared to LED, should be noted, as well as a gap between 10.0 and 13.6 MeV. The time resolution was one minute for all energy channels and all detectors.

\setlength{\tabcolsep}{1.5pt}
\begin{table}
\caption[]{Energy channels used for proton VDA. ERNE/LED and ERNE/HED span channels 1--10 and 11--20, respectively; LET spans channels 1--3, HET spans channels 4--14. Average energy is here defined as the geometric mean of the energy channel.}
\label{VDA_channels}
\centering
\begin{tabular}{l | l l l | l l l}       
\hline\hline
& \multicolumn{3}{c}{ERNE} & \multicolumn{3}{c}{LET and HET} \\
\hline
Channel & \multirow{3}{1.4cm}{Energy range [MeV]} & \multirow{3}{1.4cm}{Average energy [MeV]} & \multirow{3}{1.8cm}{Reciprocal speed [$c^\textrm{-1}$]} &\multirow{3}{1.4cm}{Energy range [MeV]} & \multirow{3}{1.4cm}{Average energy [MeV]} & \multirow{3}{1.8cm}{Reciprocal speed [$c^\textrm{-1}$]} \\    % table heading 
&&&&&&\\
&&&&&&\\
\hline                        % inserts single horizontal line
1 & 1.58\textendash 1.78 & 1.68 & 16.7 & 1.8\textendash 3.6 & 2.55 & 13.6 \\
2 & 1.78\textendash 2.16 & 1.97 & 15.5 & 4.0\textendash 6.0 & 4.90 & 9.82 \\
3 & 2.16\textendash 2.66 & 2.41 & 14.0 & 6.0\textendash 10.0 & 7.75 & 7.83 \\
\cline{5-7}
4 & 2.66\textendash 3.29 & 2.98 & 12.6 & 13.6\textendash 15.1 & 14.3 & 5.79\\
5 & 3.29\textendash 4.10 & 3.70 & 11.3 & 14.9\textendash 17.1 & 16.0 & 5.49\\
6 & 4.10\textendash 5.12 & 4.71 & 10.0 & 17.0\textendash 19.3 & 18.1 & 5.16 \\
7 & 5.12\textendash 6.42 & 5.72 & 9.10 & 20.8\textendash 23.8 & 22.2& 4.67  \\
8 & 6.42\textendash 8.06 & 7.15 & 8.15 & 23.8\textendash 26.4 & 25.1 & 4.41 \\
9 & 8.06\textendash 10.1 & 9.09 & 7.24 & 26.3\textendash 29.7 & 27.9 & 4.19 \\
10 & 10.1\textendash 12.7 & 11.4 & 6.47 & 29.5\textendash 33.4 & 31.4 & 3.96 \\
\cline{2-4}
11 & 13.8\textendash 16.9 & 15.4 & 5.59 & 33.4\textendash 35.8 & 34.6 & 3.78 \\
12 & 16.9\textendash 22.4 & 18.9 & 5.06 & 35.5\textendash 40.5 & 37.9 & 3.62 \\
13 & 20.8\textendash 28.0 & 23.3 & 4.57 & 40\textendash 60 & 49.0 & 3.21 \\
14 & 25.9\textendash 32.2 & 29.1 & 4.11 & 60\textendash 100 & 77.5 & 2.61 \\
15 & 32.2\textendash 40.5 & 36.4 & 3.69 & \\
16 & 40.5\textendash 53.5 & 45.6 & 3.32 & \\
17 & 50.8\textendash 67.3 & 57.4 & 2.99 & \\
18 & 63.8\textendash 80.2 & 72.0 & 2.70 & \\
19 & 80.2\textendash 101 & 90.5 & 2.44 & \\
20 & 101\textendash 131 & 108 & 2.26 & \\
\hline
\end{tabular}
\end{table}

In several cases an elevated background intensity from a previous event (typically at low energies) or a very small or slowly rising enhancement (typically at high energies) rendered the calculated onset time suspect for a particular energy channel or prevented onset detection by the Poisson-CUSUM method entirely. Mainly for these reasons, it was necessary to exclude one or more data points from VDA for most events. While our sample contains a total of 94 proton events (25 for SOHO/ERNE, 33 for STEREO-A, and 36 for STEREO-B) for which a precise estimate of the proton onset time is available, VDA line fitting fails in 20 of these (21\%), for instance by yielding a negative value for the apparent path length unless nearly all fit points are summarily discarded. There are also several events where a statistically plausible fit can be achieved, but it indicates an apparently unphysical value for the path length, that is, less or very much more than 1 AU. However, we have not omitted the events falling into the latter category from the tables (even though they are not included in most of the detailed statistical analysis); this is because we retain an interest in evaluating the performance of VDA.

The time-shifting analysis (TSA) involves shifting the observed proton event onset\textemdash which in this case had been previously determined using the Poisson-CUSUM method\textemdash at each spacecraft back to the Sun. If a particle species has a kinetic energy $E$, its time of solar release can be estimated as

\begin{equation}
t_{\rm rel}(E) = t_{\rm onset}(E) - 8.33 \, {\rm[min/AU]} \, L \, \beta^{-1} (E).
\end{equation}

Here, $L$ is the length of the magnetic field line which connects the particle source and the observer, while $\beta^\textrm{-1}(E)$ is the reciprocal speed of the particles, as in Equation \ref{VDA_eq}. $L$ can be derived using the speed of the solar wind $u_{\rm SW}$ recorded by the observer during the event thus:

\begin{equation}
L(u_{\rm SW}) = z(r_{\rm SC}) - z(\rm R_{\odot}).
\end{equation}

Here, $r_{\rm SC}$ and R$_{\odot}$ are the observer's radial distance from the Sun and the solar radius (6.957$\times$ 10$^5$ km), respectively. $z(r)$, the distance from the centre of the Sun along the spiral field line, is calculated as

\begin{equation}
z(r) = \frac{a}{2}\left[ \ln\left(\frac{r}{a}+\sqrt{1+\frac{r^2}{a^2}}\right)+ \frac{r}{a}\sqrt{1+\frac{r^2}{a^2}} \right],
\end{equation}
where $a$ =  $u_{\rm SW}/\omega_{\odot}$ and $2 \pi/\omega_{\odot}$ = 24.47 d, the period of the solar rotation at the equator of the Sun.

For near-Earth observations, an approximate value for $r_{\rm SC}$ was derived by subtracting the Sun-directional distance of the observer from the Earth\textendash Sun distance at the time of interest. In the case of the STEREO observations, the radial distance of both craft from the Sun was directly available.

The speed of the solar wind near the Earth was determined from data gathered by the \textit{Solar Wind Electron, Proton and Alpha Monitor} (SWEPAM) of ACE (\citealp{McComas1998}; \url{http://www.srl.caltech.edu/ACE/ASC/level2/lvl2DATA_SWEPAM.html}). For two events (numbers 10 and 20), these data were not available, so \textit{Wind/Solar Wind Experiment} (SWE) data (\citealp{Ogilvie1995}; \url{http://omniweb.gsfc.nasa.gov/ftpbrowser/wind_swe_2m.html}) were used instead. The proton bulk speed data from the \textit{Plasma and Suprathermal Ion Composition} (PLASTIC) experiment, part of the STEREO instrumentation, were used to estimate $u_{\rm SW}$ at the locations of the STEREO probes (available at \url{https://cdaweb.sci.gsfc.nasa.gov/index.html/}). In each of these cases, a 12-hour average, centered on the $t_{\rm onset}$ for the proton event at the spacecraft in question, was calculated; if the precise value of $t_{\rm onset}$ was unknown, the entire 24-hour UT calendar day on which the proton event onset most likely fell was selected for analysis. The Sun\textendash spacecraft radial distances employed in the calculations were obtained by averaging in the same manner.

A broadly similar analysis including the determination of both onset times and solar release times was performed on the ACE/EPAM and STEREO/SEPT electron data. For the former, however, the low number of available energy channels (four) and the relatively high mean speed (0.73 $c$) of the particles in the reference energy channel meant that VDA would probably have produced results of very questionable value and it was for this reason not attempted. In contrast, STEREO/SEPT offers a total of 15 energy channels, which made performing both TSA and VDA attractive in principle.\footnote{A detailed description of STEREO/SEPT electron energy channels is included in the document "STEREO/SEPT level 2 science data format specification and caveats", available at \url{http://www2.physik.uni-kiel.de/stereo/data/sept/level2/}.}

All onset times for both the ACE/EPAM and the STEREO/SEPT electron intensity data were determined using the Poisson-CUSUM method. The numerical criteria for onset determination were the same as those applied to proton data. Although the ACE/EPAM electron events listed in this work (except for events 33 and 35) were previously investigated in \citet{Paassilta2017}, they were revisited with respect to onset timing to maintain consistency with the results obtained for STEREO/SEPT events. In the majority of cases, the difference in onset times between our listing and that of \citet{Paassilta2017} only amounts to a few minutes, but in three slowly rising and comparatively small events (two of which involve probable farside flares and thus a weak magnetic connection to the near-Earth observer), the Poisson-CUSUM method results in an onset time an hour or so earlier than the method described in \citet{Paassilta2017}.

For STEREO/SEPT, the electron onset determination was initially performed in 13 separate energy channels, spanning the nominal energy range of 0.070\textendash 0.399 MeV, as well as independently for the combined 0.165\textendash 0.335 MeV channel (nominal energy 0.235 MeV), which was used as the reference and reported in the tables. The two lowest available electron channels, 0.045\textendash 0.055 MeV and 0.055\textendash 0.065 MeV, were omitted \textit{a priori} from any quantitative analysis due to their low efficiency. Both TSA and VDA were then attempted for STEREO/SEPT data, but in nearly all cases, the latter gave either very unreliable results or no results at all. The most likely reason was deemed to be energy channel crosstalk in the instrument during periods of considerable electron flux; this may cause some high-energy electron hits to be counted as low-energy ones, distorting the time profile of the particle intensities at low energies. Thus, only TSA results for electron observations are considered and reported in this work. For discussion, see Section \ref{Sec4.6}.

\subsection{Proton Fluences}
\label{Sec2.5}

We calculated the energetic proton fluences at the three locations based on intensity data and event duration separately for STEREO-A, -B, and SOHO. The energy range taken into consideration extended from $\approx$14 MeV to $\approx$100 MeV (13.6 MeV to 100 MeV for HET, 13.8 MeV to 101 MeV for ERNE).

For the purpose of fluence calculation, the SEP event durations must be known. We adopted the criterion and methods described in \citet{Paassilta2017}: essentially, an SEP event is considered to have ended as soon as the proton intensity in a medium-energy (in this case, 12.6\textendash 13.8 MeV) channel falls below twice its average pre-event value (or the quiet-time background of that channel, if the pre-event average intensity is very high or cannot be reliably calculated due to missing data) for the first time since the reference channel onset. In case a new high-energy proton event begins before the conclusion of the previous one, the end of the previous event is taken to coincide with the onset of the next. The data are also visually scanned for gaps of more than about an hour, during which substantial new proton activity may have occurred, and also for subsequent low-energy proton events that are not detectable in the energy channel used for overall event identification. If gaps or additional events are found to be present, the end of the $>$55 MeV proton event being investigated is manually adjusted to exclude them. Thus, the event durations reported in \citet{Paassilta2017} were used here for ERNE and the near-Earth proton fluences calculated based on them. The same end criterion was applied by us to STEREO/HET data, with the exception that as a proton energy channel directly corresponding to the 12.6\textendash 13.8 MeV channel of ERNE was not available, the HET 13.6\textendash 15.1 MeV channel was used instead. As most of the particle fluence in an event occurs at and near the peak and as the intensities typically return to near-background levels at higher energies earlier than at lower energies after SEP events, and with any subsequent event excluded from the time range of interest, this compromise is not expected to have a significant effect on the calculated fluences. Nevertheless, it means that the reported event durations may not be in all cases directly comparable between ERNE and HET-A and -B.

The estimated quiet-time background intensities for each energy channel were removed from the data first, and then any data gaps, including intervals with known issues with the detectors, were compensated with logarithmic interpolation which relied on the immediately surrounding good data points. This method, however, usually leads to great uncertainties and highly doubtful results if major data loss or corruption has occurred at the early stages or near the intensity peak of an event, and thus fluence estimates are omitted for all such cases.

ERNE underwent periods of saturation during event 13 (7 March 2012). The fluence value for this event was deemed reasonably reliable but somewhat uncertain, and it is given in our event listing, but it is marked to that effect and should be treated with proper caution when analysed.

\subsection{Soft X-ray, Extreme Ultraviolet, and CME Data and SEP Associations}
\label{Sec2.6}

While a list of the SEP/flare and SEP/CME associations including most of the events studied in this work has been previously published in \citet{Paassilta2017}, we took the opportunity to revisit these. Additionally, we checked our results against the event catalogue provided in \citet{Richardson2014} as far as it was possible. In particular, the information therein regarding events with a solar farside source proved valuable to our work.

The associations were deduced by investigating the available X-ray, CME, and radio data for a period of some 12 hours before the $>$55 MeV proton onset at the first spacecraft to detect a given SEP event. $>$20 MeV proton events are in general accompanied by a CME and a type III radio burst (\citealp{Cane2002}), and also typically by a solar X-ray flare. Therefore, the co-occurrence of the early expansion of a CME, the onset of a type III burst, and the onset of an X-ray flare from a few tens of minutes to a few hours before the detected onset of the SEP event at 1 AU was taken as an indication that they were possibly associated with one another and the SEP event.

The primary source of CME-related information for this work was the CDAW SOHO LASCO CME Catalog (\citealp{Gopalswamy2009}; available at \url{http://cdaw.gsfc.nasa.gov/CME_list/}), which is based on data recorded by the \textit{Large Angle and Spectrometric Coronagraph} (LASCO; \citealp{Brueckner1995}) of SOHO. These were supplemented with CME observations by the \textit{Coronagraph 1} and \textit{2} (COR1, COR2) imagers of the \textit{Sun Earth Connection Coronal and Heliospheric Investigation} (SECCHI; \citealp{Howard2000}) instrument set, carried by both STEREO spacecraft; the data are collated in the Dual-Viewpoint CME Catalog from the SECCHI/COR Telescopes (\citealp{Vourlidas2017}; available at \url{http://solar.jhuapl.edu/Data-Products/COR-CME-Catalog.php}). In conjunction with the CME data, we also made use of radio frequency observations by \textit{Wind/WAVES}\footnote{These data can be accessed at \url{http://cdaweb.gsfc.nasa.gov}; on the main page, select "Wind" and "Radio and Plasma Waves (space)", then "Wind Radio/Plasma Wave, (WAVES) Hi-Res Parameters".} (\citealp{Bougeret1995}) and combined radio spectrograms produced from both \textit{Wind/WAVES} and STEREO/\textit{WAVES} (\citealp{Bougeret2008}) data at the Radio Monitoring website (\url{http://secchirh.obspm.fr/index.php}).

For information on recorded solar X-ray flare activity, we relied mostly on \textit{Geostationary Operational Environmental Satellite} (GOES) soft X-ray intensity data, together with United States National Oceanic and Atmospheric Administration (NOAA) flare listings based on GOES observations (available online at \url{http://www.ngdc.noaa.gov/stp/space-weather/solar-data/solar-features/solar-flares}). Typically, the NOAA listings directly provided the pertinent flare parameters (time of onset, classification, and location), aside from the time derivative maxima of the soft X-ray intensity, which were calculated by us from GOES one-minute X-ray intensity data smoothed with five-minute sliding average. The coordinates for the SEP-associated flare were not directly available in the NOAA X-ray or H-alpha listings for events 1, 2, 3, 7, 20, 28, 31, 37, 38, and 40. For these events, the position given in the SolarSoft Latest Events Archive (\url{http://www.lmsal.com/solarsoft/latest_events_archive.html}) was substituted.

The moment of the peak of the initial acceleration of energetic particles in magnetic loops at the Sun was assumed to coincide with the local maximum of the time derivative of the soft X-ray flux. This interpretation is based on the mechanism of the so-called Neupert effect (after \citealp{Neupert1968}; for a more modern treatment and details, see for instance \citealp{Veronig2002}), whereby non-thermal accelerated electrons lose energy through bremsstrahlung when they encounter dense surrounding plasma, which is then heated up quickly. If open magnetic field lines are present at the site of the acceleration, the electrons are released into interplanetary space. We assume that proton release generally also occurs at or after this time.

The CDAW SOHO LASCO CME Catalog was initially searched for CME activity that had occurred during the \textit{circa} 12 hour period preceding the estimated particle event onset at 1 AU. If a visual scan of \textit{Wind/WAVES} radio frequency data for the relevant time period confirmed the presence of a substantial type III burst during the (notional) solar surface departure or early expansion of the CME, as estimated with linear and quadratic fits to observed CME distances from the solar centre, the CME was tentatively chosen for analysis. The NOAA flare listings were scanned for concomitant X-ray flare activity. If a flare rated as C1.0 or greater in the NOAA classification system with the soft X-ray intensity time derivative maximum approximately coinciding (time difference of not more than $\approx$20 minutes) with the type III burst was recorded, it was considered a match, providing further support for the associated CME and radio event identification. Whenever possible, the CME and radio burst information were checked against those given in the Dual-Viewpoint CME Catalog from the SECCHI/COR Telescopes and the Radio Monitoring website, respectively, with the latter primarily used to identify such type III bursts detected by the STEREO/\textit{WAVES} instruments that might point to separate particle injections as sources for the intensity enhancements at different spacecraft. If such bursts were found to be present, the CME, radio frequency, and X-ray observations probably related to the event were revisited in the light of this information and the event discarded from the listing if it was found to be doubtful as a multi-spacecraft case.

The NOAA X-ray flare lists, however, only provide information about flares that occur on the visible solar disk or over either the eastern or the  western solar limb. If the longitude of the event source region is considerably more than some 100 degrees in either direction, the flare cannot be seen directly from the vicinity of the Earth and is therefore not included in the listings. For this reason, solar farside flare activity 
had to be considered separately in this work. In several cases, SEP event-related flare activity on the solar farside was deemed likely due to a lack of identified GOES flares close to the estimated solar release time of the SEPs, small or absent SEP intensity enhancement at spacecraft magnetically connected to the visible solar disk, weak type III radio burst detected by near-Earth spacecraft (in comparison to other observers), and other similar considerations. Such events were investigated by identifying the CME likely associated with the SEP event using the other data sources mentioned above and then referring to the combined STEREO/SECCHI/COR1 and STEREO/SECCHI/\textit{Extreme Ultraviolet Imager} (EUVI) time-lapse movies included in the STEREO COR1 CME Catalog (available at \url{https://cor1.gsfc.nasa.gov/catalog/}). Using the STEREO/SECCHI EUVI images on the 30.4 nm bandpass, we attempted to identify the extreme ultraviolet (EUV) brightening indicating the farside X-ray flare associated with those CME/SEP events for which no GOES flare appeared to be a reasonably likely match. The flare/CME/SEP association in these cases was deduced based on the timing of the image showing a large and dynamic EUV brightening, \textit{i.e.} the brightening was required to have occurred before but close to the first detection of the CME by either SOHO/LASCO or STEREO/COR1, as well as near the onset of the SEP event-associated type III radio burst, if applicable.

When a likely EUV brightening candidate was identified, we attempted to locate its centre and obtained its coordinates in pixels from the bitmap image which was deemed to show the active region of interest in clearest detail during the flare activity. The picture coordinates were then  transformed into heliographic coordinates, with the location of the observing spacecraft also being taken into account. However, the observed feature was always assumed to be near the surface of the Sun and so any projection effects caused by its height were neglected.

While our listing of the solar associations of the SEP events is, as expected, very similar to that given in \textit{e.g.} \citet{Richardson2014}, some differences may be highlighted here. Aside from the fact that the authors of that article relied on the CACTUS LASCO CME database (http://sidc.oma.be/cactus/) and real-time LASCO halo CME reports for events after June 2013, resulting in different estimated CME speeds from those given in the CDAW SOHO LASCO CME Catalog, the reported longitudes for a few of the event-related flares are not identical. Most notably, there is a disagreement as to the longitudes of the flares associated with events 26 (NOAA: W15, SolarSoft: W87; \citealp{Richardson2014}: W70) and 30 (NOAA: E44, SolarSoft: E87; \citealp{Richardson2014}: E96) in the sources used by us. For consistency, our catalogue shows the location given in the NOAA GOES X-ray flare listing. A similar situation, albeit with a much lesser difference, exists for event 7.

\subsection{The Event Catalogue}
\label{Sec2.8}

The catalogue is given in Tables \ref{solar_obs} through \ref{sep_electrons_1}. For ease of reference, every table contains the event identification number (ID) and the calendar day of the event (Date). The entries other than ID, Date, and solar observations are repeated for each observing spacecraft. All times are UT. The longitudinal distance between the event flare and observer is given as positive if the flare is west of the observer and negative if it is east of the observer; similarly, the distance between event flare and the footpoint of the nominal Parker spiral connected to the observer, \textit{i.e.} the connection angle, is positive if the flare is west of the footpoint and negative if it is east of the footpoint. The tables are structured as follows.

\begin{itemize}
\item Table \ref{solar_obs}: onset time, time of steepest increase of the soft X-ray intensity (Max. d$I$/d$t$), latitude (Lat., in degrees), longitude (Long., in degrees), and NOAA classification of the event-related solar flare; the time of first observation (1st obs.), estimated radial speed ($v$), width and position angle (in degrees) of the event-related CME (from the CDAW SOHO LASCO CME Catalog); and the time of first observation (1st obs.) and estimated radial speed ($v$) of the CME derived from combined observations (see further below for explanation), together with the name of the spacecraft (S/C) which recorded the data used as the basis of the combined results.
\item Table \ref{sep_protons_1}: time of proton event onset, longitudinal distance between the event flare and the observing spacecraft ($\Delta \phi$, in degrees), proton event maximum intensity ($I_{\rm max}$, in pfu MeV$^{-1}$), proton event rise time (in hours, not listed for SOHO/ERNE), proton event duration (Dur., in hours), and $\approx$14 MeV to $\approx$100 MeV proton fluence during the event (in cm$^\textrm{-2}$ sr$^\textrm{-1}$);
\item Table \ref{sep_protons_2}: time of proton event onset, connection angle ($\phi_{\rm C}$, in degrees), VDA apparent path length ($s$, in AU), VDA particle release time ($t_0$, with light travel time of 500 seconds added), the square of the sample correlation coefficient $R^2$ for the VDA fit, and time of steepest increase of the soft X-ray intensity (Max. d$I$/d$t$) of the event-related flare;
\item Table \ref{sep_protons_3}: time of proton event onset, connection angle ($\phi_{\rm C}$, in degrees), spiral field line length $L$ (in AU), TSA proton release time $t_{\rm rel}$, and time of steepest increase of the soft X-ray intensity (Max. d$I$/d$t$) of the event-related flare;
\item Table \ref{sep_electrons_1}: time of electron event onset, longitudinal distance between the event flare and the observing spacecraft ($\Delta \phi$, in degrees), electron event maximum intensity ($I_{\rm max}$, in pfu MeV$^{-1}$), electron event rise time (in hours), the spiral field line length $L$ (in AU), and the TSA electron release time $t_{\rm rel}$; also electron intensity data type used for ACE/EPAM (LF = LEFS60 or DE = DE30).
\end{itemize}

Both VDA and TSA particle release times as well as event rise times are given with error limits. In the case of VDA, the limits were derived from the standard errors of the linear fit parameters. For TSA and the event rise times, we estimated the error due to event onset timing uncertainty by deriving two values for the onset time in the reference energy channel, using both two-sigma and 2.5-sigma shift criteria with the Poisson-CUSUM method (see \citealp{Huttunen-Heikinmaa2005}, p. 675, for explanation of the onset criterion), calculating their difference, and multiplying it by 2. This estimate was then combined with the worst case error in the particle travel time, $\Delta L/v$, where $\Delta L$ is the uncertainty in the spiral field line length $L$, taken to be $\Delta L \approx \pm$0.4 AU, and $v$ is the mean speed of the particle species in the energy channel of interest. $\Delta L/v$ is thus $\approx \pm$5 minutes for all electron observations and $\approx \pm$10 minutes for all proton observations (rounding up to the nearest integer). Event rise time error limits combine the onset time uncertainty with $\pm$5 minutes due to a sliding 5-minute average having been used to locate the event maximum intensity.

In total, our catalogue comprises 46 multi-spacecraft events. However, not all spacecraft observed all of these events. What is more, in some cases an intensity enhancement for one particle species was detected and classified as an SEP event at a given location while the accompanying intensity enhancement for the other species was either masked by previous SEP activity or was too small to fulfill the selection criterion. A typical such case is event 26 (22 May 2013), in which proton and electron events are identified for near-Earth observatories and STEREO-A, but only the SEPT electron intensities point to the same event having occurred at STEREO-B, the HET proton intensity in the 40\textendash 100 MeV energy range remaining below our event classification criterion at that location for the duration of the event. Conversely, a proton-electron event is known to have occurred near the Earth on 8 July 2012, and a proton event is indicated for STEREO-A; however, an earlier electron intensity rise recorded by STEREO-A/SEPT on that date must, on the basis of timing and radio observations, be related to a different particle injection, and the electron component of the actual multi-spacecraft event, if present in the data, is masked by this earlier rise. Nevertheless, since a criterion for SEP events to be included in this study is that significant proton enhancements were detected at a minimum of two spacecraft (see Section \ref{Sec2.1}), no events with only electron enhancements are present in our catalogue.

Of the 46 listed proton events, 16 were detected at all three locations and the remainder at two locations: nine both near the Earth and at STEREO-A, eight near the Earth and at STEREO-B, and 12 at both STEREO-A and -B but not near the Earth. The one remaining event, that of 15 May 2013, was observed near the Earth by SOHO/ERNE and possibly at STEREO-B; it is included due to electron data confirming an SEP event at STEREO-B on that date. For electron events, the numbers are 17 (three locations), four (near the Earth and STEREO-A), eight (near the Earth and STEREO-B), and 12 (at both STEREO-A and -B). For both particle species, the proportion of three-observer events to all events is thus somewhat more than 1/3.

Considering the observations by individual spacecraft, 34 proton events were recorded by SOHO/ERNE, 37 by STEREO-A, and 36 by STEREO-B. In addition to this, there were two cases (one for each STEREO-A and -B) where the proton event may have been present but was obscured by previous SEP activity. Two events (2 and 7 November 2013) could not be investigated for SOHO/ERNE because of a data gap, and another two for STEREO-B (13 December 2014 and 1 July 2015) due to the same reason. In nine cases for SOHO/ERNE and four cases for STEREO-A, a proton event was found to have occurred but its precise onset time could not be determined due to partially missing data. SOHO/ERNE data, considered in isolation, show a total of 62 proton events in the 55\textendash 80 MeV energy channel during the same time period (\citealp{Paassilta2017}), so some 55\% of these were in fact multi-spacecraft events with a large longitudinal width, according to the definition adopted in this work (\textit{i.e.} $>$45\textdegree).

A similar tally for electron events, again considered individually by spacecraft and independently of the proton events, gives 33 cases for ACE/EPAM, 34 for STEREO-A/SEPT, and 37 for STEREO-B/SEPT. An elevated background may have masked an electron event in seven cases for ACE/EPAM, in four cases for STEREO-A/SEPT, and in three cases for STEREO-B/SEPT. No data were available for five possible electron events at STEREO-A and two at STEREO-B; however, at least four of those at STEREO-A may have been actual electron events since notable proton intensity enhancements, which electron intensity enhancements usually appear to accompany, were detected at those times.

All the listed events, aside from event 46, are reasonably securely associated with an identified solar flare. In 26 cases, the flare occurred on the visible solar disk and has a GOES classification; the other 19 events involve a farside flare. For the latter, the exact times of onset and soft X-ray flux time derivative maximum, along with magnitude estimates, are not available in our catalogue. GOES soft X-ray data are completely missing for the event of 1 July 2015 (event 46), so no conclusive determination of the event-associated flare was made. However, it is noted that the CME first appearing over the western solar limb, together with lack of EUV activity on the visible disk, suggest a farside flare in this case, as well.

A probable CME match was found for every event. While our event listing mainly relies on CME information from the CDAW SOHO LASCO CME Catalog, the observations of the same CME from multiple vantage points were combined whenever pertinent information was available in the Dual-Viewpoint CME Catalog. The combination was done so that the earliest moment of first detection and the highest radial speed estimate from the three observing instruments (SOHO/LASCO, STEREO-A/SECCHI/COR, STEREO-B/SECCHI/COR) were selected for each entry. However, as the methods applied in deriving CME width and position angle estimates differ somewhat between the CDAW SOHO LASCO CME Catalog and the Dual-Viewpoint CME Catalog, these were omitted. Both the parameters from the CDAW SOHO LASCO CME Catalog and the combined CME parameters are given and analysed (separately) in the following.

The results introduced in this article may be compared to those reported in \citet{Paassilta2017}. The inclusion of data collected by the STEREO spacecraft naturally allows a detailed examination of the behaviour of peak intensities and fluences as functions of angular distance from the particle source for single events, as well as a more certain identification of the SEP event-related flares and CMEs. Of special value is the fact that the farside SEP-associated flares can be located with fair confidence, even if they cannot be examined in quite the same detail as the visible disk flares. On the other hand, the multi-spacecraft events of Solar Cycle 24, as defined in this work, are about 1/4 fewer in number than the near-Earth events listed in the aforementioned earlier paper, leading\textemdash expectedly\textemdash to poorer statistics with respect to unique particle injections.

Our catalogue suggests that there have been at least two 55\textendash 80 MeV proton events near the Earth not included in previous ERNE SEP event listings (but featured in \citealp{Richardson2014}). The multi-spacecraft events of 2 and 7 November 2013 (events 33 and 34) occurred during a long ERNE data gap that spanned the time between 29 October and 4 December that year. A visual scan of GOES intensity data for $>$10 MeV and $>$50 MeV protons shows intensity rises corresponding to the SEP events detected by the STEREO spacecraft, and so it is likely that the comparable ERNE data, were it available during the period in question, would also indicate these events. A more marginal case is that of the STEREO-A/B proton event of 11 October 2013 (event 30): it has so far been omitted from ERNE event listings due to its maximum intensity being close to the lower limit of selection and the fact that the slowness of the intensity rise precludes a reliable estimate for the time of onset with the Poisson-CUSUM method. It is, nevertheless, a candidate for inclusion in future updates and revisions of ERNE SEP event lists. We give 18:00 UT as a visually estimated earliest possible time for its onset. Additionally, while the multi-spacecraft SEP event of 26 December 2013 (event 35) does not qualify as an ERNE proton event according to our criteria, it is recognized as an energetic near-Earth electron event forming part of a larger multi-spacecraft event and is listed as such in this work.

\section{Example Events}
\label{Sec3}

In this section, the analysis performed on the multi-spacecraft SEP events is demonstrated from the viewpoint of deducing the associations between solar and particle observations. Three qualitatively different events, each involving either three or two widely separated observing spacecraft, are presented as examples, and a brief discussion of each example event in context is included. The light travel time, approximated as 500 seconds for all observer locations, is added to all particle release times mentioned in the following. The VDA apparent path lengths given for STEREO-A and -B have been scaled to the radial distance of 1 AU for the purpose of comparison with SOHO; in contrast to this, the spiral field line lengths given are the actual values for each observer location, based on the measured solar wind speed.

\subsection{Event 39 (25 February 2014)}
\label{Sec3.1}

\begin{figure}
\centering
\includegraphics[width=0.8\columnwidth]{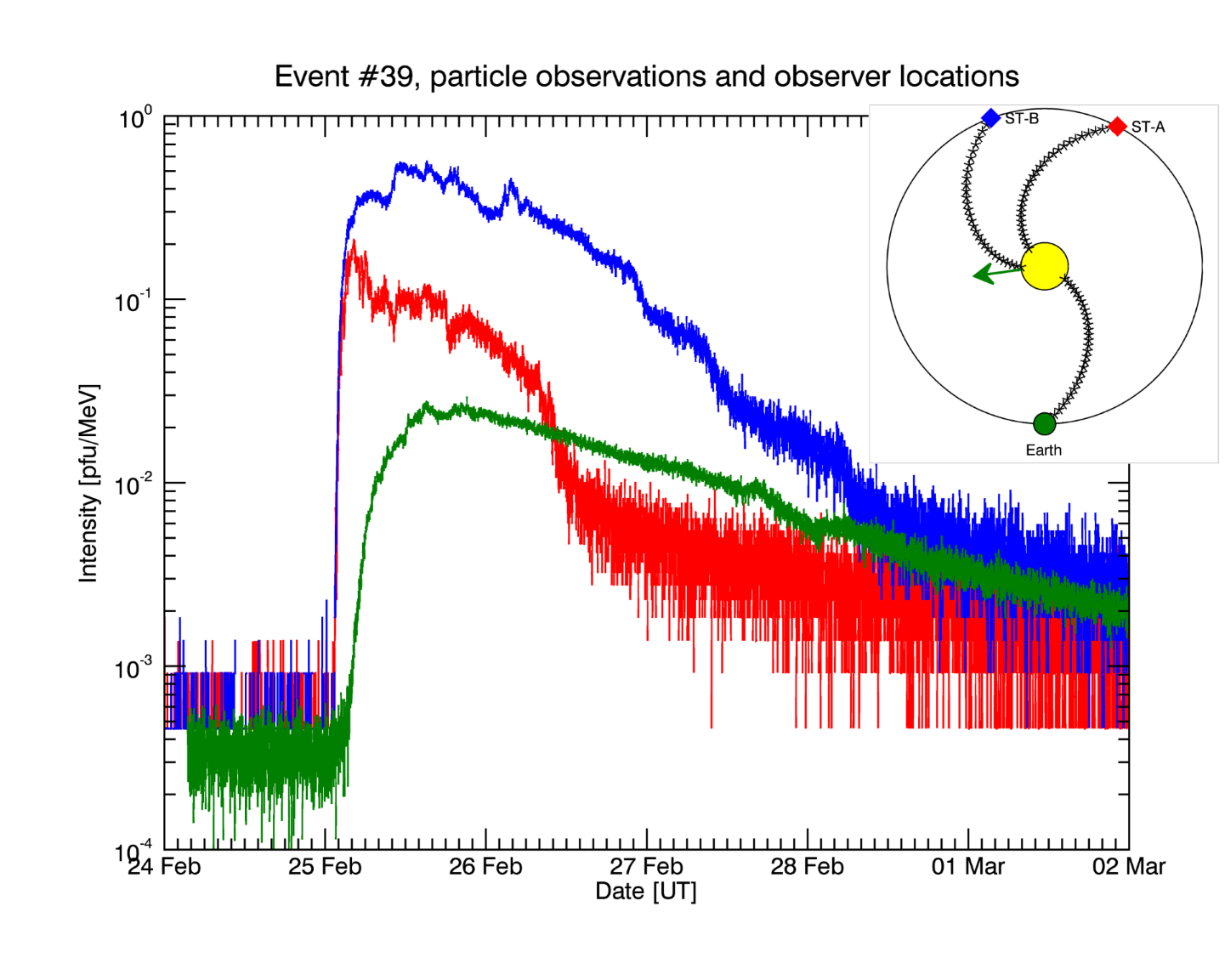}
\caption{\small Detected proton intensities (red = STEREO-A/HET, blue = STEREO-B/HET, green = SOHO/ERNE) during event 39 (25 February 2014). Inset: relative locations of the STEREO spacecraft and the Earth during the event. The arrow pointing out from the Sun shows the location of the event-related flare (at E82), and the asterisks mark the nominal Parker spiral magnetic field lines connecting each observer to the Sun.}
\label{ev_39_pos}
\end{figure}

A visual scan of the one-minute SOHO/ERNE 55\textendash 80 MeV proton intensity data and of the one-minute STEREO/HET 40\textendash 100 MeV and 60\textendash 100 MeV proton data revealed a considerable intensity enhancement that occurred nearly simultaneously at all three spacecraft during the early hours of 25 February 2014. At that time, STEREO-A was located 152.5 degrees west of the Earth, while STEREO-B was 160.1 degrees east; the 1-minute proton intensity data recorded during the event and relative positions of the Earth and the STEREOs are presented in Figure \ref{ev_39_pos}.

The Poisson-CUSUM onset determination method yielded the times of proton event onset as 3:05 UT for SOHO, 1:35 UT for STEREO-A, and 1:36 UT for STEREO-B. Consulting the CDAW SOHO LASCO CME Catalog, we find that a halo-type CME with an estimated speed of 2147 km s$^{-1}$ was first observed by SOHO/LASCO on the C3 coronagraph at 1:26 UT; a linear fit to observations, given in the Catalog, suggests that it first emerged at about 0:30 UT, this being approximately the time when the fit indicates zero altitude above the solar surface for the CME. The Dual-Viewpoint CME Catalog lists the same CME as first observed at 1:24 UT at both STEREO-A and -B, with speed estimates of 1950 km s$^{-1}$ and 2045 km s$^{-1}$, respectively, based on data from these spacecraft. This CME coincides with both an X4.9 class flare (onset at 0:39 UT, intensity time derivative maximum at 0:46 UT) located at E82 and considerable type III radio activity (commencing at about 0:46 UT) detected by \textit{Wind/WAVES}, as well as STEREO-A and -B/\textit{WAVES}. An associated type II burst is apparently also present in the summary radio data plots. While the soft X-ray intensity, as measured by GOES, remains elevated for several hours during the early part of the day, it is noteworthy that no new CMEs or substantial type III radio bursts are recorded by any instrument until close to midday. On the other hand, a C5 class flare occurred at 21:31 UT on the previous day, followed by a (possibly unrelated) CME and relatively faint type III activity about one to one and a half hours later. However, all of these are clearly too early to account plausibly for the SEP event.

A VDA fit could be obtained for all three instruments, even though the ten lowest energy channels had to be discarded from the analysis in the case of SOHO/ERNE (no onset time was obtained), as well as the lowest two (LET) channels in the case of STEREO-B/HET and STEREO-B/LET (onset times were considered too late to be certainly related to the same particle injection). The apparent path lengths indicated by VDA are 4.23 $\pm$ 0.10 AU for SOHO, 1.11 $\pm$ 0.07 AU for STEREO-A, and 1.63 $\pm$ 0.11 AU for STEREO-B; these correspond to solar release times of 1:55 $\pm$ 0:11 UT, 1:29 $\pm$ 0:04 UT, and 1:14 $\pm$ 0:05 UT for these spacecraft, respectively. The apparent path length for SOHO is fairly large and is best regarded with some caution. With the calculated spiral field line lengths (based on the solar wind speed at each spacecraft) of 1.12 AU, 1.14 AU, and 1.36 AU, again respectively, for SOHO, STEREO-A, and STEREO-B, TSA yields solar release times of 2:48 $\pm$ 00:45 UT, 1:16 $\pm$ 00:10 UT, and 1:12 $\pm$ 00:10 UT. This information shows the near-Earth observer as having the weakest magnetic connection to the source region, and the majority of the evidence would suggest the solar release of energetic protons as having occurred some time between 1:10 UT and 1:30 UT.

A very similar picture emerges when the electron observations are considered. ACE/EPAM data show an onset time at 1 AU of 2:38 UT, while STEREO-A/SEPT and STEREO-B/SEPT indicate energetic electron onsets at 1:11 UT and 1:17 UT, respectively. TSA gives the estimated solar injection times as 2:34 $\pm$ 00:06 UT, 1:06 $\pm$ 00:05 UT, and 1:10 $\pm$ 00:05 UT, again pointing at the particle release having commenced at about 1:10 UT on well connected field lines, with a delay of around 80 minutes on field lines connected to the Earth.

Thus, the logical conclusion seems to be that the SEP intensity enhancements at the three spacecraft represent a single event associated with the CME, X-ray flare, and the type III burst mentioned above. This result also agrees with that given in \citet{Paassilta2017}.

As the data coverage of this event is overall very good, parameters such as proton fluences, rise times, and maximum SEP intensities could be determined for all spacecraft. Thus, event 39 was included in various statistical considerations, presented in detail in Section \ref{Sec4}. Its measured particle intensities as a function of the connection angle can be modelled with a Gaussian curve so that the maximum height of the Gaussian corresponds to the maximum intensity of the event and the standard deviation  $\sigma$ characterises the width of the event, both at 1 AU (see Section \ref{Sec4.3} for detailed explanation and discussion). For proton intensities, the modelling results suggest $\sigma$ = 50\textdegree, together with the event centre, \textit{i.e.} the peak of the Gaussian function, appearing at a connection angle of $-$13\textdegree, \textit{i.e.} the field line footpoint is to the west of the parent flare. For electron intensities, the corresponding values are 62\textdegree \,and $-$25\textdegree; for proton fluences, 54\textdegree \,and $-$43\textdegree, respectively, all pointing to a considerable width and the tendency of the location that experiences the greatest particle intensities to be connected to the west of the event flare (negative connection angles).

\citet{Lario2016} have performed a detailed analysis of this event, concluding that particle release occurred at $\gtrsim$ 2 R$_{\odot}$ above the surface of the Sun. This accounts for the fact that the event, despite originating from a relatively limited area as determined by electromagnetic observations (EUV signatures), extended over a wide range in longitude. This event is investigated also in \citet{Klassen2016}.

\subsection{Event 42 (1 September 2014)}
\label{Sec3.2}

\begin{figure}
\centering
\includegraphics[width=0.8\columnwidth]{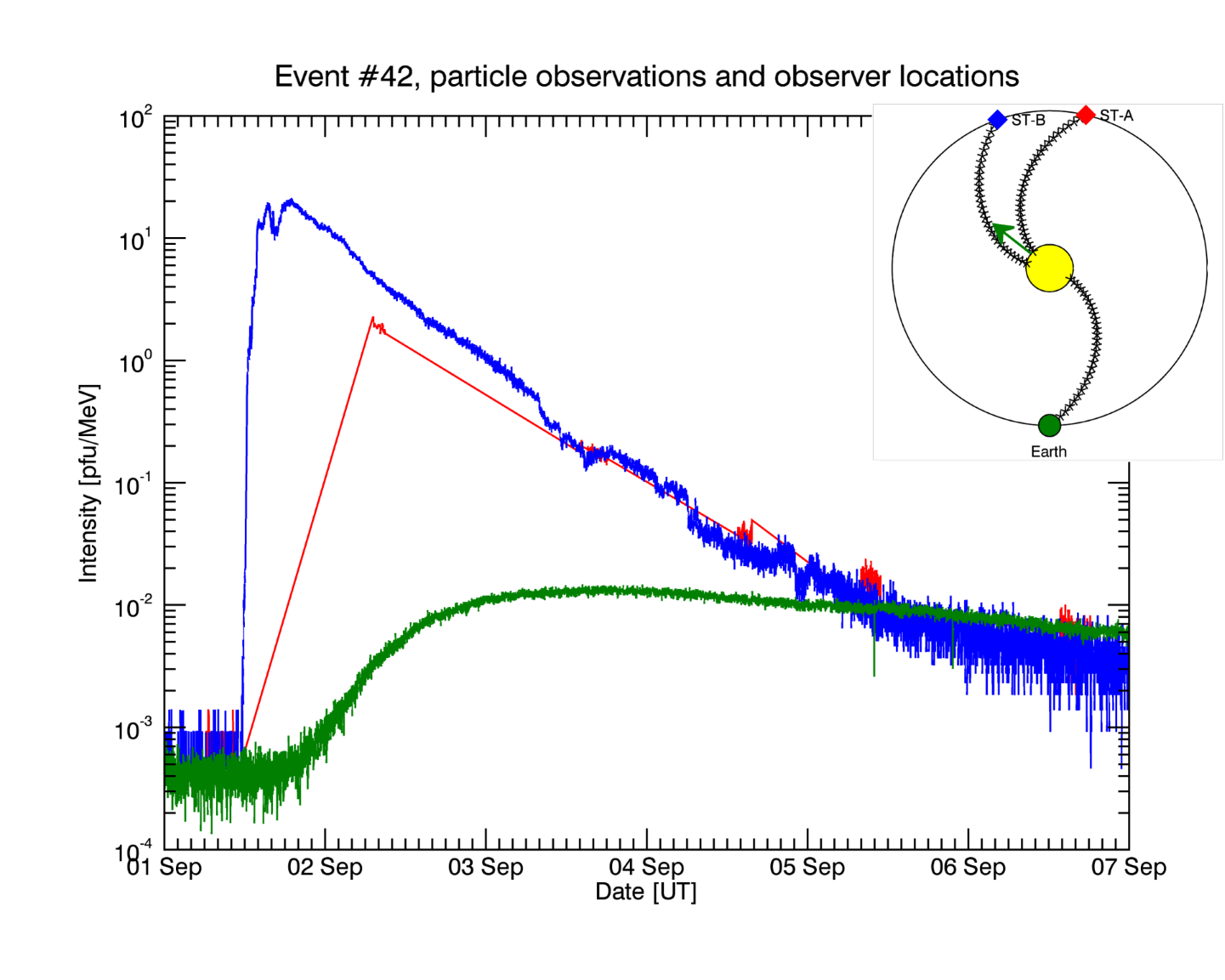}
\caption{\small Detected proton intensities (red = STEREO-A/HET, blue = STEREO-B/HET, green = SOHO/ERNE) during event 42 (1 September 2014). Note the  gaps in STEREO-A/HET proton intensity data. Inset: relative locations of the STEREO spacecraft and the Earth during the event. The arrow pointing out from the Sun shows the location of the event-related flare (at E128), and the asterisks mark the nominal Parker spiral magnetic field lines connecting each observer to the Sun.}
\label{ev_42_pos}
\end{figure}

STEREO-A was 166.7 degrees west and STEREO-B 160.7 degrees east of the Earth on 1 September 2014. Proton intensity data indicate an event on that day; onset for SOHO/ERNE is determined to have occurred at 20:41 UT and at 11:49 UT for STEREO-B/HET. There are extensive gaps in STEREO-A/HET proton data during the event, making an exact time of onset impossible to determine, and all STEREO-A/SEPT electron data for this event are unavailable. Nonetheless, an intensity enhancement some time during the day in question is clearly visible in STEREO-A/HET data, as well. Figure \ref{ev_42_pos} shows the 1-minute proton intensity data recorded by the three spacecraft, as well as the relative positions of the observing platforms.

A fast (1901 km s$^{-1}$) halo CME first detected by SOHO/LASCO at 11:21 UT is the most plausible match for this event (no valid results for the CME parameters are given in the Dual-Viewpoint CME Catalog for this event). Radio observations of type III bursts by STEREO/\textit{WAVES} and \textit{Wind/WAVES} lend support to this association, as the next earlier radio burst and CME had occurred more than 1.5 hours previously. No clear indications of substantial new radio bursts are present in STEREO-B/\textit{WAVES} or \textit{Wind/WAVES} data until 19:00 UT, after which time \textit{Wind/WAVES} data are partially missing (STEREO-A/\textit{WAVES} data are unavailable for the whole period of interest). Another CME was recorded at about 16:00 UT, but the radio data show little activity at this time.

\begin{figure}
\centering
\includegraphics[width=1.0\columnwidth]{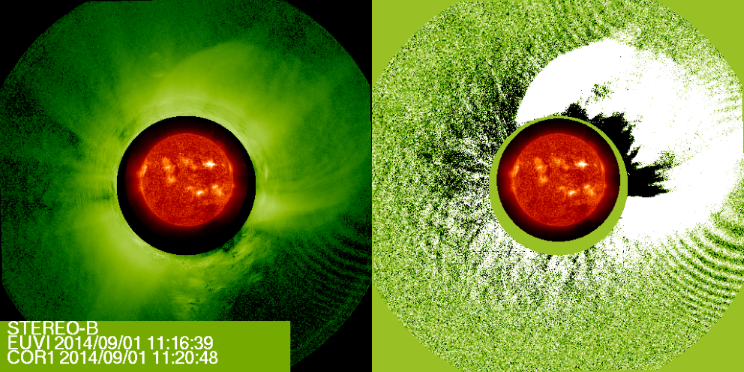}
\caption{\small Composite image recorded by STEREO-B/SECCHI/EUVI (extreme ultaviolet; centre) and STEREO-B/SECCHI/COR1, showing the farside flare associated with event 42. The right-hand panel shows a difference image of the coronagraph data. The flaring region can be seen as a large bright spot above and to the right of the solar disk centre.}
\label{EUVI_42}
\end{figure}

NOAA flare listings feature a C1.7 class flare at E05 commencing at 15:27 UT and ending at 17:11 UT. It could be related to the later CME mentioned above but very likely not to the SEP event. The EUVI movies from STEREO-B show a clear brightening and eruption at about 11:17 UT on the solar farside, somewhat beyond the east limb as seen from the Earth (E128); this coincides well with the 11:21 UT halo CME, and also with the STEREO-B particle observations. At this location, a particle source would be expected to be poorly connected with the Earth, accounting for the slow and delayed SEP event onset at both SOHO and ACE. The flare location and the CME are pictured in Figure \ref{EUVI_42}.

VDA and TSA were possible for both ERNE and HET-B but not for HET-A. VDA yields the apparent path lengths of 9.20 $\pm$ 0.27 AU and 2.42 $\pm$ 0.39 AU near the Earth and at STEREO-B, respectively, while the measured solar wind speed gives 1.17 AU and 1.18 AU as the spiral field line lengths at the same observer locations. Five energy channels of ERNE and four channels of HET-B were rejected from the VDA due to unreliable onset times; however, the VDA result for ERNE is still poor due to its very large apparent path length. The resulting proton injection times, determined using both VDA and TSA, are 11:30 $\pm$ 0:17 UT (STEREO-B, VDA), 11:29 $\pm$ 00:10 UT (STEREO-B, TSA), 17:39 $\pm$ 0:31 UT (SOHO, VDA), and 20:23 $\pm$ 00:19 UT (SOHO, TSA). For electron data, the onset at ACE is indicated at 18:07 UT and at STEREO-B at 11:32 UT; TSA yields the injection times near the Sun as 18:02 $\pm$ 00:34 UT and 11:27 $\pm$ 00:05 UT, respectively. There is thus considerable variation in the estimated solar departure times of the particles. However, the slow intensity rise in ERNE proton data suggests a delayed release of particles on the field lines connected to the near-Earth observer. We therefore give more weight to the STEREO-B data and presume that the first-detected SEPs involved in this event left the Sun some time close to 11:30 UT.

\subsection{Event 35 (26 December 2013)}
\label{Sec3.3}

\begin{figure}
\centering
\includegraphics[width=0.8\columnwidth]{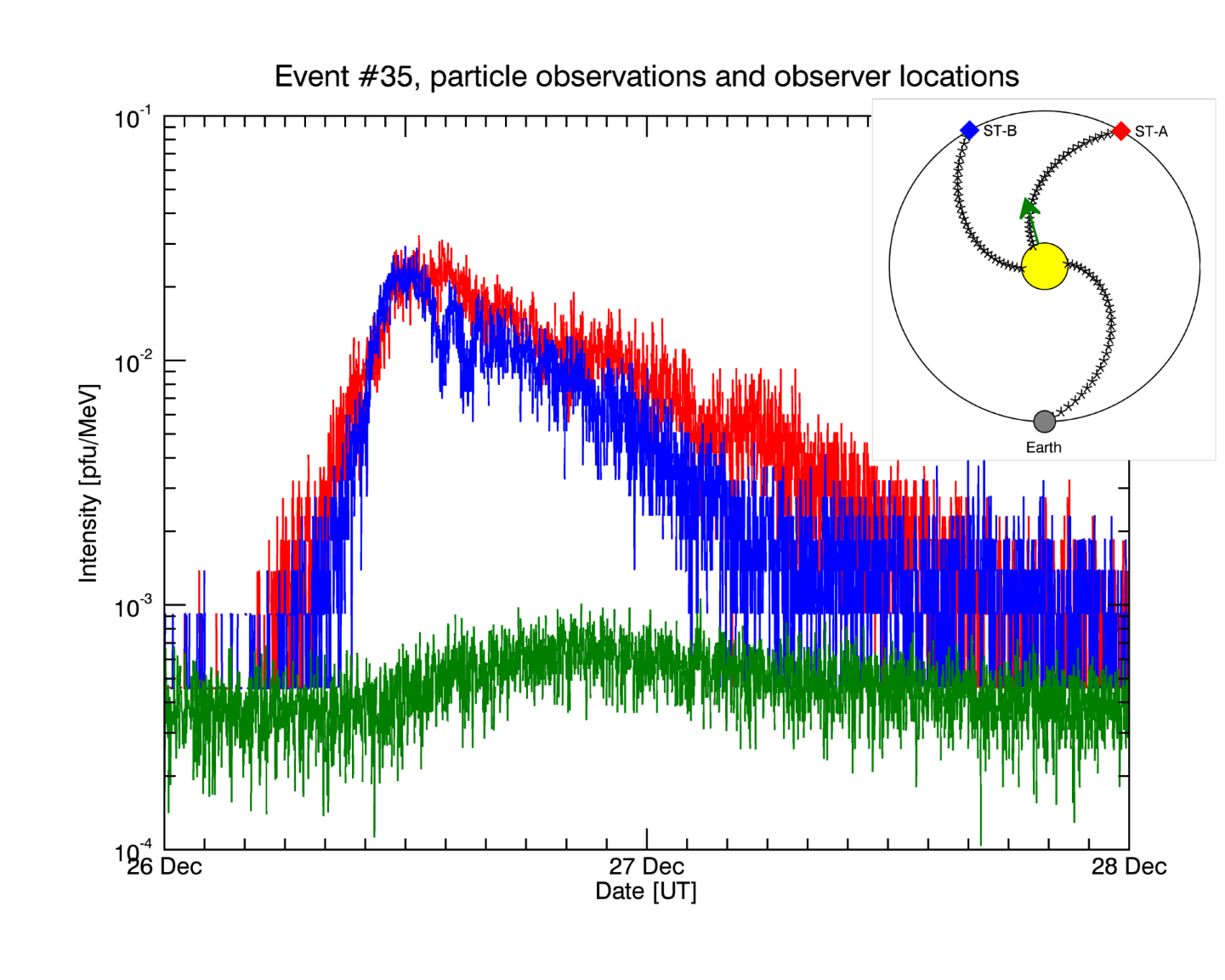}
\caption{\small Detected proton intensities (red = STEREO-A/HET, blue = STEREO-B/HET, green = SOHO/ERNE) during event 35 (26 December 2013). ERNE data show a weak enhancement that does not qualify as an event according to our selection criterion. Inset: relative locations of the STEREO spacecraft and the Earth during the event. The arrow pointing out from the Sun shows the location of the event-related flare (at E164), and the asterisks mark the nominal Parker spiral magnetic field lines connecting each observer to the Sun.}
\label{ev_35_pos}
\end{figure}

STEREO-A and -B were located 150.5 degrees west and 151.1 degrees east, respectively, at the time of this event. Proton event onset was indicated at the former at 6:04 UT and at 7:59 UT at the latter. A minor proton intensity enhancement is visible in ERNE data on this date, but it does not constitute an unambiguous SEP event according to our criteria and is therefore omitted from the listing. ACE/EPAM electron data do show an event, with onset at 5:07 UT. The corresponding electron event onset times at STEREO-A and -B were determined as 3:49 UT and 4:10 UT, respectively. The proton intensities measured by the observing spacecraft and their relative positions are shown in Figure \ref{ev_35_pos}.

\begin{figure}
\centering
\includegraphics[width=1.0\columnwidth]{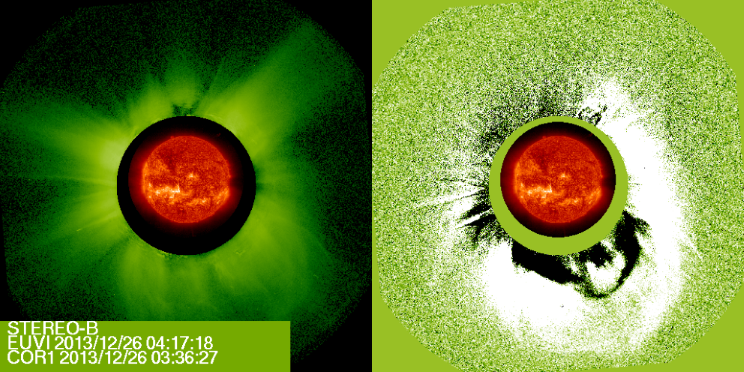}
\caption{\small Composite image recorded by STEREO-B/SECCHI/EUVI (extreme ultaviolet; centre) and STEREO-B/SECCHI/COR1, showing the farside flare associated with event 35. The right-hand panel shows a difference image of the coronagraph data. The flaring region can be seen as an elongated brightening slightly below and to the left of the solar disk centre.}
\label{EUVI_35}
\end{figure}

A halo CME was recorded by both SOHO/LASCO and STEREO-A/COR at 3:24 UT, with its speed estimated as 1336 km s$^{-1}$ (the CDAW SOHO LASCO CME Catalog, SOHO/LASCO data) or 1399 km s$^{-1}$ (the Dual-Viewpoint CME Catalog, STEREO-A/COR data); a partial halo CME immediately preceded it. Radio data from \textit{Wind} and the STEREOs show considerable type III burst activity at about 3 UT, followed by a period of no prominent radio bursts until about 6:45 UT, after which the next CME was detected. It is also noted that there seems to have been no GOES flare activity that could match a large SEP release between 3:00 UT and 4:00 UT. However, a farside brightening appears in STEREO/EUVI images between 2:17 UT and 4:17 UT (seen in Figure \ref{EUVI_35} as a thin, elongated shape near the centre of the solar disk) at about E164, and the EUVI-COR movie available at the STEREO COR1 CME Catalog would appear to support the association with this EUV feature and the halo CME mentioned above. The same identification for this event is also given in \citet{Richardson2014}.

VDA fails for both STEREOs in this case. As for TSA, spiral field line length calculation yields 1.10 AU (both ACE and STEREO-A) and 1.33 AU (STEREO-B), implying in turn injection times of 5:46 $\pm$ 00:14 UT (STEREO-A) and 7:36 $\pm$ 00:12 UT (STEREO-B) for protons and 6:23 $\pm$ 00:09 UT (ACE), 3:45 $\pm$ 00:05 UT (STEREO-A), and 4:03 $\pm$ 00:06  UT (STEREO-B) for electrons. There is thus a fair degree of uncertainty as to the time of SEP solar release, but it would seem to have occurred no later and probably not much earlier than about 3:30 UT.

The longitudinal distances between the flare and the observers imply that the event is likely to have been very weak at the Earth (if detected at all) and moderate at best at STEREO-B, but STEREO-A, in contrast, should have been relatively well connected to the source region. In the case of $>$55 MeV protons, this expectation is proven broadly correct; however, a significant electron intensity enhancement still reaches the near-Earth space.\footnote{Although SOHO/ERNE did record a minor 55\textendash 80 MeV proton intensity enhancement on 26 December 2013, it was not substantial enough to be classified as a proton event according to our criteria. We note that in lower proton energies, however, the near-Earth SEP activity on that date could have qualified as an event.} In fact, due to the electron maximum intensities being well defined, a similar analysis based on modelling the intensities with a Gaussian curve as that performed for event 39 can be carried out in this case. Event 35, considered as a whole, extended at least some 210 degrees in longitude; its maximum intensities show what might be considered a greater-than-average longitudinal spreading for three-spacecraft electron events in our energy range of interest (event width parameter $\sigma$ = 54\textdegree). Based on the Gaussian model, the event centre was at 1 AU connected to a location 33\textdegree \,west of the parent solar flare.

This event is considered in detail in \citet{Dresing2018} (submitted manuscript). The authors conclude that it is a complex case of different SEP acceleration mechanisms, highlighting the fact that relativistic electrons (which are not considered in this article) and near-relativistic electrons detected during the event form two populations, with those of higher energy arriving several hours after the lower-energy particles. It is also pointed out that the partial halo CME and the halo CME, mentioned above, interact. Such a scenario is proposed to have resulted in trapping of high-energy electrons, which were then released due to later solar activity.

\section{Statistical Results and Discussion}
\label{Sec4}

\subsection{Event Widths}
\label{Sec4.2}

Due to sparse observer coverage, the exact in situ measurement of the longitudinal width of an SEP event is currently impossible. However, to gain at least a qualitative understanding of the spatial extent of the events listed in this work, we estimated the lower limit of their width by considering for each case the relative positions of the Earth, both STEREO spacecraft, and the solar flare associated with the particle injection. A coarse estimate for the event width at $\approx$1 AU was derived by assuming that events extend longitudinally from the easternmost to the westernmost observer, with respect to the event-related flare.

Applying this method, the widest observed SEP events in our work are event 38 (7 January 2014, $>$304 degrees) and 32 (28 October 2013, $>$291 degrees), with five other events reaching widths of at least between 240 and 250 degrees. It should be noted that at least some of the events presented in this work, including events 32 and 38 mentioned above, may actually have extended a full 360 degrees in longitude.

\subsection{Longitudinal Dependence of SEP Maximum Intensities and Proton Fluences}
\label{Sec4.3}

The longitudinal distribution of SEP event maximum intensities at 1 AU can be modelled with a Gaussian function (\citealp{Lario2006}):

\begin{equation}
\label{gaussian}
I(\phi_{\rm C}) = I_0 \exp \left(- \frac{\left ( \phi_{\rm C} - \phi_0 \right)^2}{2 \sigma^2} \right),
\end{equation}
where $I_0$ is the maximum intensity at the point connected to the particle source by the Parker spiral, $\sigma$ is the standard deviation, here related to the width of the event, and $\phi_0$ is the longitude of the centre of the Gaussian distribution. $\phi_{\rm C}$ is the longitudinal angle between the flare and the Parker spiral footpoint on the surface of the Sun, also called the connection angle; it is defined here so that $\phi_{\rm C} = \phi_{\rm flare} - \phi_{\rm foot}$, with positive and negative values indicating that the flare is to the solar west and east of the footpoint, respectively. We calculate $\phi_{\rm foot}$ as

\begin{equation}
\phi_{\rm foot} = \omega_{\odot} \frac{r_{\rm SC}}{u_{\rm SW}},
\end{equation}
where $\omega_{\odot}$ is the angular speed of solar rotation at the equator of the Sun, $r_{\rm SC}$ is the radial distance of the observer from the Sun, and $u_{\rm SW}$ is the measured solar wind speed.

Events involving all three observer locations were selected for this part of our analysis, excluding, however, two cases for protons (events 13 and 32) where only a lower limit estimate is available for ERNE maximum intensity, as well as five cases for electrons (events 3, 11, 15, 18, and 36) where ion contamination has likely or possibly occurred in the electron channels of one or both of the SEPT-A and SEPT-B instruments during the event peak phase. Additionally, the maximum intensity could not be determined with any certainty due to data gaps in events 38 and 42 for STEREO-A/HET and in event 7 for ACE/EPAM, and these events were likewise omitted. This left nine proton and 13 electron three-spacecraft events available for study.

\begin{figure}
\centering
\includegraphics[width=0.7\columnwidth]{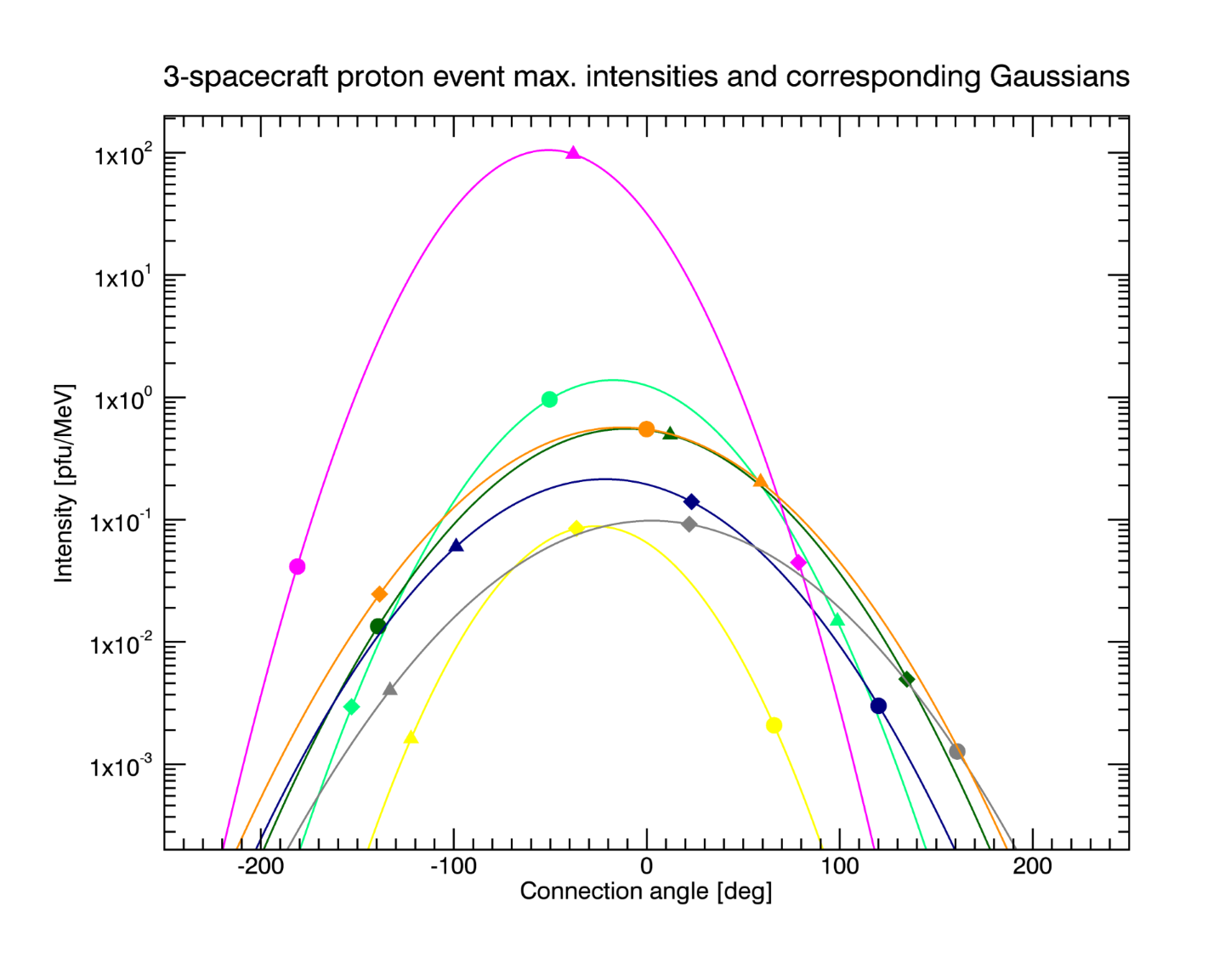}
\caption{\small Proton event maximum intensities as a function of connection angle (seven events shown). Positive abscissa values indicate that the 
particle source (event flare) is to the west of the Parker spiral footpoint connected to the spacecraft. The different symbols represent the three 
spacecraft (diamond = SOHO, triangle = STEREO-A, circle = STEREO-B), and observations of a given event are indicated with the same colour (event 5 = yellow, event 7 = light green, event 9 = dark green, event 11 = dark blue, event 18 = magenta, event 37 = dark gray, event 39 = orange). The lines show the Gaussian curves that correspond to the observed maximum intensities. Two events for which the curve parameter results were considered unreasonable (see text) are omitted.}
\label{proton_max_ints_conn_angle}
\end{figure}

\begin{figure}
\centering
\includegraphics[width=0.7\columnwidth]{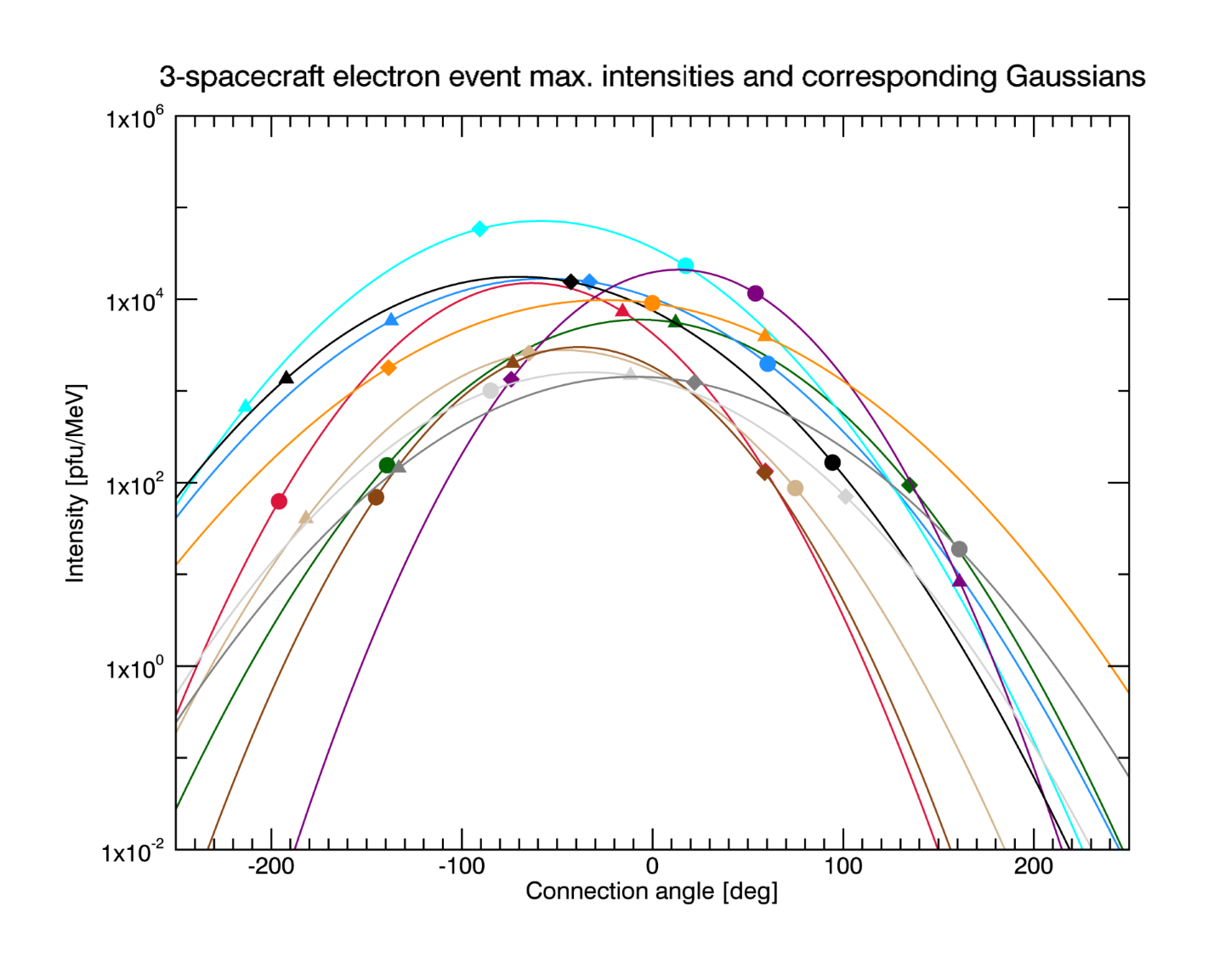}
\caption{\small Electron event maximum intensities as a function of connection angle (11 events shown). Positive abscissa values indicate that the particle source (event flare) is to the west of the Parker spiral footpoint connected to the spacecraft. The different symbols represent the three spacecraft (diamond = ACE, triangle = STEREO-A, circle = STEREO-B), and observations of a given event are indicated with the same colour (event 4 = crimson, event 9 = dark green, event 10 = light blue, event 13 = cyan, event 23 = purple, event 26 = black, event 28 = tan, event 33 = dark brown, event 35 = light gray, event 37 = dark gray, event 39 = orange). The lines show the Gaussian curves that correspond to the observed maximum intensities. Events for which the curve parameter results were considered unreasonable (see text) are omitted.}
\label{electron_max_ints_conn_angle}
\end{figure}

While the proton intensity observations were not intercalibrated between spacecraft and instruments due to a relatively good correspondence between 
the instruments on one hand and the fact that deriving a reliable  intercalibration factor is not straightforward on the other (see Appendix \ref{App1_proton_intercal}), the quiet-time background intensity was subtracted before analysis. For the ERNE 55\textendash 80 MeV proton channel, it varied between 5.5$\times 10^{-4}$ pfu MeV$^{-1}$ and 4.0$\times 10^{-4}$ pfu MeV$^{-1}$, while the corresponding values for HET-A and -B were 3.0$\times 10^{-4}$ pfu MeV$^{-1}$ and 2.0$\times 10^{-4}$ pfu MeV$^{-1}$. As for electron data, ACE/EPAM maximum intensities were multiplied by 0.84 (LEFS60) or 0.65 (DE30), depending on the data type, after an estimated yearly quiet-time background was first subtracted. This ranged from 5.0 pfu MeV$^{-1}$ to 4.0 pfu MeV$^{-1}$ for LEFS60 and from 21 pfu MeV$^{-1}$ to 15 pfu MeV$^{-1}$ for DE30. In the case of both SEPT-A and SEPT-B, the background intensity was estimated to be 2.0 pfu MeV$^{-1}$ and assumed to have remained approximately constant throughout the period of interest; it was also subtracted from the analysed maximum intensities.

\begin{figure}
\centering
\begin{tabular}{c c}
\includegraphics[width=0.4\columnwidth]{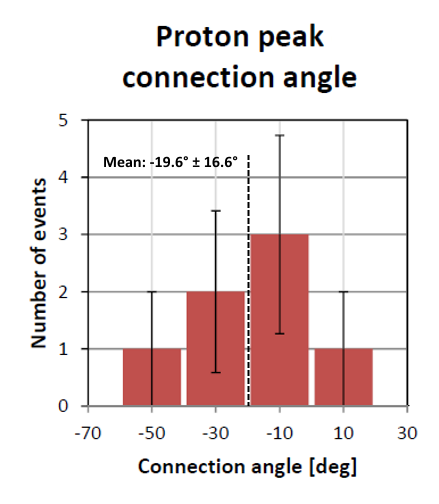} & \includegraphics[width=0.4\columnwidth]{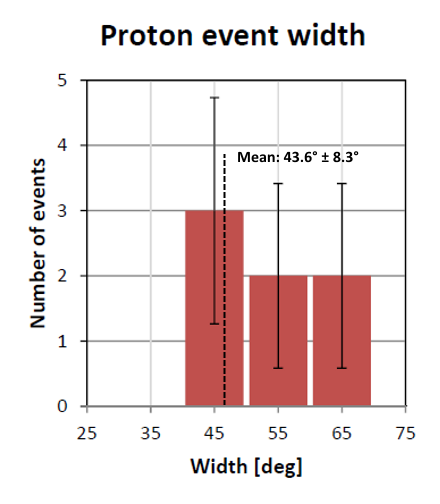}\\
\includegraphics[width=0.4\columnwidth]{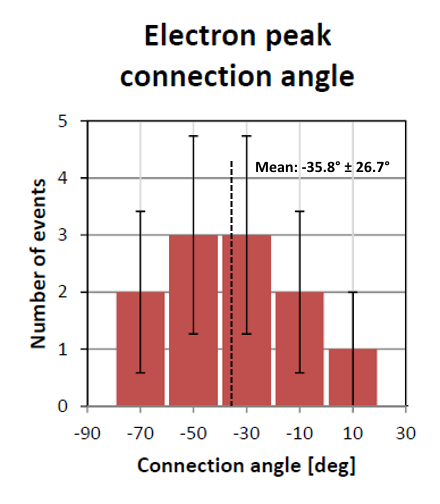} &
\includegraphics[width=0.4\columnwidth]{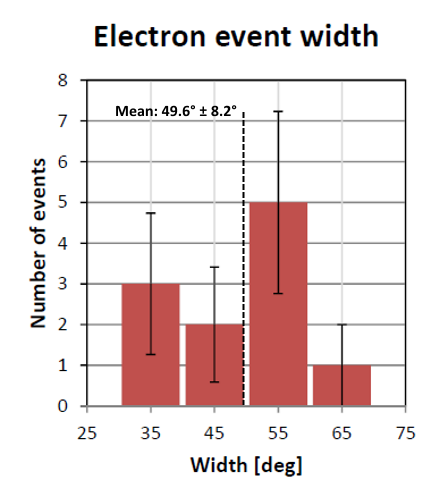}\\
\end{tabular}
\caption{\small Distributions of the connection angle of event centre and the event width for protons (top row) and for electrons (bottom row), based on Gaussian curves corresponding to the observed maximum intensities. Included here are the three-spacecraft events in Figures \ref{proton_max_ints_conn_angle} (protons) and \ref{electron_max_ints_conn_angle} (electrons). The error bars denote the statistical error, and the values on the abscissae denote the centre of each bin.}
\label{phi_sigma_distributions}
\end{figure}

In two proton and four electron events, Equation \ref{gaussian} could not be solved for all three parameters, and in two electron events, it yielded $\lvert \phi_0 \rvert >$ 90\textdegree, indicating a poor result. The connection angle of one data point in each of these events was shifted to the 
corresponding opposite longitude in an attempt to obtain more meaningful results. This was done for proton event 18 (STEREO-B: $\phi_{\rm C}$ =  179\textdegree \,to $-$181\textdegree) and 36 (STEREO-B: $\phi_{\rm C}$ = $-$141\textdegree \,to 219\textdegree), as well as electron events 4 (STEREO-B: $\phi_{\rm C}$ = 164\textdegree \,to $-$196\textdegree), 13 (STEREO-A: $\phi_{\rm C}$ = 146\textdegree \,to $-$214\textdegree), 26 (STEREO-A: $\phi_{\rm C}$ = 168\textdegree \,to $-$192\textdegree), 28 (STEREO-A: $\phi_{\rm C}$ = 178\textdegree \,to $-$182\textdegree), 30 (ACE: $\phi_{\rm C}$ = $-$108\textdegree \,to 252\textdegree), and 38 (STEREO-A: $\phi_{\rm C}$ = 151\textdegree \,to $-$209\textdegree). Nevertheless, the Gaussian curves for  proton events 3 (with $\phi_0 \gg$ 90\textdegree) and 36 ($\sigma \approx$ 92\textdegree) were still deemed doubtful and were omitted from further  analysis, as were the Gaussians for the electron events 30 ($\phi_0 \approx$ 94\textdegree) and 38 ($\phi_0 \approx -$87\textdegree).

Figure \ref{proton_max_ints_conn_angle} shows the maximum intensities and the corresponding Gaussian curves for the seven proton events where the parameter determination was considered successful. The arithmetic mean and standard deviation of the parameters yield $\phi_0$ = $-$19.6\textdegree \,$\pm$ 16.6\textdegree \,and $\sigma$ = 43.6\textdegree \,$\pm$ 8.3\textdegree. Figure \ref{electron_max_ints_conn_angle} similarly illustrates the electron events (11 cases) for which a meaningful solution to Equation \ref{gaussian} could be derived, and Figure \ref{phi_sigma_distributions} shows the distributions of $\phi_0$ and $\sigma$, as well as their mean values, for both protons and electrons. The mean and standard deviation for the electron parameters are $\phi_0$ = $-$35.8\textdegree \,$\pm$ 26.7\textdegree \, and $\sigma$ = 49.6\textdegree \,$\pm$ 8.2\textdegree. (If one slightly outlying data point\textemdash that represents event 26\textemdash is omitted from the electron data set, these become $\phi_0$ = $-$32.4\textdegree \,$\pm$ 25.4\textdegree \,and $\sigma$ = 49.2\textdegree \,$\pm$ 8.5\textdegree.) The width of the Gaussian $\sigma$ varies a great deal between events of the two particle species, but a small bias favouring negative values of $\phi_0$, and thus situations where the parent flare is located to the east of the Parker spiral footpoint, seems to be present for protons, as well as for electrons.

\begin{figure}
\centering
\includegraphics[width=0.7\columnwidth]{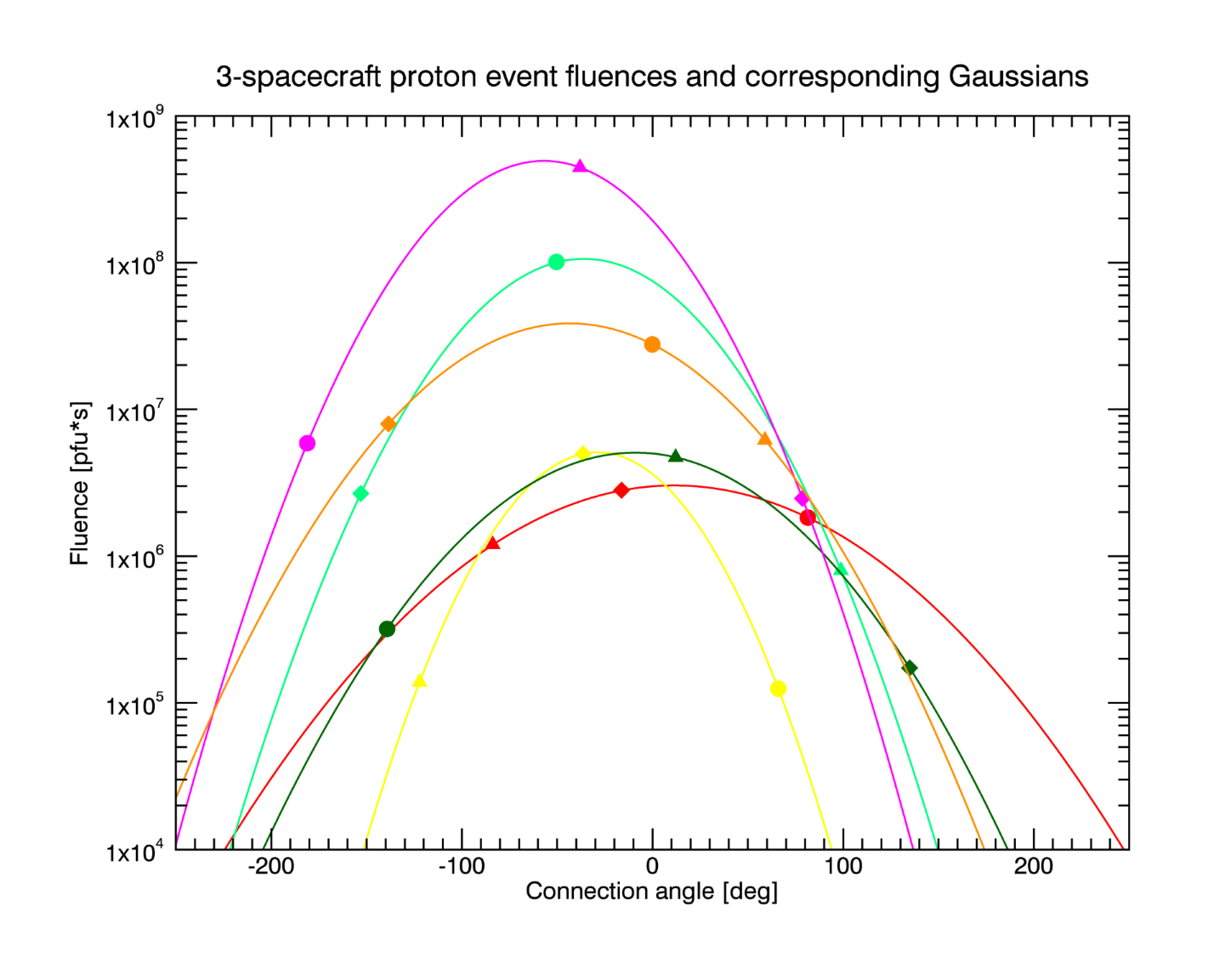}
\caption{\small Proton event fluences as a function of connection angle. Positive abscissa values indicate that the particle source (event flare) is  to the west of the Parker spiral footpoint connected to the spacecraft. The different symbols represent the three spacecraft (diamond = SOHO, triangle = STEREO-A, circle = STEREO-B), and observations of a given event are indicated with the same colour (event 3 = red, event 5 = yellow, event 7 = light green, event 9 = dark green, event 11 = dark blue, event 18 = magenta, event 39 = orange). The lines show the Gaussian curves that correspond to the observed fluences.}
\label{proton_fluences_conn_angle}
\end{figure}

A similar analysis was performed on proton fluences. Our listing includes six three-spacecraft events (3, 5, 7, 9, 18, and 39) where reasonably reliable estimates for the total proton fluences in the energy range of $\approx$14 MeV to $\approx$100 MeV were available. If the connection angle for STEREO-B in event 18 is again shifted to the opposite longitude as described above, the arithmetic mean and standard deviation for the parameters yield $\phi_0$ = $-$27.1\textdegree \,$\pm$ 22.5\textdegree \,and $\sigma$ = 49.7\textdegree \,$\pm$ 12.5\textdegree. Poorer statistics and greater 
scatter notwithstanding, these are fairly similar to the corresponding values derived for maximum proton intensities. The results are shown in Figure \ref{proton_fluences_conn_angle}.

Previously, \citet{Lario2013}, who investigated multi-spacecraft SEP events that occurred during the early part of Solar Cycle 24 (late 2009 to late 2012), obtained $\phi_0$ = $-$12\textdegree \,$\pm$ 3\textdegree \,and $\sigma$ = 43\textdegree \,$\pm$ 2\textdegree \,for 15\textendash 40 MeV proton peak  intensities and $\phi_0$ = $-$12\textdegree \,$\pm$ 3\textdegree \, and $\sigma$ = 45\textdegree \,$\pm$ 1\textdegree \,for 25\textendash 53 MeV proton peak intensities. \citet{Richardson2014}, in turn, found $\phi_0$ = $- $15.1\textdegree \,$\pm$ 35.2\textdegree, $\sigma$ = 43\textdegree \,$\pm$ 
13\textdegree \,for 14\textendash 24 MeV proton peak intensities in 2009\textendash 2013. In addition, \citet{Lario2013} report $\phi_0$ = $-
$16\textdegree, $\sigma$ = 49\textdegree \,$\pm$ 2\textdegree \,for multi-spacecraft 71\textendash 112 keV electron event maximum intensities during 
2009\textendash 2012, while \citet{Dresing2014} derive $\phi_0$ = 11\textdegree, $\sigma$ = 39.1\textdegree \,(when the results are adjusted to the same method as that used by \citealp{Lario2013}) for 55\textendash 105 keV electron maximum intensities in 21 widespread electron events that  occurred between late 2009 and mid-2013. Even though our work deals with somewhat higher particle energies, our results appear broadly similar to those of the earlier studies quoted above. In particular, the mean of the intensity $\sigma$ parameter of wide proton events lies in a fairly narrow range of $\approx$43 to $\approx$45 degrees. Both positive and negative mean values for $\phi_0$, based on maximum intensities, are reported in literature; here, a negative value, albeit with large margins of uncertainty, is obtained for protons and electrons. What is more, while the limitations of statistics force us to adopt a cautious approach to making any truly firm conclusions, there nevertheless seems to be no strong dependence of the intensity distribution parameters $\phi_0$ and $\sigma$ on the kinetic energy of the SEPs.

\subsection{Longitudinal Dependence of Event Rise Times}
\label{Sec4.4}

The event rise times, \textit{i.e.} the delay between the onset and the maximum SEP intensity, were next considered as functions of the connection angle for protons and electrons. As the maximum particle intensity calculation method employed for the other spacecraft and detectors could not be used as such for SOHO/ERNE (see Section \ref{Sec2.3}), the ERNE proton events were omitted from this part of the analysis.

\begin{figure}
\centering
\includegraphics[width=0.7\columnwidth]{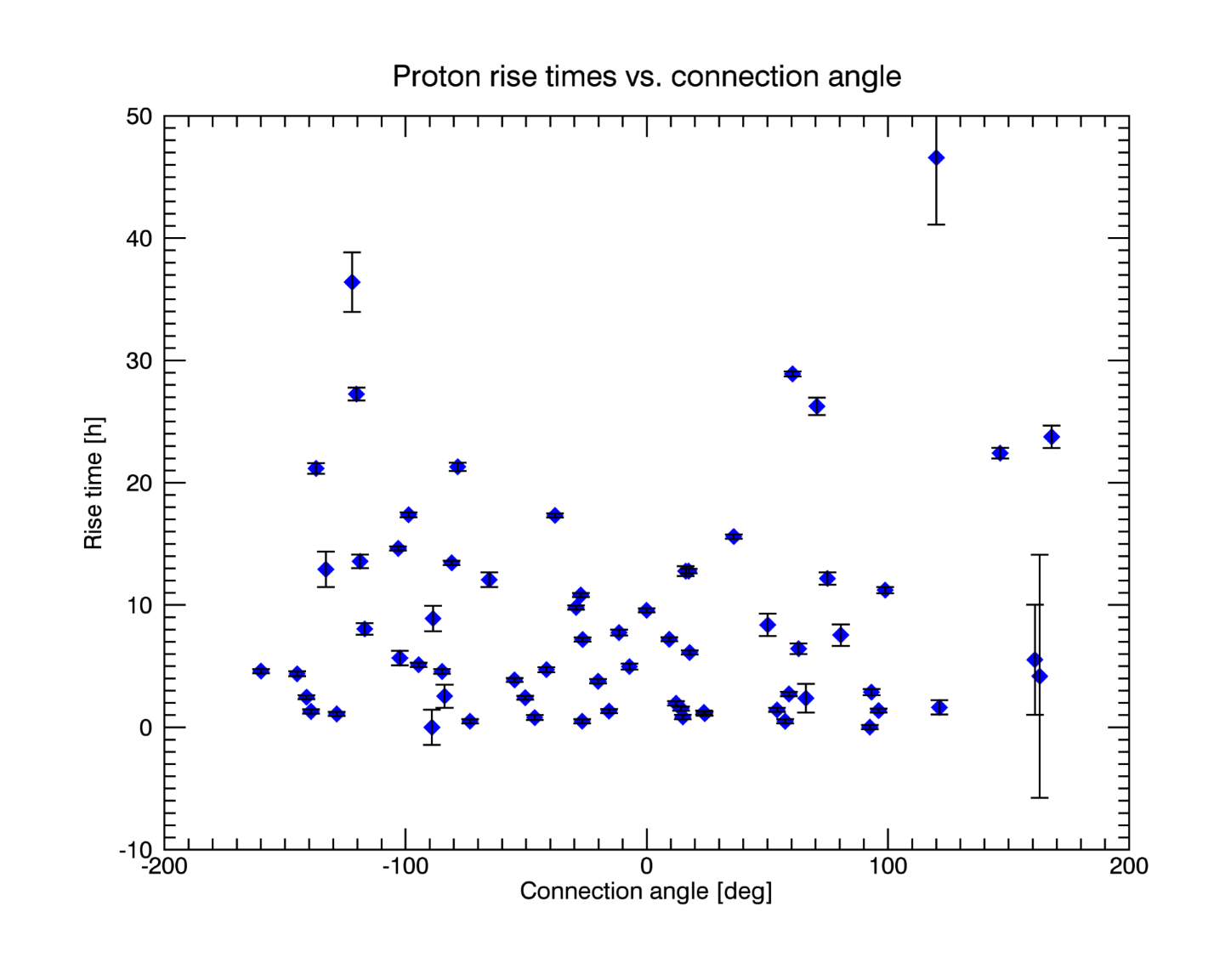}
\caption{\small STEREO-A and -B/HET proton event rise times as a function of connection angle. Positive abscissa values indicate that the particle source (event flare) is to the west of the Parker spiral footpoint connected to the spacecraft. One event with a rise time $>$48 hours is omitted.}
\label{proton_rises_vs_conn_angle}
\end{figure}

\begin{figure}
\centering
\includegraphics[width=0.7\columnwidth]{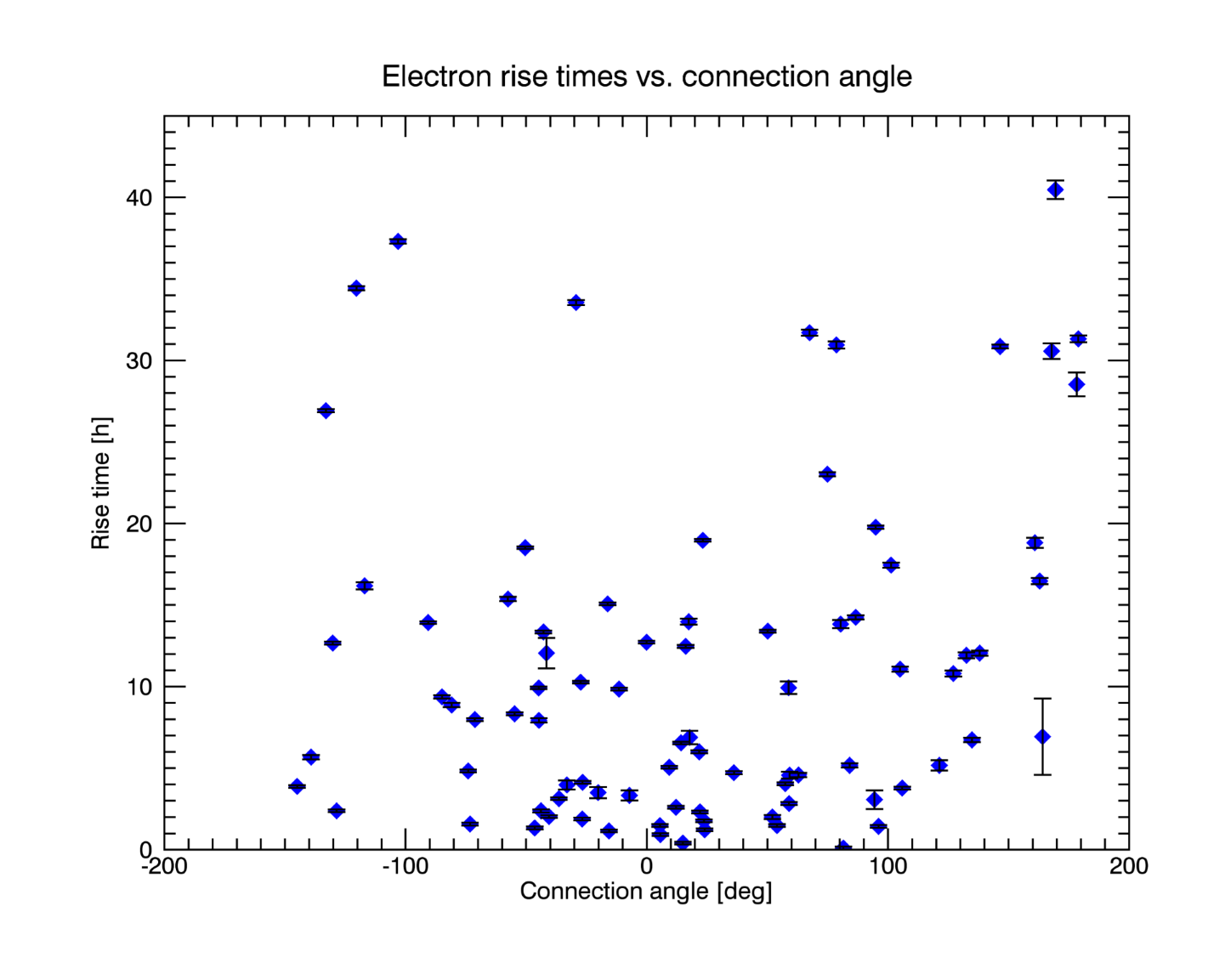}
\caption{\small Electron event rise times as a function of connection angle. Positive abscissa values indicate that the particle source (event flare) is to the west of the Parker spiral footpoint connected to the spacecraft. 20 events with ion contamination during onset or peak phase, a suspect onset time or a rise time $>$48 hours are omitted.}
\label{electron_rises_vs_conn_angle}
\end{figure}

Figure \ref{proton_rises_vs_conn_angle} shows the STEREO/HET proton events where the rise time is less than 48 hours (one event omitted); Figure \ref{electron_rises_vs_conn_angle} shows the electron events with the same upper limit for the rise time, additionally excluding those events in which ion contamination has occurred in the SEPT detectors during event onset or peak, as well as event 38 for SEPT-A, where the onset timing is suspect due to elevated pre-event background (84 events shown, 20 in total omitted). The rise time errors include the propagated uncertainty ($\pm$5 minutes) due to a sliding average having been used to locate the event peak.  The shortest rise times, broadly speaking, have a tendency to occur near the connection angle of zero degrees, but a great deal of scatter is apparent for both particle species.

While the dependence between the connection angle and event rise time seems to be slightly greater for electrons than it is for protons, it does not appear to be particularly clear or strong for either species as a whole. Rather, the distribution of event rise times seems to widen as the absolute value of the connection angle increases: the rise times appear to remain very small for a number of poorly connected events but become considerably large for others. This is further discussed in Section \ref{Sec4.6} in the context of solar release times of SEPs.

\subsection{VDA Results}
\label{Sec4.5}

Considering the observation locations separately, we investigated a total of 94 events. A VDA result was found for 74 of these.

The onset time determination, which has to be carried out with several energy channels to make the VDA possible, relies on a method that is subject to failure when, for instance, there is a high pre-event background or when the intensities of interest are low throughout the early part of the event, fluctuate considerably just prior to onset, or undergo a slow, prolonged rise. Such conditions may often lead to one or more energy channel onset times being untrustworthy. As the onset times are used as data points in the linear fit process that produces the VDA parameters, excluding suspect channel onset times can yield physically more significant fits; however, this also means that the investigator performing the analysis must make the choice of including or rejecting a given data point, which inescapably introduces a certain amount of subjectivity to the final results.

The cases with no VDA result were not included in any of the detailed statistical considerations. In addition to these, the analysis yielded 18 cases of very long ($>$5 AU) and three cases of very short ($<$1 AU) apparent path length among the 74 events. In 16 of the former and in one of the latter events, a considerable number of energy channels (at least five for STEREO/HET and LET, at least eight for SOHO/ERNE) had to be discarded from VDA because the indicated onset time was either unavailable or judged to be probably erroneous. Typically, a proton event with an exceedingly long apparent path length featured a slow and very weak intensity enhancement, especially in high energies.

\begin{figure}
\centering
\includegraphics[width=0.7\columnwidth]{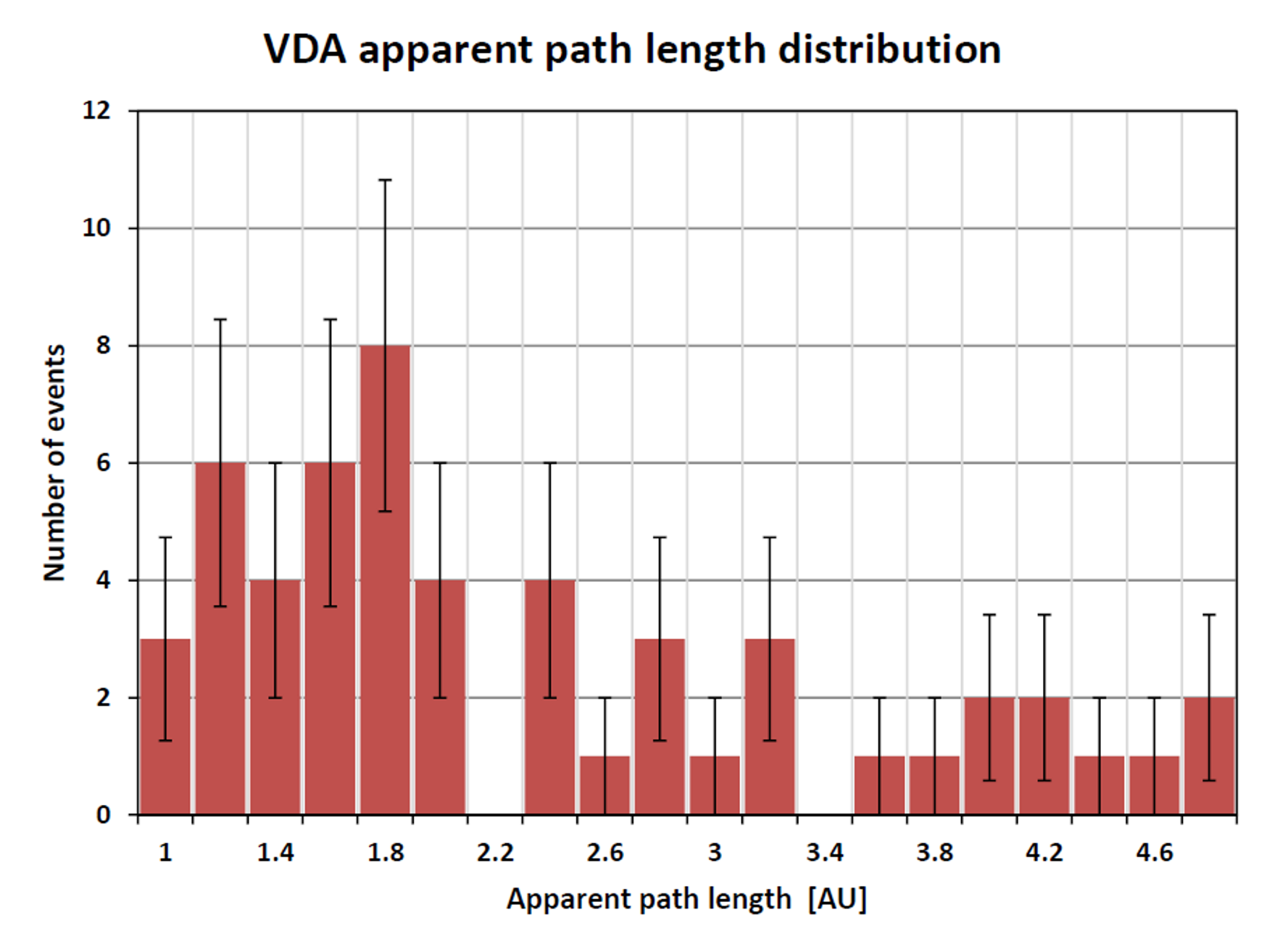}
\caption{\small The distribution of VDA apparent path length $s$, with 21 cases of $s <$ 1 AU and $s >$ 5 AU omitted. The error bars denote the statistical error, and the values on the abscissa denote the lower limit of each path length bin.}
\label{proton_VDA_histogram}
\end{figure}

The average apparent path length and its standard deviation yield $s$ = 4.31 $\pm$ 4.19 AU for all events with a VDA result, highlighting the very large scatter. If the 21 events with $s <$ 1 AU or $s >$ 5 AU are omitted from analysis, we obtain $s$ = 2.42 $\pm$ 1.10 AU. A comparison to the spiral field line length values from TSA shows that in all but five of the events with a VDA result, the apparent path length is greater than the spiral field line length. The distribution of $s$ values is shown as a histogram in Figure \ref{proton_VDA_histogram}, with cases of very small and very large values omitted.

It may be noted here that the VDA results are, on the whole, poorer than those reported by \citet{Vainio2013} or by \citet{Paassilta2017}, who performed this type of analysis only on SOHO/ERNE proton events. In the latter paper, the proportion of events with a meaningful VDA result \textit{versus} all events (including ones for which an exact time of onset could not be derived) is given as 161/176 = 91\%, while in this work the respective numbers are 74/109 = 68\%. Very long apparent path lengths are also relatively more common here. The main reason for this is the fact that the requirement for large longitudinal event width adds cases where some of the observing spacecraft are not well connected to the particle source region, resulting in delayed event onsets and low intensities, which in turn tend to make VDA considerably more challenging than in situations where the observer records a rapid and large intensity enhancement.

\subsection{Flares and Solar Release Times of Energetic Particles}
\label{Sec4.6}

An associated solar soft X-ray flare was identified for a total of 26 events. Of these, 12 belonged to the X class, 10 to the M class, and four to the C class. Additionally, there were 19 farside flares for which no GOES classification is available. GOES X-ray data do not indicate a significant flare for event 46 (1 July 2015), so this event is likely associated with farside flaring activity, as well.

A comparison to \citet{Paassilta2017}, where the same portion of Solar Cycle 24 is covered as in this paper, underscores that wide-angle SEP events tend to be associated with large flares. \citet{Paassilta2017} identified 43 GOES-classified solar flares for 62 SOHO/ERNE 55\textendash 80 MeV proton events in 2009\textendash 2016, with 12 of the flares being X class, 24 M class, and seven C class; further study of two of the events listed in that work adds two more X class flares for a total of 45 visible solar disk flares. 27 of the 62 events do not meet the multi-observer criteria, having apparently been too narrow in longitude to reach the STEREO spacecraft. These narrow events involve two X class, 14 M class, and five C class flares, \textit{i.e.} the majority of the identified moderate and weak event-related flares in 2009\textendash 2016 but only a small minority of the large ones. (Note, however, that these numbers may be slightly affected by the loss of all STEREO-B data from 1 October 2014 onward; nine SEP events listed in \citet{Paassilta2017} occurred between October 2014 and the end of 2016. Three of them are present in our catalogue.)

\begin{figure}
\centering
\includegraphics[width=0.7\columnwidth]{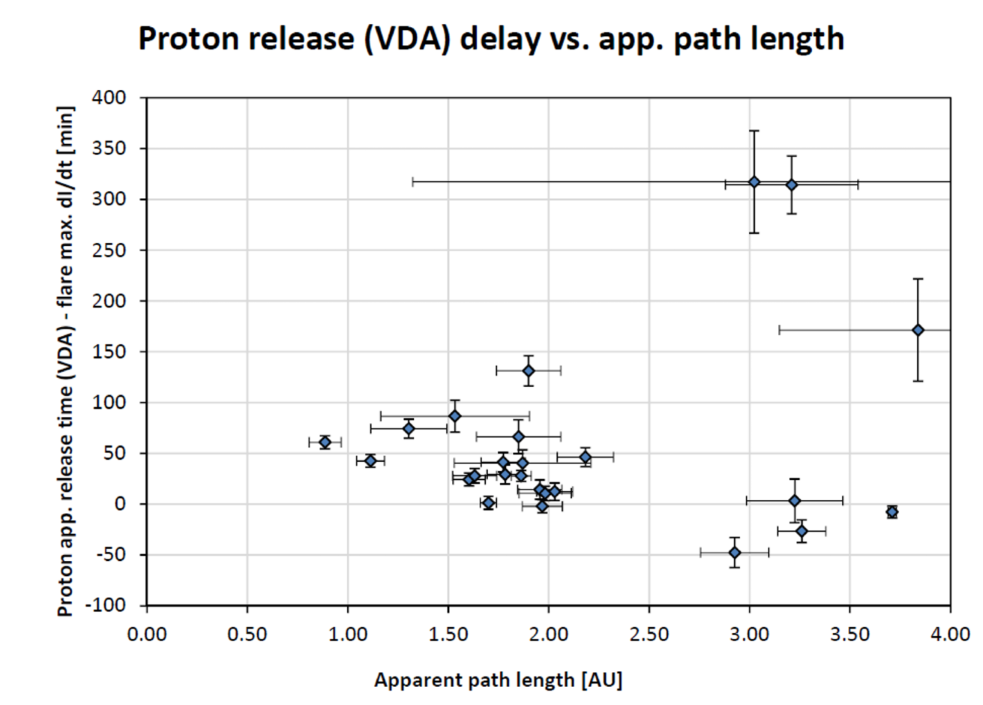}
\caption{\small Proton event release delay (from VDA; see text) as a function of VDA apparent path length $s$; events for which $s <$ 0.8 AU or $s >$ 4 AU are excluded. The release delay errors include the propagated uncertainty ($\pm$5 minutes) due to a sliding average having been used to locate the maximum slope of the flare X-ray intensity.}
\label{Proton_delay_vs_s_selection}
\end{figure}

Figure \ref{Proton_delay_vs_s_selection} shows the difference between apparent proton release time with light travel time of 8.33 minutes added, derived from VDA, and the time derivative maximum of soft X-ray intensity of the event-associated flare (here called release delay for short) as a function of apparent path length $s$. Events with $s <$ 0.8 AU or $s >$ 4 AU are omitted, leaving 27 events in total. The data points display strong scatter. It can nevertheless be seen that the events where $s$ falls into a reasonable range, \textit{i.e.} 1 AU $\leq s \leq$ 3 AU, typically show the shortest VDA release delays\textemdash that is, the closest match between the VDA- and flare soft X-ray flux-based estimates for particle release time at the Sun\textemdash as well as the least amount of scatter. This result is essentially in agreement with those reported for VDA in both \citet{Vainio2013} and \citet{Paassilta2017} (see also references in both), even though the data points available for analysis in this work are considerably fewer in number than in either of the earlier studies mentioned.

\begin{figure}%[H]
\centering
\includegraphics[width=0.7\columnwidth]{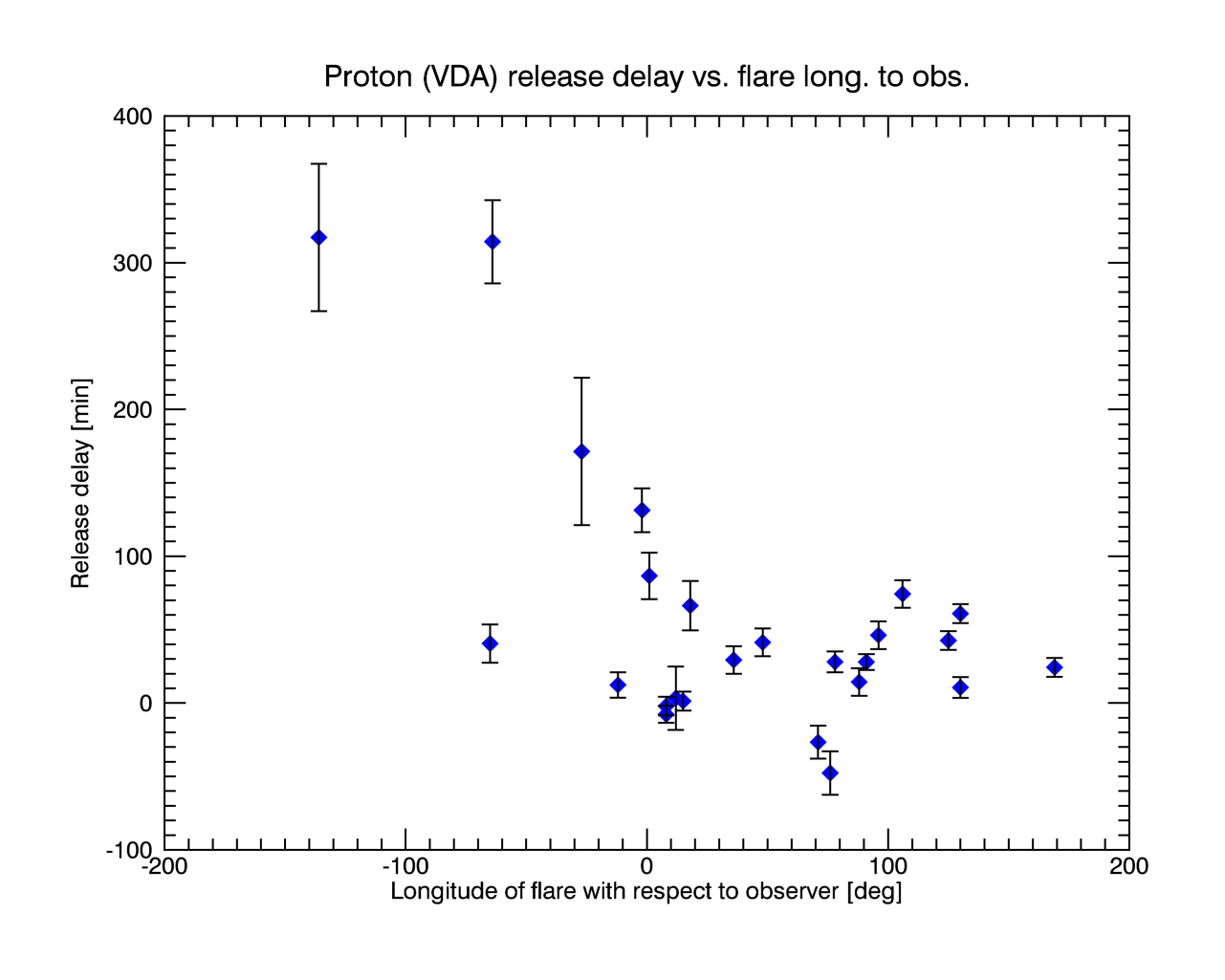}
\caption{\small Proton event release delay (from VDA) as a function of the longitudinal distance between the flare and the observer. The release delay errors include the propagated uncertainty ($\pm$5 minutes) due to a sliding average having been used to locate the maximum slope of the flare X-ray intensity. Events for which $s <$ 0.8 AU or $s >$ 4 AU are excluded.}
\label{Proton_delay_vs_phi}
\end{figure}

\begin{figure}%[H]
\centering
\includegraphics[width=0.7\columnwidth]{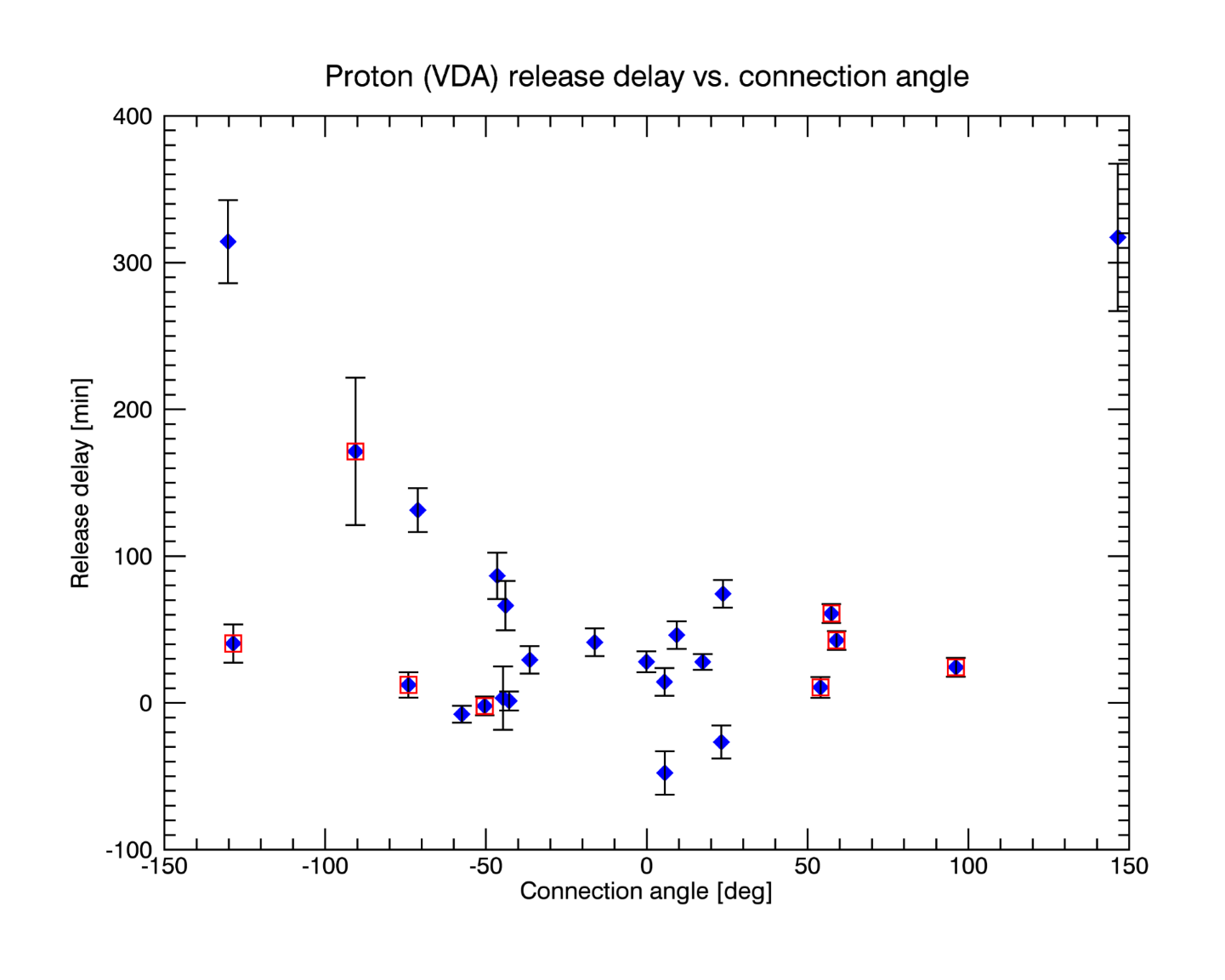}
\caption{\small Proton event release delay (from VDA) as a function of the connection angle. The release delay errors include the propagated uncertainty ($\pm$5 minutes) due to a sliding average having been used to locate the maximum slope of the flare X-ray intensity. Events for which $s <$ 0.8 AU or $s >$ 4 AU are excluded. Events 13 and 23 for SOHO/ERNE, 30, 39, and 40 for STEREO-A/HET, and 7, 17, and 23 for STEREO-B/HET are highlighted; see text for discussion.}
\label{Proton_delay_vs_conn_angle}
\end{figure}

Considering only the events included in Figure \ref{Proton_delay_vs_s_selection}, the proton VDA release delay as a function of the longitudinal distance between the event-associated flare and the observer is presented in Figure \ref{Proton_delay_vs_phi}; Figure \ref{Proton_delay_vs_conn_angle} replaces the longitudinal distance with the connection angle. Considerable scatter is very apparent. In both cases, the data sets appear to suggest the shortest release delays for events with connection angles close to zero, as might be expected.

\begin{figure}%[H]
\centering
\includegraphics[width=0.7\columnwidth]{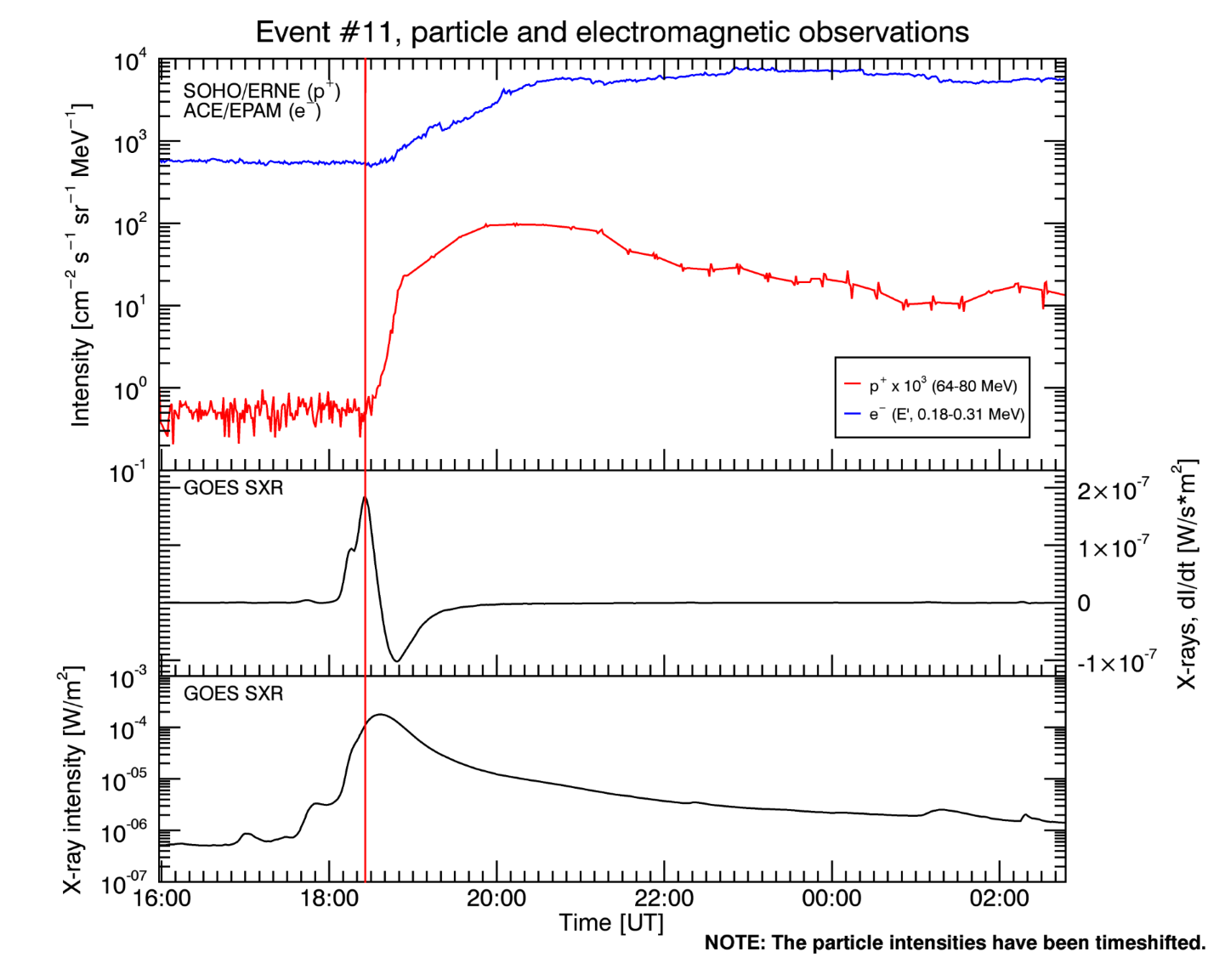}
\caption{\small Proton, electron, and X-ray observations made near the Earth during event 11 (27 January 2012). The upper panel: SOHO/ERNE proton (red, multiplied by 10$^3$) and ACE/EPAM electron intensities (blue); the middle panel: the time derivative of GOES soft X-ray intensity; the lower panel: GOES soft X-ray intensity. The vertical red line marks the time of the maximum of d$I$/d$t$ for X-ray intensity. The particle intensities have been backshifted by ($L$/$v$ - 500) seconds, where $L$ = 1.07 AU and $v$ = the mean speed of the particle species in question. Note the two-staged rise of the X-ray intensity.}
\label{Ev_11_EM}
\end{figure}

In both Figure \ref{Proton_delay_vs_phi} and \ref{Proton_delay_vs_conn_angle} there appear two outlying events\textemdash numbers 11 and 15, both observed by SOHO/ERNE\textemdash which are expected to be well connected to the observer (with connection angles of 5.6\textdegree \,and 23.2\textdegree, respectively) and which exhibit a negative release delay of tens of minutes. While both have a reasonable VDA apparent path length of $\approx$3 AU, their associated soft X-ray flares rise relatively slowly; that of event 11 features two peaks, with the first reaching a flux equivalent to about C3 on the NOAA classification scale and the main peak, following after a plateau of some 20 minutes, to X1.7. The near-Earth particle and GOES X-ray observations related to event 11 are shown Figure \ref{Ev_11_EM}. In this case it is probable that part of the proton release occurred during the early stages of the flare, before the maximum of the soft X-ray flux time derivative was reached. The flare related to event 15 is somewhat similar, albeit with a much less pronounced and slower early rise and lower global peak intensity (M5.1).

\begin{figure}%[H]
\centering
\includegraphics[width=0.7\columnwidth]{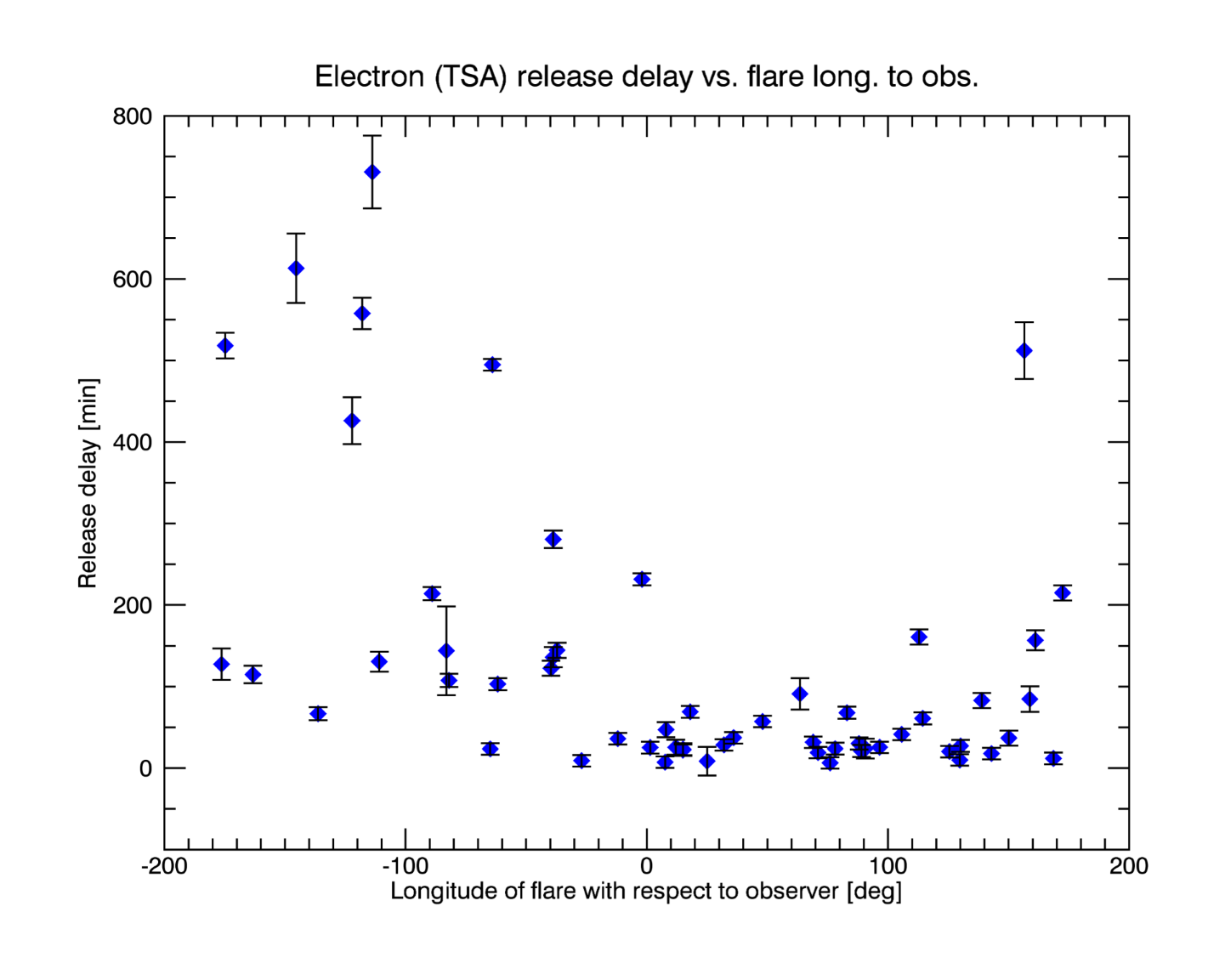}
\caption{\small Electron event release delay (from TSA) as a function of the longitudinal distance between the flare and the observer. The release delay errors include the propagated uncertainty ($\pm$5 minutes) due to a sliding average having been used to locate the maximum slope of the flare X-ray intensity. Event 38 for STEREO-A/SEPT is excluded; see text.}
\label{Electron_delay_vs_phi}
\end{figure}

\begin{figure}%[H]
\centering
\includegraphics[width=0.7\columnwidth]{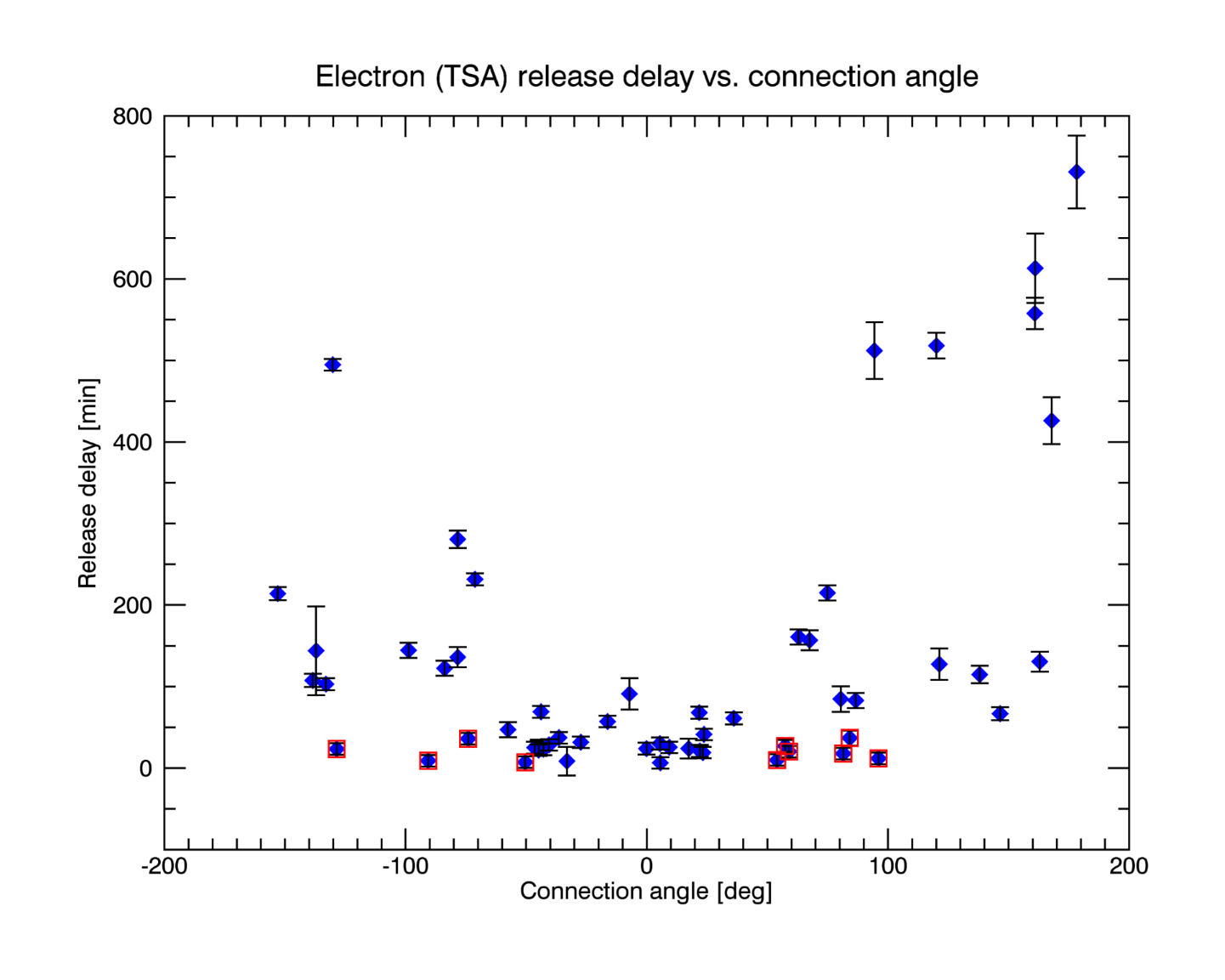}
\caption{\small Electron event release delay (from TSA) as a function of the connection angle. The release delay errors include the propagated uncertainty ($\pm$5 minutes) due to a sliding average having been used to locate the maximum slope of the flare X-ray intensity. Events 13 and 23 at ACE/EPAM, 30, 39, and 40 for STEREO-A/SEPT, and 3, 7, 17, 20, and 23 for STEREO-B/SEPT are highlighted, and event 38 for STEREO-A/SEPT is excluded; see text for discussion.}
\label{Electron_delay_vs_conn_angle}
\end{figure}

Our listing contains 52 electron events with both an identified visible disk solar flare and a meaningful TSA result. Figures \ref{Electron_delay_vs_phi} and \ref{Electron_delay_vs_conn_angle} display these events, the former showing the electron TSA release delay as a function of the longitudinal distance between flare and observer and the latter as a function of the connection angle. While the shortest delay times predictably tend to cluster around longitudinal distances between 0 and 90 degrees as well as connection angles with a small absolute value, there are some cases where a large positive connection angle ($>$120\textdegree) still gives a fairly short release delay that exhibits no strong dependence on the connection angle. Overall, clear trends seem to be absent at longitudinal distances between 0 and $\lesssim$ 150 degrees, as well as at connection angles between $\approx -50$ and $\approx$ 50 degrees.

A TSA-derived electron release delay of less than 40 minutes and $\lvert \phi_{\rm C} \rvert >$ 50\textdegree \,co-occur for events 13 and 23 at ACE, 30, 39, and 40 for STEREO-A, and 3, 7, 17, 20, and 23 for STEREO-B. These are highlighted in Figure \ref{Electron_delay_vs_conn_angle}. A VDA-derived proton injection delay (with $s <$ 4 AU) is available for eight of the corresponding proton events: 13 and 23 at SOHO, 30, 39, and 40 for STEREO-A, and 7, 17, and 23 for STEREO-B. Excepting the proton event 13 for SOHO/ERNE, in which the VDA fit is only moderately good ($s$ = 3.84 AU and $R^2$ = 0.637, resulting in a standard error of $\pm$50 minutes for the release time estimate), it can be seen from Figure \ref{Proton_delay_vs_conn_angle}\textemdash where the same events are again highlighted\textemdash that these proton events tend to exhibit a relatively short release delay with respect to the connection angle, as do the corresponding electron events.

One plausible interpretation is that the dominant SEP acceleration and transport mechanisms in each event are largely responsible for the spreading of the release times. The events where the release delay appears to increase rapidly as a function of the absolute value of the connection angle would, in that case, be most likely consistent with primarily diffusive transport from a compact source region, resulting in a slow longitudinal spreading of the particles, while the other group that only shows a weak dependence could be dominated by shock acceleration in a laterally expanding coronal shock wave. The variation in the longitudinal extent of the particle source, in conjunction with the transport effects, could also be responsible for some of the differentiation between events. However, the events with weak dependence may actually contain cases of rapidly opening magnetic field lines near the parent active region, accounting for a very small and virtually constant release delay at large absolute values of the connection angle. Such rapid opening could be a result of expansion of open active region fields in the corona (\textit{e.g.}, \citealp{Wiedenbeck2013}) or turbulent field line meandering in the corona and interplanetary medium (\citealp{Laitinen2016}). The spreading of solar magnetic field lines has been studied by \citet{Wiedenbeck2013}, who concluded that it is possible for field lines originating from inside a circle of 10\textdegree \,in radius at the solar photosphere to reach up to some 130\textdegree \,in longitudinal extent. While relatively rare, such cases could correspond to SEP events where a very rapid intensity enhancement is detected over a large range of longitudes with respect to the event source. At any rate, this sort of phenomenon would conceivably be competing with the presence of extended acceleration regions in shocks, and the observational methods at our disposal do not allow us to make a distinction between them.

A handful of outlying data points are present in both Figure \ref{Electron_delay_vs_phi} and Figure \ref{Electron_delay_vs_conn_angle}. In six events (25 at ACE; 23 and 28 at STEREO-A; 11, 26, and 37 at STEREO-B), the TSA-derived release delay for electrons is greater than 500 minutes. Of these, 23 (STEREO-A) and 37 (STEREO-B) are extremely weak events, with observed maximum electron intensity of the order of 20 pfu MeV$^{-1}$, whereas the others feature a considerably elevated pre-event background. The absolute values of connection angles for all of these events are in excess of 90 degrees, exceeding 150 degrees in three events, which leads to slow intensity rises. Thus, it would seem that very large release delays may be partially attributable to difficulties in deriving a reasonable estimate for the event onset time at the observer.

A negative value for the delay between flare X-ray flux time derivative maximum and the TSA-derived injection time is obtained in event 38 (7 January 2014) at STEREO-A. The SEPT-A electron intensity is generally elevated before and during the period of interest, and it shows a gradual rise of steps and plateaus over several days, probably indicating multiple electron injections. Thus, while the identity of the electron event itself is not in serious doubt, the estimate for its onset time at STEREO-A is likely affected by previous SEP activity, and it is excluded from the Figures mentioned above.

The results for the proton and electron release delays are also noteworthy in that they are qualitatively consistent with the results of Section \ref{Sec4.4} and not inconsistent with those of Section \ref{Sec4.3}: all suggest that the most effective (both in terms of speed and maximum intensity) SEP transport from the Sun to the observer tends to occur when the observer is magnetically connected to solar regions near the event-related flare. Based on the connection angle results for event centres, shown in Section \ref{Sec4.3}, a small bias favouring negative connection angles might be expected to occur here, as well; however, as the points of all the data sets considered in this section are strongly scattered, such a bias, whether or not actually present, could easily be masked.

\subsection{SEP Event-related CMEs}
\label{Sec4.7}
Considering the SOHO/LASCO observations, the strong dominance of halo-type CMEs, as they are defined in the CDAW SOHO LASCO CME Catalog, is a prominent feature of the events examined in this work; this is in agreement with a number of previous studies, such as \citet{Cane1998}. Only two events out of 46 were associated with a non-halo CME, meaning that halo CMEs occurred in 96\% of the cases in our present catalogue. While halo CMEs (\textit{per} the CDAW SOHO LASCO CME Catalog) are indeed common in high-energy SEP events, they appear to be proportionately more common still in wide events: \citet{Paassilta2017} report that they occurred in 81\% (48/59) of the near-Earth 55\textendash 80 MeV proton events during 2009\textendash 2016 and in 72\% (72/101) during 1996\textendash 2016. Of interest in this context is the work of \citet{Kwon2014}, who have concluded that the outermost front of a halo-type CME (as seen from several viewpoints) may be a bubble-shaped structure, with the internal flux rope forming the inner front. Such a structure might expand over field lines connected to widely spaced observers in a short period of time, accounting for relatively fast event rise times even when the nominal connection angles of the observers are not small.

We list two estimates for the radial speed of the event-associated CMEs: one derived from SOHO/LASCO observations only, and the other representing the maximum of the three values listed for SOHO/LASCO (in the CDAW SOHO LASCO CME Catalog) and STEREO-A and -B/SECCHI/COR (in the Dual-Viewpoint CME Catalog). This is motivated by the fact that as CMEs in general are observed as two-dimensional projections against the sky plane, the speed estimate obtained is necessarily always equal to or less than the true radial speed in three dimensions, and so the highest speed estimate may be assumed to be the closest approximation to the actual value. (In any case, it should be noted that the methods used to derive the CME speeds are not identical between the catalogues, so the combined values must be regarded as approximative in that regard, as well. Furthermore, a full listing of CME parameters is available in the Dual-Viewpoint CME Catalog in only 26 of the investigated events for STEREO-A and in 22 for STEREO-B.)

\begin{figure}%[H]
\centering
\includegraphics[width=0.7\columnwidth]{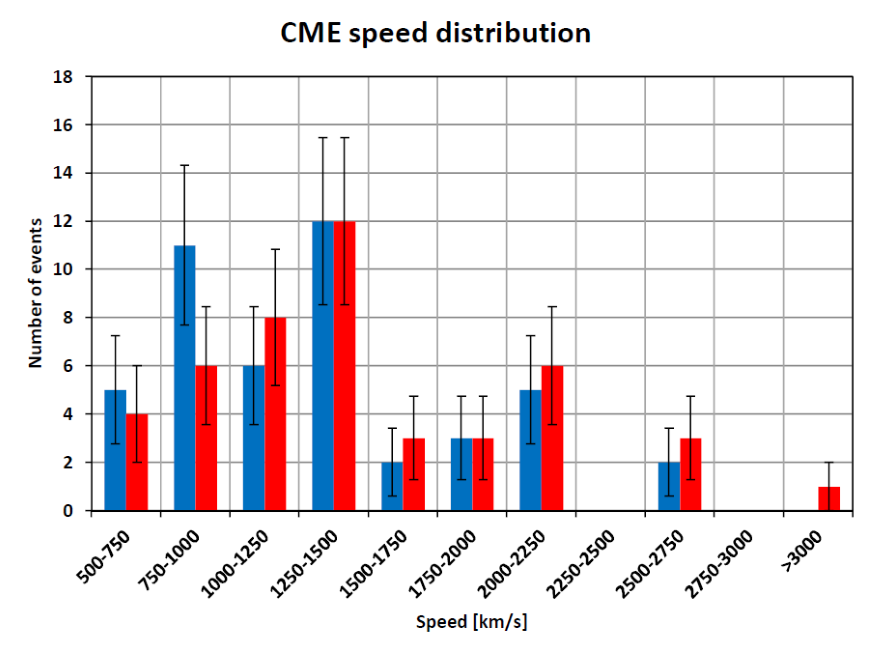}
\caption{\small The speed distribution of event-related CMEs. The left-hand (blue) bars represent the SOHO/LASCO observations (obtained from the CDAW SOHO LASCO CME Catalog), and the right-hand (red) bars the combined observations from all three spacecraft (obtained from the CDAW SOHO LASCO CME Catalog and the Dual-Viewpoint CME Catalog), whenever available. The error bars denote the statistical error.}
\label{CME_speeds}
\end{figure}

\begin{figure}%[H]
\centering
\includegraphics[width=0.7\columnwidth]{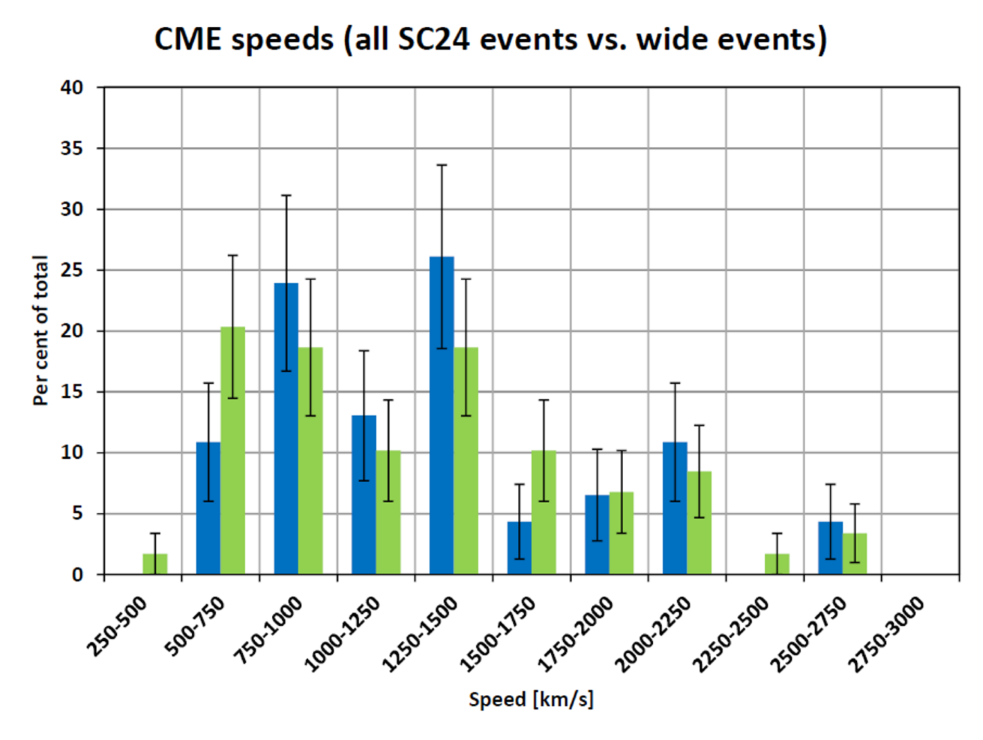}
\caption{\small The proportional speed distribution of CMEs related to events in 2009\textendash 2016, according to the CDAW SOHO LASCO CME Catalog. The left-hand (blue) bars represent the CMEs of wide events (in this work; 46 in total) and the right-hand (green) bars the CMEs of all recorded SOHO/ERNE 55\textendash 80 MeV proton events (from \citealp{Paassilta2017}; 59 in total). The error bars denote the statistical error.}
\label{CME_speeds_comp}
\end{figure}

The SOHO/LASCO observations yield mean and median speeds of 1323 km s$^{-1}$ and 1291 km s$^{-1}$, respectively, for the CMEs in this work; the combined  observations, in turn, yield 1484 km s$^{-1}$ and 1376 km s$^{-1}$, respectively. These results do not differ greatly from those given in \textit{e.g.} \citet{Papaioannou2016} ($\approx$ 1390 km s$^{-1}$ mean CME speed for $>$10 MeV GOES proton events in 1996\textendash 2013) or \citet{Paassilta2017} (mean 1276  km s$^{-1}$ for the 59 near-Earth 2009\textendash 2016 proton events mentioned above, based on SOHO/LASCO only). 

Figure \ref{CME_speeds} shows the distribution of the radial speeds of the event-related CMEs for both SOHO/LASCO-only and combined observations, and Figure \ref{CME_speeds_comp} juxtaposes the speed distribution of the wide events considered in this work with that of all SOHO/ERNE 55\textendash 80 MeV proton events detected at the Earth, as given in \citet{Paassilta2017}. The distribution of wide events exhibits considerable width: the slowest CME included in our catalogue is estimated to have had a radial speed of 575 km s$^{-1}$ (SOHO/LASCO, event 6), while the fastest attained 2684 km s$^{-1}$ (event 13) according to the CDAW SOHO LASCO CME Catalog; if combined observations are considered, the highest speed is 3706 km s$^{-1}$, measured from STEREO-B/SECCHI/COR data in event 18 (CDAW SOHO LASCO CME Catalog reports 2003 km s$^{-1}$ for the same CME). While slow ($<$ 500 km s$^{-1}$) CMEs are absent from our sample of events, there are nevertheless a few with moderate or relatively low speeds (between 500 km s$^{-1}$ and 750 km s$^{-1}$). A comparison to the CME speed distribution presented in \citet{Paassilta2017} suggests that although the averages of the speeds of CMEs associated with wide and narrow SEP events in Solar Cycle 24 are fairly similar, CMEs in wide events appear to be proportionally somewhat less common in the speed range of 500\textendash 750 km s$^{-1}$ and conversely more common in the range 1000\textendash 1500 km s$^{-1}$. Again, the small number of events hinders reaching firm conclusions, but aside from this slight preference for higher speeds in wide events, no major differences between SEP events of narrow or wide longitudinal extent during the same time period are readily apparent.

In addition, we carried out a brief investigation into whether there is a correlation between the particle release height, here assumed to be approximately proportional to CME height at the inferred proton release time (from VDA), and the VDA apparent path length. If one of these values could be shown to depend on the other, this might possibly suggest that the proton pitch-angle scattering, as inferred from the VDA path length, is affected by the release height in wide events. Using the CME height-time data points derived from observations, available in the CDAW SOHO LASCO CME Catalog, we fitted quadratic curves to the observations to model the early propagation of the CMEs in events where VDA results likely to be trustworthy, \textit{i.e.} with the apparent path length falling between some 1 AU and 3 AU, are available. This method assumes that the CME is first accelerated instantaneously at the earliest stage of its expansion and undergoes constant acceleration afterwards. The data points appear very strongly scattered, and no correlation was discovered; the results are not shown here.

\section{Conclusions and Outlook}
\label{Sec5}
Having examined particle data from SOHO, ACE, and STEREO-A and -B from the period of 2009\textendash 2016, we identified 46 high-energy SEP events that were detected by at least two spacecraft with a longitudinal separation of $>$45 degrees. Considering each spacecraft individually, there were a total of 107 $>$55 MeV proton events and 104 0.18\textendash 0.31 MeV electron events. Every multi-spacecraft event could be reasonably securely associated with a CME; similarly, a solar flare or an EUV brightening signifying an associated eruption was identified for every event, aside from one case of poor data coverage.

Where exact estimates for event onset times were available, VDA and TSA were performed for the proton events and TSA for electron events; VDA yielded 74 cases of successful fits out of 94. Compared to the single-spacecraft observations from SOHO/ERNE, we estimate that some 55\% of the near-Earth proton events (\citealp{Paassilta2017}) were also wide enough in longitude to be detected at either one or both of the STEREOs during the time period mentioned above.

Maximum intensities of protons and electrons, proton fluences, release delay times for both protons (from VDA and TSA) and electrons (from TSA) as a function of observer longitude with respect to the flare site and the connection angle, and CME speeds were investigated using the derived event parameters. While the statistics of this study, due to the overall rather small number of events, are perhaps weaker than what would be desirable for presenting definite conclusions, certain tendencies seem to be present.

\begin{enumerate}[(i)]
\item There is some evidence that the most efficient particle transport in wide-longitude SEP events, in terms of maximum particle intensity, occurs when the observer at $\approx$1 AU is magnetically connected to a solar region located somewhat to the west of the flare. This is suggested by modelling the SEP maximum intensities and proton fluences as functions of the connection angle with Gaussian curves. The longitudinal distance of the best-connected region from the inferred injection site seems to fall, in average, between some 20 and 36 degrees, depending on the particle species. However, it should be noted that there are rather wide error margins in all of these results due to the relatively small numbers of available data points. \citet{Lario2013}, who obtained similar results also by applying Gaussian modelling to event peak intensities (see Section \ref{Sec4.3}), suggested that the negative connection angles of event centre  are an indication of the prompt SEP component forming a considerable distance away from the solar surface. Discounting the possibility of large systematic errors in determining the location of the event-related flares and the magnetic footpoints connected to observers, an explanation involving the features of interplanetary particle transport does appear plausible.
\item If the dependence of the three-spacecraft event maximum intensities on the connection angle is modelled with Gaussian curves (see above), the mean event widths, considered as corresponding to the mean of the standard deviation $\sigma$, are 43.6\textdegree \,$\pm$ 8.3\textdegree \,for $>$55 MeV protons and 49.6\textdegree \,$\pm$ 8.2\textdegree for 0.18\textendash 0.31 MeV electrons. These results, especially the one obtained for protons, are close to those derived in prior works that focus, for the most part, on events of particles with lower mean kinetic energies: 43\textdegree \,$\pm$ 2\textdegree \,for 15\textendash 40 MeV protons (\citealp{Lario2013}), 45\textdegree \,$\pm$ 1\textdegree \,for 25\textendash 53 MeV protons (\citealp{Lario2013}), and 43\textdegree \,$\pm$ 13\textdegree \,for 14\textendash 24 MeV protons (\citealp{Richardson2014}; all values based on peak intensities). In the case of electrons, our results are similar to those of comparable earlier studies (\citealp{Dresing2014}: 39\textdegree \,for 55\textendash 105 keV electrons; \citealp{Lario2013}: 49\textdegree \,$\pm$ 2\textdegree \,for 71\textendash 112 keV electrons) when the statistical uncertainty is taken into account. This implies that the event width is not strongly dependent on SEP energy.
\item Proton and electron event rise times and release delays, considered as functions of the connection angle, indicate that there is no simple relationship between these quantities: rise times and release delays increase considerably when then absolute value of the connection angle increases in some events, while in others there is only a very weak dependence (if any). Our interpretation is that this is likely to be the result of the dominant SEP acceleration and transport mechanisms involved in each event, either diffusive transport from a compact source (the strongly dependent cases) or emission from a laterally expanding coronal shock (the weakly dependent cases). However, contribution from a compact source via rapidly expanding coronal and interplanetary field lines in the latter event type is also a possibility. It may be noted that the fast SEP intensity rises and short release delays recorded by nominally weakly connected observers, occurring in some events, underscore the importance of understanding how different particle transport effects operate in interplanetary space. The expectation would be different for diffusive perpendicular transport from a compact source, predicting rapidly increasing delay times as a function of source longitude, and transport resulting from ballistic propagation along meandering field lines, predicting much faster access to field lines distant from the source than mere diffusion (see, \textit{e.g.}, \citealp{Laitinen2016}; \citealp{Laitinen2017}; \citealp{LaitinenDallaMarriott2017}). The mode of field line transport, whether due to turbulence or large-scale coronal magnetic structures, would undoubtedly  add its fingerprint to the onset delay and peak intensities observed far from the source, as well. Also relevant to this consideration is the longitudinal extent of the particle source and injection region; very wide sources can potentially reach across the magnetic footpoints leading to several observers separated by a considerable angular distance.
\item While the estimated mean speeds and speed distributions of SEP event-related CMEs probably do not differ significantly between events of large longitudinal extent and all near-Earth events detected during the same time period, the proportion of halo-type CMEs (as defined in the CDAW SOHO LASCO CME Catalog) is greater in wide-longitude than in narrow-longitude events: they occur in some 96\% of the events studied in this work as opposed to some 81\% of all comparable near-Earth events during the same time period. At the SEP energies considered in this work, wide-longitude events appear almost exclusively associated with halo CMEs. This can be viewed as suggestive of wide particle sources, as well (see \citealp{Kwon2014}).
\end{enumerate}

Including observations from spacecraft at various radial distances from the Sun could be of help in investigating (i) how significant the bias for negative connection angles found at the particle energies considered in this study is in a broader context, and (ii) whether and how it arises in the 
interplanetary space during particle transport. Electron data gathered during the 2009\textendash 2016 period are available from the MESSENGER (\textit{Mercury Surface, Space Environment, Geochemistry and Ranging}) and \textit{Juno} missions, but similar, directly comparable data sets for $>$55 MeV protons, recorded at distances other than $\approx$1 AU during Solar Cycle 24, unfortunately do not exist. A major increase in SEP activity, in any case, is not very likely until the rising phase of the next solar cycle, making a marked improvement in SEP event statistics improbable overall in the near future. It therefore appears that pending results from upcoming missions, such as the \textit{Parker Solar Probe}, further simulation studies of energetic particle transport in interplanetary space (see, for instance, \citealp{Luhmann2017}) might offer the fastest path to a more refined understanding of event dynamics in this respect.

\addtolength{\textheight}{1.5cm}
\begin{landscape}

% Solar observations:
\scriptsize
\setlength{\tabcolsep}{2pt}
\begin{longtable}{llp{0.5pt}lllllp{0.5pt}lrrrp{0.5pt}lrl}
\caption{\label{solar_obs} Solar flare and CME observations for $>$55 MeV proton and 0.18--0.31 MeV electron multi-spacecraft events in 2009--2016.}\\
\hline\hline
\noalign{\smallskip}
ID & Date &&\multicolumn{5}{c}{GOES or STEREO/EUVI flare observations}&&\multicolumn{4}{c}{SOHO/LASCO CME observations}&&\multicolumn{3}{c}{Combined SOHO/LASCO \&}\\
&&&&&&&&&&&&&&\multicolumn{3}{c}{STEREO/COR CME observations}\\
\cline{4-8}
\cline{10-13}
\cline{15-17}
\noalign{\smallskip}
&&& Onset & \multirow{2}{1.6cm}{Max. d$I$/d$t$ [UT]}& Class & \ Lat. [\textdegree]& Long. [\textdegree]&& 1st obs.&$v$[km s$^{-1}$]& Width [\textdegree] & PA [\textdegree]&&1st obs.&$v$[km s$^{-1}$]&S/C\\
&&&[UT]&&&&&&[UT]&&&&&[UT]&&\\
\hline
\endfirsthead
\caption{continued.}\\
\hline\hline
\noalign{\smallskip}
ID & Date &&\multicolumn{5}{c}{GOES or STEREO/EUVI flare observations}&&\multicolumn{4}{c}{SOHO/LASCO CME observations}&&\multicolumn{3}{c}{Combined SOHO/LASCO \&}\\
&&&&&&&&&&&&&&\multicolumn{3}{c}{STEREO/COR CME observations}\\
\cline{4-8}
\cline{10-13}
\cline{15-17}
\noalign{\smallskip}
&&& Onset & \multirow{2}{1.6cm}{Max. d$I$/d$t$ [UT]}& Class & \ Lat. [\textdegree]& Long. [\textdegree]&& 1st obs.&$v$[km s$^{-1}$]& Width [\textdegree] & PA [\textdegree]&&1st obs.&$v$[km s$^{-1}$]&S/C\\
&&&[UT]&&&&&&[UT]&&&&&[UT]&&\\
\hline%\hline
\noalign{\smallskip}
\endhead
\noalign{\smallskip}
\hline
\endfoot
\noalign{\smallskip}
1&2011 Jan. 28&&00:44&00:58&M1.3&N16&W88&&01:26&606&119&288&&01:26&606&SOHO\\
2&2011 Feb. 15&&01:44&01:52&X2.2&S20&W12&&02:24&669&360&halo&&02:24&1161&STEREO-A\\
3&2011 Mar. 07&&19:43&19:56&M3.7&N30&W48&&20:00&2125&360&halo&&20:00&2125&SOHO\\
4&2011 Mar. 21&&\multicolumn{3}{l}{Farside flare (EUVI)}&N16&W130&&02:24&1341&360&halo&&02:24&1341&SOHO\\
5&2011 Aug. 04&&03:41&03:53&M9.3&N19&W36&&04:12&1315&360&halo&&04:12&1808&STEREO-A\\
6&2011 Sep. 06&&22:12&22:18&X2.1&N14&W18&&23:06&575&360&halo&&22:39&730&STEREO-B\\
7&2011 Sep. 22&&10:29&10:44&X1.4&N09&E89&&10:48&1905&360&halo&&10:48&1905&SOHO\\
8&2011 Oct. 04&&\multicolumn{3}{l}{Farside flare (EUVI)}&N33&E151&&13:26&1101&360&halo&&13:26&1101&SOHO\\
9&2011 Nov. 03&&\multicolumn{3}{l}{Farside flare (EUVI)}&N12&E152&&23:30\textsuperscript{C3}&991&360&halo&&22:39&991&SOHO\\
10&2012 Jan. 23&&03:38&03:49&M8.7&N18&W25&&04:00&2175&360&halo&&03:24&2175&SOHO\\
11&2012 Jan. 27&&17:37&18:26&X1.7&N27&W71&&18:28&2508&360&halo&&18:28&2508&SOHO\\
12&2012 Mar. 05&&02:30&03:46&X1.1&N16&E54&&04:00&1531&360&halo&&02:54&1531&SOHO\\
13&2012 Mar. 07&&00:02&00:18&X5.4&N17&E27&&00:24&2684&360&halo&&00:24&2684&SOHO\\
14&2012 Mar. 24&&\multicolumn{3}{l}{Farside flare (EUVI)}&N11&E166&&00:24&1152&360&halo&&00:24&1417&STEREO-A\\
15&2012 May 17&&01:25&01:36&M5.1&N11&W76&&01:48&1582&360&halo&&01:48&1582&SOHO\\
16&2012 Jul. 08&&16:23&16:29&M6.9&S14&W83&&16:54\textsuperscript{C3}&1495&157&212&&16:54&1495&SOHO\\
17&2012 Jul. 12&&15:37&16:32&X1.4&N13&W15&&16:48&885&360&halo&&16:48&1508&STEREO-B\\
18&2012 Jul. 23&&\multicolumn{3}{l}{Farside flare (EUVI)}&S23&W137&&02:36&2003&360&halo&&02:36&3706&STEREO-B\\
19&2012 Sep. 20&&\multicolumn{3}{l}{Farside flare (EUVI)}&S11&E159&&15:12&1202&360&halo&&15:12&2531&STEREO-A\\
20&2012 Sep. 28&&23:36\textsuperscript{*}&23:43\textsuperscript{*}&C3.7&N09&W32&&00:12&947&360&halo&&23:54\textsuperscript{*}&1137&STEREO-A\\
21&2012 Nov. 08&&\multicolumn{3}{l}{Farside flare (EUVI)}&S06&W164&&11:00&972&360&halo&&11:00&1386&STEREO-B\\
22&2013 Mar. 05&&\multicolumn{3}{l}{Farside flare (EUVI)}&N10&E145&&03:48&1316&360&halo&&03:48&1316&SOHO\\
23&2013 Apr. 11&&06:55&07:10&M6.5&N09&E12&&07:24&861&360&halo&&07:24&861&SOHO\\
24&2013 Apr. 24&&\multicolumn{3}{l}{Farside flare (EUVI)}&N10&W167&&22:12&594&360&halo&&22:12&594&SOHO\\
25&2013 May 15&&01:25&01:42&X1.2&N12&E64&&01:48&1366&360&halo&&01:48&1366&SOHO\\
26&2013 May 22&&13:08&13:17&M5.0&S18&W15&&13:26&1466&360&halo&&13:24&2106&STEREO-A\\
27&2013 Aug. 20&&\multicolumn{3}{l}{Farside flare (EUVI)}&N12&W178&&23:12\textsuperscript{*}&877&360&halo&&23:12\textsuperscript{*}&877&SOHO\\
28&2013 Sep. 30&&21:43\textsuperscript{*}&21:48\textsuperscript{*}&C1.2&N10&W33&&22:12\textsuperscript{*}&1179&360&halo&&22:12\textsuperscript{*}&1179&SOHO\\
29&2013 Oct. 05&&\multicolumn{3}{l}{Farside flare (EUVI)}&S24&E121&&07:10&964&360&halo&&07:10&964&SOHO\\
30&2013 Oct. 11&&07:01&07:15&M1.5&S21&E44&&07:24&1200&360&halo&&07:24&1200&SOHO\\
31&2013 Oct. 25&&07:53&07:59&X1.7&S08&E73&&08:12&587&360&halo&&08:12&587&SOHO\\
32&2013 Oct. 28&&15:07&15:11&M4.4&S08&E28&&15:36&812&360&halo&&15:36&812&SOHO\\
33&2013 Nov. 02&&\multicolumn{3}{l}{Farside flare (EUVI)}&N06&W133&&04:48&828&360&halo&&04:48&963&STEREO-B\\
34&2013 Nov. 07&&\multicolumn{3}{l}{Farside flare (EUVI)}&N02&E151&&10:36&1405&360&halo&&10:24&1405&SOHO\\
35&2013 Dec. 26&&\multicolumn{3}{l}{Farside flare (EUVI)}&S10&E164&&03:24&1336&360&halo&&03:24&1399&STEREO-A\\
36&2013 Dec. 28&&\multicolumn{3}{l}{Farside flare (EUVI)}&S13&W125&&17:36&1118&360&halo&&17:36&1118&SOHO\\
37&2014 Jan. 06&&7:27&07:43&C2.1&S15&W89&&08:00&1402&360&halo&&08:00&1402&SOHO\\
38&2014 Jan. 07&&18:04&18:18&X1.2&S12&W08&&18:24&1830&360&halo&&18:24&2240&STEREO-A\\
39&2014 Feb. 25&&00:39&00:46&X4.9&S12&E82&&01:26\textsuperscript{C3}&2147&360&halo&&01:24&2147&SOHO\\
40&2014 Mar. 04&&18:04&18:22&C5.5&N12&W88&&18:48&794&360&halo&&18:48&1067&STEREO-A\\
41&2014 Mar. 05&&\multicolumn{3}{l}{Farside flare (EUVI)}&N12&E179&&13:48&828&360&halo&&13:48&1045&STEREO-A\\
42&2014 Sep. 01&&\multicolumn{3}{l}{Farside flare (EUVI)}&N14&E128&&11:12&1901&360&halo&&11:12&1901&SOHO\\
43&2014 Sep. 10&&17:21&17:33&X1.6&N14&E2&&18:00&1267&360&halo&&17:54&1267&SOHO\\
44&2014 Sep. 25&&\multicolumn{3}{l}{Farside flare (EUVI)}&N13&E175&&21:30\textsuperscript{*,C3}&1350&360&halo&&21:24\textsuperscript{*}&1350&SOHO\\
45&2014 Dec. 13&&\multicolumn{3}{l}{Farside flare (EUVI)}&S25&W141&&14:24&2222&360&halo&&14:24&2222&SOHO\\
46&2015 Jul. 01&&N/A&&&\multicolumn{2}{l}{N/A}&&14:36&1435&360&halo&&14:36&1435&SOHO\\
\end{longtable}
\raggedright
\noindent *: The time given refers to the previous calendar day.\\
\noindent C3: The first observation of the CME occurred on the LASCO C3 coronagraph.\\
Items marked with "N/A" or "No data!" could not be determined due to insufficient or completely missing data, respectively. All times are  rounded to the nearest minute.
\normalsize

\newpage
\addtolength{\textheight}{1.5cm}

\scriptsize
\setlength{\tabcolsep}{2pt}
\begin{longtable}{llp{0.5pt}lrlrrp{0.5pt}lrllrrp{0.5pt}lrllrr}
\caption{\label{sep_protons_1} Proton event onset times, flare-to-observer longitudinal distances, event rise times and total durations, and proton fluences for $>$55 MeV multi-spacecraft proton events in 2009--2016.}\\
\hline\hline
\noalign{\smallskip}
ID & Date &&\multicolumn{5}{c}{SOHO/ERNE 55--80 MeV proton flux}&&\multicolumn{6}{c}{STEREO-A/HET 40--100 MeV proton flux}&&\multicolumn{6}{c}{STEREO-B/HET 40--100 MeV proton flux}\\
\cline{4-8}
\cline{10-15}
\cline{17-22}
\noalign{\smallskip}
&& & Onset & $\Delta\phi$ [\textdegree]& \multirow{3}{1cm}{$I_{\rm max}$ [pfu MeV$^\textrm{-1}$]} & \multirow{2}{0.7cm}{Dur. [h]} & \multirow{3}{1.2cm}{p$^+$ fluence [cm$^\textrm{-2}$ sr$^\textrm{-1}$]}&& Onset & $\Delta\phi$ [\textdegree]& \multirow{3}{1cm}{$I_{\rm max}$ [pfu MeV$^\textrm{-1}$]} & \multirow{3}{0.8cm}{Rise time [h]} & \multirow{2}{0.7cm}{Dur. [h]} & \multirow{3}{1.2cm}{p$^+$ fluence [cm$^\textrm{-2}$ sr$^\textrm{-1}$]}&& Onset & $\Delta\phi$ [\textdegree]& \multirow{3}{1cm}{$I_{\rm max}$ [pfu MeV$^\textrm{-1}$]} & \multirow{3}{0.8cm}{Rise time [h]} & \multirow{2}{0.7cm}{Dur. [h]} & \multirow{3}{1.2cm}{p$^+$ fluence [cm$^\textrm{-2}$ sr$^\textrm{-1}$]}\\
&&&[UT]&&&&&&[UT]&&&&&&&[UT]&&\\
&&&&&&&&&&&&&&&&&&\\
\hline
\endfirsthead
\caption{continued.}\\
\hline\hline
\noalign{\smallskip}
ID & Date &&\multicolumn{5}{c}{SOHO/ERNE 55--80 MeV proton flux}&&\multicolumn{6}{c}{STEREO-A/HET 40--100 MeV proton flux}&&\multicolumn{6}{c}{STEREO-B/HET 40--100 MeV proton flux}\\
\cline{4-8}
\cline{10-15}
\cline{17-22}
\noalign{\smallskip}
&& & Onset & $\Delta\phi$ [\textdegree]& \multirow{3}{1cm}{$I_{\rm max}$ [pfu MeV$^\textrm{-1}$]} & \multirow{2}{0.7cm}{Dur. [h]} & \multirow{3}{1.2cm}{p$^+$ fluence [cm$^\textrm{-2}$ sr$^\textrm{-1}$]}&& Onset & $\Delta\phi$ [\textdegree]& \multirow{3}{1cm}{$I_{\rm max}$ [pfu MeV$^\textrm{-1}$]} & \multirow{3}{0.8cm}{Rise time [h]} & \multirow{2}{0.7cm}{Dur. [h]} & \multirow{3}{1.2cm}{p$^+$ fluence [cm$^\textrm{-2}$ sr$^\textrm{-1}$]}&& Onset & $\Delta\phi$ [\textdegree]& \multirow{3}{1cm}{$I_{\rm max}$ [pfu MeV$^\textrm{-1}$]} & \multirow{3}{0.8cm}{Rise time [h]} & \multirow{2}{0.7cm}{Dur. [h]} & \multirow{3}{1.2cm}{p$^+$ fluence [cm$^\textrm{-2}$ sr$^\textrm{-1}$]}\\
&&&[UT]&&&&&&[UT]&&&&&&&[UT]&&\\
&&&&&&&&&&&&&&&&&&\\
\hline%\hline
\noalign{\smallskip}
\endhead
\noalign{\smallskip}
\hline
\endfoot
\noalign{\smallskip}
1&2011 Jan. 28&&01:56&88&3.5$\times 10^{-3}$&72.4&7.5$\times 10^{4}$&&02:37&1&3.1$\times 10^{-3}$&0.8$\pm$0.2&55.5&7.4$\times 10^{4}$&&\ldots&\ldots&\ldots&\ldots&\ldots&\ldots\\
2&2011 Feb. 15&&03:39&10&1.2$\times 10^{-3}$&76.4&5.3$\times 10^{4}$&&\ldots&\ldots&\ldots&\ldots&\ldots&\ldots&&03:18&104&8.6$\times 10^{-3}$&1.2$\pm$0.2&64.2&2.0$\times 10^{5}$\\
3&2011 Mar. 07&&21:06&48&8.0$\times 10^{-3}$&170.7&2.8$\times 10^{6}$&&19:14\textsuperscript{+}&-40&2.2$\times 10^{-3}$&2.5$\pm$0.9&82.1&1.2$\times 10^{6}$&&21:46&143&4.6$\times 10^{-2}$&0.0$\pm$0.2&122.7&1.8$\times 10^{6}$\\
4&2011 Mar. 21&&03:27&130&1.1$\times 10^{-2}$&151.6&5.8$\times 10^{5}$&&02:56&42&2.2$\times 10^{0}$&1.3$\pm$0.2&182.4&2.5$\times 10^{7}$&&\ldots&\ldots&\ldots&\ldots&\ldots&\ldots\\
5&2011 Aug. 04&&04:40&36&8.5$\times 10^{-2}$&109.8&5.0$\times 10^{6}$&& 06:49\textsuperscript{+}&-65&1.9$\times 10^{-3}$&36.4$\pm$2.4&89.2&1.4$\times 10^{5}$&&20:26&128&2.4$\times 10^{-3}$&2.4$\pm$1.2&99.6&1.3$\times 10^{5}$\\
6&2011 Sep. 06&&23:27&28&1.8$\times 10^{-2}$&178.0&3.9$\times 10^{5}$&&\ldots&\ldots&\ldots&\ldots&\ldots&\ldots&&04:59\textsuperscript{+}&123&2.8$\times 10^{-3}$&6.4$\pm$0.4&65.7&8.4$\times 10^{4}$\\
7&2011 Sep. 22&&N/A&-89&3.5$\times 10^{-3}$&526.1&2.7$\times 10^{6}$&&15:35&167&1.5$\times 10^{-2}$&11.2$\pm$0.3&259.4&8.0$\times 10^{5}$&&11:18&8&9.6$\times 10^{-1}$&2.4$\pm$0.2&60.7&1.0$\times 10^{8}$\\
8&2011 Oct. 04&&\ldots&\ldots&\ldots&\ldots&\ldots&&18:35&105&3.7$\times 10^{-3}$&8.4$\pm$0.9&85.7&2.3$\times 10^{5}$&&15:06&-53&2.1$\times 10^{-2}$&5.1$\pm$0.2&161.5&9.0$\times 10^{5}$\\
9&2011 Nov. 03&&23:39&-152&5.5$\times 10^{-3}$&229.3&1.7$\times 10^{5}$&&23:10&103&4.9$\times 10^{-1}$&2.0$\pm$0.2&171.0&4.7$\times 10^{6}$&&23:51&-50&1.4$\times 10^{-2}$&1.3$\pm$0.2&120.1&3.2$\times 10^{5}$\\
10&2012 Jan. 23&&N/A&25&$>$1.0$\times 10^{-3}$&N/A&N/A&&14:28&-83&1.9$\times 10^{-2}$&21.2$\pm$0.4&102.7&5.0$\times 10^{6}$&&06:48&139&4.7$\times 10^{-2}$&28.9$\pm$0.2&145.6&4.8$\times 10^{6}$\\
11&2012 Jan. 27&&18:48&71&1.4$\times 10^{-1}$&N/A&N/A&&21:08&-37&6.0$\times 10^{-2}$&17.4$\pm$0.2&339.5&2.5$\times 10^{7}$&&08:26\textsuperscript{++}&-175&3.3$\times 10^{-3}$&46.6$\pm$5.5&333.3&1.0$\times 10^{6}$\\
12&2012 Mar. 05&&09:57&-54&1.0$\times 10^{-3}$&40.2&N/A&&\ldots&\ldots&\ldots&\ldots&\ldots&\ldots&&06:27&64&3.9$\times 10^{-2}$&5.0$\pm$0.2&42.9&1.9$\times 10^{6}$\\
13&2012 Mar. 07&&02:11&-27&$>$2.0$\times 10^{-1}$&159.7&2.1$\times 10^{7}$\textsuperscript{S}&&05:57&-136&3.2$\times 10^{-2}$&22.4$\pm$0.4&N/A&N/A&&01:21&91&3.6$\times 10^{0}$&12.8$\pm$0.2&N/A&N/A\\
14&2012 Mar. 24&&\ldots&\ldots&\ldots&\ldots&\ldots&&00:56&84&3.3$\times 10^{-1}$&1.1$\pm$0.2&202.5&1.5$\times 10^{6}$&&06:08&-48&7.2$\times 10^{-3}$&8.0$\pm$0.5&65.9&5.3$\times 10^{5}$\\
15&2012 May 17&&01:54&76&1.1$\times 10^{-1}$&228.6&N/A&&12:28&-39&4.1$\times 10^{-3}$&21.3$\pm$0.3&93.7&1.7$\times 10^{5}$&&\ldots&\ldots&\ldots&\ldots&\ldots&\ldots\\
16&2012 Jul. 08&&N/A&83&$>$1.0$\times 10^{-2}$&N/A&N/A&&18:38&-37&1.2$\times 10^{-2}$&5.7$\pm$0.6&101.4&5.4$\times 10^{5}$&&\ldots&\ldots&\ldots&\ldots&\ldots&\ldots\\
17&2012 Jul. 12&&N/A&15&4.5$\times 10^{-2}$&135.1&1.9$\times 10^{6}$&&\ldots&\ldots&\ldots&\ldots&\ldots&\ldots&&17:35&130&2.5$\times 10^{-2}$&0.5$\pm$0.2&73.4&4.6$\times 10^{5}$\\
18&2012 Jul. 23&&05:57&137&4.5$\times 10^{-2}$&513.4&2.5$\times 10^{6}$&&03:28&16&9.7$\times 10^{1}$&17.3$\pm$0.2&334.1&4.4$\times 10^{8}$&&23:34&-108&4.2$\times 10^{-2}$&63.3$\pm$0.2&355.7&5.9$\times 10^{6}$\\
19&2012 Sep. 20&&\ldots&\ldots&\ldots&\ldots&\ldots&&16:42&76&4.2$\times 10^{-1}$&12.8$\pm$0.4&163.3&1.1$\times 10^{7}$&&20:19&-42&1.6$\times 10^{-2}$&13.6$\pm$0.6&163.5&2.2$\times 10^{6}$\\
20&2012 Sep. 28&&N/A&32&$>$5.0$\times 10^{-3}$&N/A&N/A&&07:29&-93&1.5$\times 10^{-2}$&4.6$\pm$0.2&133.3&5.8$\times 10^{6}$&&02:38&150&1.8$\times 10^{-2}$&2.9$\pm$0.2&141.6&4.3$\times 10^{5}$\\
21&2012 Nov. 08&&N/A&164&$>$1.5$\times 10^{-3}$&N/A&N/A&&11:22&37&5.5$\times 10^{-2}$&0.5$\pm$0.2&128.4&1.8$\times 10^{6}$&&\ldots&\ldots&\ldots&\ldots&\ldots&\ldots\\
22&2013 Mar. 05&&\ldots&\ldots&\ldots&\ldots&\ldots&&03:57&84&2.8$\times 10^{0}$&1.5$\pm$0.2&188.2&5.2$\times 10^{7}$&&07:10&-5&2.4$\times 10^{-2}$&14.6$\pm$0.2&212.8&3.7$\times 10^{6}$\\
23&2013 Apr. 11&&08:10&-12&9.0$\times 10^{-2}$&231.8&3.7$\times 10^{6}$&&\ldots&\ldots&\ldots&\ldots&\ldots&\ldots&&07:52&130&1.0$\times 10^{0}$&1.4$\pm$0.2&134.7&8.6$\times 10^{6}$\\
24&2013 Apr. 24&&22:53&167&2.5$\times 10^{-3}$&122.3&3.2$\times 10^{4}$&&23:16&32&1.8$\times 10^{-2}$&3.8$\pm$0.2&87.9&2.9$\times 10^{5}$&&\ldots&\ldots&\ldots&\ldots&\ldots&\ldots\\
25&2013 May 15&&10:19&-64&3.5$\times 10^{-3}$&171.5&3.1$\times 10^{6}$&&\ldots&\ldots&\ldots&\ldots&\ldots&\ldots&&BG&N/A&5.3$\times 10^{-3}$&N/A&N/A&N/A\\
26&2013 May 22&&13:47&15&1.2$\times 10^{-1}$&271.8&9.1$\times 10^{6}$&&02:55 \textsuperscript{+}&-122&5.2$\times 10^{-3}$&23.7$\pm$0.9&254.0&1.3$\times 10^{6}$&&\ldots&\ldots&\ldots&\ldots&\ldots&\ldots\\
27&2013 Aug. 20&&\ldots&\ldots&\ldots&\ldots&\ldots&&01:41&34&1.1$\times 10^{-1}$&9.8$\pm$0.2&200.9&2.2$\times 10^{7}$&&08:12&-44&1.2$\times 10^{-2}$&27.2$\pm$0.5&198.7&2.9$\times 10^{6}$\\
28&2013 Sep. 30&&N/A&33&4.0$\times 10^{-3}$&551.5&N/A&&\ldots&\ldots&\ldots&\ldots&\ldots&\ldots&&09:56&172&2.5$\times 10^{-3}$&12.2$\pm$0.5&63.2&9.9$\times 10^{4}$\\
29&2013 Oct. 05&&\ldots&\ldots&\ldots&\ldots&\ldots&&08:24&92&4.2$\times 10^{-2}$&6.1$\pm$0.2&118.0&1.2$\times 10^{6}$&&21:46&19&2.8$\times 10^{-3}$&12.1$\pm$0.6&85.7&1.6$\times 10^{5}$\\
30&2013 Oct. 11&&$\gtrsim$18:00&-44&MR&N/A&N/A&&07:58&169&8.6$\times 10^{-1}$&1.4$\pm$0.2&160.3&4.9$\times 10^{6}$&&08:29&96&8.9$\times 10^{-2}$&7.2$\pm$0.2&176.4&2.0$\times 10^{6}$\\
31&2013 Oct. 25&&13:49&-73&1.2$\times 10^{-3}$&64.5&4.3$\times 10^{4}$&&\ldots&\ldots&\ldots&\ldots&\ldots&\ldots&&09:41&69&1.0$\times 10^{-1}$&10.8$\pm$0.2&79.3&1.6$\times 10^{6}$\\
32&2013 Oct. 28&&18:49&-28&$>$4.5$\times 10^{-3}$&N/A&N/A&&20:05&-176&2.0$\times 10^{-3}$&1.6$\pm$0.6&13.1&1.6$\times 10^{4}$&&17:01&114&2.5$\times 10^{-2}$&15.6$\pm$0.2&102.0&4.6$\times 10^{5}$\\
33&2013 Nov. 02&&No data!&133&N/A&N/A&N/A&&05:02&-15&3.9$\times 10^{-1}$&0.5$\pm$0.2&125.9&8.1$\times 10^{6}$&&07:24&-84&5.9$\times 10^{-3}$&4.4$\pm$0.2&40.6&1.3$\times 10^{5}$\\
34&2013 Nov. 07&&No data!&-151&N/A&N/A&N/A&&10:54&60&3.5$\times 10^{-1}$&0.8$\pm$0.2&150.9&2.6$\times 10^{6}$&&11:11&-7&2.8$\times 10^{-1}$&3.9$\pm$0.2&201.9&3.9$\times 10^{7}$\\
35&2013 Dec. 26&&\ldots&\ldots&\ldots&\ldots&\ldots&&06:04&45&2.8$\times 10^{-2}$&7.7$\pm$0.2&176.5&2.2$\times 10^{6}$&&07:59&-13&2.4$\times 10^{-2}$&4.6$\pm$0.2&65.5&2.1$\times 10^{6}$\\
36&2013 Dec. 28&&18:32&125&1.2$\times 10^{-2}$&169.6&N/A&&06:18\textsuperscript{+}&-26&2.6$\times 10^{-3}$&0.0$\pm$1.4&104.3&1.4$\times 10^{5}$&&01:31\textsuperscript{+}&-83&3.8$\times 10^{-3}$&2.5$\pm$0.2&109.7&1.5$\times 10^{5}$\\
37&2014 Jan. 06&&N/A&89&9.2$\times 10^{-2}$&N/A&N/A&&12:42&-62&4.2$\times 10^{-3}$&12.9$\pm$1.4&36.0&N/A&&00:58\textsuperscript{+}&-118&1.5$\times 10^{-3}$&5.5$\pm$4.5&N/A&N/A\\
38&2014 Jan. 07&&19:25&8&2.0$\times 10^{-1}$&466.9&1.3$\times 10^{7}$&&00:41\textsuperscript{+}&-143&$>$1.3$\times 10^{-2}$&N/A&256.6&N/A&&01:31\textsuperscript{+}&161&8.7$\times 10^{-3}$&26.3$\pm$0.7&260.0&7.5$\times 10^{5}$\\
39&2014 Feb. 25&&03:05&-82&2.5$\times 10^{-2}$&533.2&7.9$\times 10^{6}$&&01:35&125&2.0$\times 10^{-1}$&2.7$\pm$0.2&178.4&6.2$\times 10^{6}$&&01:36&78&5.5$\times 10^{-1}$&9.6$\pm$0.2&178.4&2.8$\times 10^{7}$\\
40&2014 Mar. 04&&\ldots&\ldots&\ldots&\ldots&\ldots&&19:24&-65&9.0$\times 10^{-3}$&1.1$\pm$0.2&19.4&1.1$\times 10^{5}$&&10:35\textsuperscript{+}&-111&2.0$\times 10^{-3}$&4.2$\pm$9.9&11.2&1.8$\times 10^{4}$\\
41&2014 Mar. 05&&\ldots&\ldots&\ldots&\ldots&\ldots&&14:48&28&3.7$\times 10^{-2}$&4.7$\pm$0.2&142.5&6.8$\times 10^{5}$&&21:49&-18&4.6$\times 10^{-3}$&8.9$\pm$1.0&142.5&2.1$\times 10^{5}$\\
42&2014 Sep. 01&&20:41&-128&1.3$\times 10^{-2}$&214.8&1.6$\times 10^{6}$&&N/A&65&N/A&N/A&N/A&N/A&&11:49&33&2.0$\times 10^{1}$&7.2$\pm$0.2&225.8&2.0$\times 10^{8}$\\
43&2014 Sep. 10&&19:28&-2&4.6$\times 10^{-2}$&244.5&2.5$\times 10^{6}$&&BG&N/A&N/A&N/A&N/A&N/A&&21:39&159&5.8$\times 10^{-3}$&7.5$\pm$0.9&248.9&3.9$\times 10^{5}$\\
44&2014 Sep. 25&&00:05&-175&1.3$\times 10^{-3}$&383.9&1.2$\times 10^{5}$&&N/A&17&N/A&N/A&N/A&N/A&&22:47\textsuperscript{*}&-14&4.1$\times 10^{-1}$&13.5$\pm$0.2&$>$65.7&$>$1.4$\times 10^{7}$\\
45&2014 Dec. 13&&N/A&141&2.0$\times 10^{-4}$&N/A&N/A&&N/A&N/A&3.3$\times 10^{0}$&N/A&N/A&N/A&&No data!&N/A&N/A&N/A&N/A&N/A\\
46&2015 Jul. 01&&15:32&N/A&7.5$\times 10^{-4}$&99.4&N/A&&N/A&N/A&N/A&N/A&N/A&N/A&&No data!&N/A&N/A&N/A&N/A&N/A\\
\end{longtable}
\raggedright
\noindent *: The time given refers to the previous calendar day.\\
\noindent +: The time given refers to the following calendar day.\\
\noindent ++: The time given refers to the second calendar day after the date given.\\
\noindent S: Value is uncertain/unreliable due to saturation of ERNE/HED during the event.\\
Items marked with "N/A" or "No data!" could not be determined due to insufficient or completely missing data, respectively. The ellipsis (\ldots) indicates that the event in question was not detected. "BG" signifies that an event may have been present but is masked by elevated particle intensity after previous event(s); "MR" signifies that the event is marginal but probably present. The onset time preceded by $\gtrsim$ is a visually estimated lower limit. All times are rounded to the nearest minute. Proton fluence is given for the energy range $\approx$14 MeV to $\approx$100 MeV.
\normalsize
\newpage

% Proton events with VDA params (but not TSA) + flare max. dI/dt time:

\scriptsize
\setlength{\tabcolsep}{2pt}
\begin{longtable}{llp{0.5pt}lrlllp{0.5pt}lrlllp{0.5pt}lrlllp{0.5pt}l}
\caption{\label{sep_protons_2} Proton event onset times, connection angles, and selected \textbf{VDA} parameters with the maximum of the soft X-ray flux time derivative of the associated solar flare for $>$ 55 MeV multi-spacecraft proton events in 2009--2016.}\\
\hline\hline
\noalign{\smallskip}
ID & Date &&\multicolumn{5}{c}{SOHO/ERNE 55--80 MeV proton flux}&&\multicolumn{5}{c}{STEREO-A/HET 40--100 MeV proton flux}&&\multicolumn{5}{c}{STEREO-B/HET 40--100 MeV proton flux}&&\multirow{2}{1.2cm}{X-ray flare max. d$I$/d$t$}\\
\cline{4-8}
\cline{10-14}
\cline{16-20}
\noalign{\smallskip}
&& & Onset & $\phi_{\rm C}$ [\textdegree]&\multirow{2}{0.8cm}{VDA $s$ [AU]}&\multirow{2}{1cm}{VDA $t_0$ + 500 s}&$R^2$&& Onset & $\phi_{\rm C}$ [\textdegree]& \multirow{2}{0.8cm}{VDA $s$ [AU]}& \multirow{2}{1cm}{VDA $t_0$ + 500 s}&$R^2$&& Onset & $\phi_{\rm C}$ [\textdegree]& \multirow{2}{0.8cm}{VDA $s$ [AU]}&\multirow{2}{1cm}{VDA $t_0$ + 500 s}&$R^2$&&\\
&&&[UT]&&&&&&[UT]&&&&&&[UT]&&&&&\\
&&&&&&[UT]&&&&&&[UT]&&&&&&[UT]&&&[UT]\\
\hline
\endfirsthead
\caption{continued.}\\
\hline\hline
\noalign{\smallskip}
ID & Date &&\multicolumn{5}{c}{SOHO/ERNE 55--80 MeV proton flux}&&\multicolumn{5}{c}{STEREO-A/HET 40--100 MeV proton flux}&&\multicolumn{5}{c}{STEREO-B/HET 40--100 MeV proton flux}&&\multirow{2}{1.2cm}{X-ray flare max. d$I$/d$t$}\\
\cline{4-8}
\cline{10-14}
\cline{16-20}
\noalign{\smallskip}
&& & Onset & $\phi_{\rm C}$ [\textdegree]&\multirow{2}{0.8cm}{VDA $s$ [AU]}&\multirow{2}{1cm}{VDA $t_0$ + 500 s}&$R^2$&& Onset & $\phi_{\rm C}$ [\textdegree]& \multirow{2}{0.8cm}{VDA $s$ [AU]}& \multirow{2}{1cm}{VDA $t_0$ + 500 s}&$R^2$&& Onset & $\phi_{\rm C}$ [\textdegree]& \multirow{2}{0.8cm}{VDA $s$ [AU]}&\multirow{2}{1cm}{VDA $t_0$ + 500 s}&$R^2$&&\\
&&&[UT]&&&&&&[UT]&&&&&&[UT]&&&&&\\
&&&&&&[UT]&&&&&&[UT]&&&&&&[UT]&&&[UT]\\
\hline%\hline
\noalign{\smallskip}
\endhead
\noalign{\smallskip}
\hline
\endfoot
\noalign{\smallskip}
1&2011 Jan. 28&&01:56&5&1.96$\pm$0.11&01:12$\pm$00:08&0.951&&02:37&-46&1.53$\pm$0.37&02:25$\pm$00:15&0.776&&\ldots&\ldots&\ldots&\ldots&\ldots&&00:58\\
2&2011 Feb. 15&&03:39&-47&3.23$\pm$0.24&01:55$\pm$00:21&0.917&&\ldots&\ldots&\ldots&\ldots&\ldots&&03:18&22&1.30$\pm$0.19&03:06$\pm$00:08&0.880&&01:52\\
3&2011 Mar. 07&&21:06&-16&1.77$\pm$0.11&20:37$\pm$00:08&0.944&&19:14\textsuperscript{+}&-84&N/A&N/A&N/A&&21:46&81&0.38$\pm$0.11&21:45$\pm$00:04&0.625&&19:56\\
4&2011 Mar. 21&&03:27&59&2.41$\pm$0.09&02:39$\pm$00:07&0.978&&02:56&-16&1.16$\pm$0.08&02:43$\pm$00:04&0.953&&\ldots&\ldots&\ldots&\ldots&\ldots&&N/A\\
5&2011 Aug. 04&&04:40&-36&1.78$\pm$0.09&04:22$\pm$00:08&0.853&& 06:49\textsuperscript{+}&-122&18.78$\pm$3.53&09:35$\pm$02:23&0.905&&20:26&66&7.92$\pm$2.45&17:36$\pm$01:43&0.723&&03:53\\
6&2011 Sep. 06&&23:27&-34&1.85$\pm$0.21&23:24$\pm$00:16&0.819&&\ldots&\ldots&\ldots&\ldots&\ldots&&04:59\textsuperscript{+}&73&N/A&N/A&N/A&&22:18\\
7&2011 Sep. 22&&N/A&-153&N/A&N/A&N/A&&15:35&99&6.88$\pm$1.02&14:10$\pm$00:39&0.821&&11:18&-50&1.97$\pm$0.10&10:42$\pm$00:04&0.979&&10:44\\
8&2011 Oct. 04&&\ldots&\ldots&\ldots&\ldots&\ldots&&18:35&50&N/A&N/A&N/A&&15:06&-95&1.66$\pm$0.21&14:48$\pm$00:08&0.884&&N/A\\
9&2011 Nov. 03&&23:39&135&2.83$\pm$0.14&22:24$\pm$00:10&0.962&&23:10&12&1.23$\pm$0.09&23:00$\pm$00:05&0.953&&23:51&-139&2.55$\pm$0.19&23:05$\pm$00:08&0.954&&N/A\\
10&2012 Jan. 23&&N/A&-33&N/A&N/A&N/A&&14:28&-137&N/A&N/A&N/A&&06:48&60&N/A&N/A&N/A&&03:49\\
11&2012 Jan. 27&&18:48&23&3.26$\pm$0.12&17:59$\pm$00:10&0.953&&21:08&-99&6.91$\pm$0.99&19:23$\pm$00:35&0.860&&08:26\textsuperscript{++}&120&N/A&N/A&N/A&&18:26\\
12&2012 Mar. 05&&09:57&-124&N/A&N/A&N/A&&\ldots&\ldots&\ldots&\ldots&\ldots&&06:27&-7&14.96$\pm$2.79&02:03$\pm$02:50&0.828&&03:46\\
13&2012 Mar. 07&&02:11&-91&3.84$\pm$0.69&03:09$\pm$00:50&0.637&&05:57&146&3.02$\pm$1.70&05:35$\pm$00:50&0.444&&01:21&17&1.86$\pm$0.05&00:46$\pm$00:02&0.993&&00:18\\
14&2012 Mar. 24&&\ldots&\ldots&\ldots&\ldots&\ldots&&00:56&24&1.53$\pm$0.05&00:38$\pm$00:03&0.988&&06:08&-117&11.38$\pm$1.80&02:13$\pm$01:08&0.852&&N/A\\
15&2012 May 17&&01:54&6&2.93$\pm$0.17&00:48$\pm$00:14&0.893&&12:28&-78&13.37$\pm$2.96&09:14$\pm$01:46&0.804&&\ldots&\ldots&\ldots&\ldots&\ldots&&01:36\\
16&2012 Jul. 08&&N/A&22&N/A&N/A&N/A&&18:38&-102&10.66$\pm$4.84&15:34$\pm$02:15&0.619&&\ldots&\ldots&\ldots&\ldots&\ldots&&16:29\\
17&2012 Jul. 12&&N/A&-45&N/A&N/A&N/A&&\ldots&\ldots&\ldots&\ldots&\ldots&&17:35&57&0.89$\pm$0.08&17:33$\pm$00:04&0.925&&16:32\\
18&2012 Jul. 23&&05:57&79&2.85$\pm$0.29&05:39$\pm$00:22&0.841&&03:28&-38&1.28$\pm$0.13&03:07$\pm$00:06&0.918&&23:34&179&N/A&N/A&N/A&&N/A\\
19&2012 Sep. 20&&\ldots&\ldots&\ldots&\ldots&\ldots&&16:42&16&N/A&N/A&N/A&&20:19&-119&N/A&N/A&N/A&&N/A\\
20&2012 Sep. 28&&N/A&-41&N/A&N/A&N/A&&07:29&-160&N/A&N/A&N/A&&02:38&84&0.40$\pm$0.14&02:54$\pm$00:06&0.594&&23:43\textsuperscript{*}\\
21&2012 Nov. 08&&N/A&105&N/A&N/A&N/A&&11:22&-27&1.58$\pm$0.11&11:00$\pm$00:04&0.959&&\ldots&\ldots&\ldots&\ldots&\ldots&&N/A\\
22&2013 Mar. 05&&\ldots&\ldots&\ldots&\ldots&\ldots&&03:57&14&1.21$\pm$0.06&03:43$\pm$00:03&0.977&&07:10&-103&4.12$\pm$0.74&06:21$\pm$00:29&0.777&&N/A\\
23&2013 Apr. 11&&08:10&-74&2.03$\pm$0.09&07:22$\pm$00:07&0.969&&\ldots&\ldots&\ldots&\ldots&\ldots&&07:52&54&1.98$\pm$0.13&07:21$\pm$00:05&0.965&&07:10\\
24&2013 Apr. 24&&22:53&106&1.33$\pm$0.03&22:30$\pm$00:04&0.833&&23:16&-20&1.12$\pm$0.51&23:51$\pm$00:22&0.376&&\ldots&\ldots&\ldots&\ldots&\ldots&&N/A\\
25&2013 May 15&&10:19&-130&3.21$\pm$0.33&06:56$\pm$00:28&0.852&&\ldots&\ldots&\ldots&\ldots&\ldots&&BG&26&N/A&N/A&N/A&&01:42\\
26&2013 May 22&&13:47&-43&1.70$\pm$0.04&13:18$\pm$00:04&0.969&&02:55\textsuperscript{+}&168&N/A&N/A&N/A&&\ldots&\ldots&\ldots&\ldots&\ldots&&13:17\\
27&2013 Aug. 20&&\ldots&\ldots&\ldots&\ldots&\ldots&&01:41&-29&4.45$\pm$0.37&00:17$\pm$00:13&0.944&&08:12&-120&9.12$\pm$1.29&04:42$\pm$00:51&0.878&&N/A\\
28&2013 Sep. 30&&N/A&-65&N/A&N/A&N/A&&\ldots&\ldots&\ldots&\ldots&\ldots&&09:56&75&6.63$\pm$1.46&07:22$\pm$00:59&0.748&&21:48\textsuperscript{*}\\
29&2013 Oct. 05&&\ldots&\ldots&\ldots&\ldots&\ldots&&08:24&18&2.48$\pm$0.36&07:54$\pm$00:13&0.857&&21:46&-65&N/A&N/A&N/A&&N/A\\
30&2013 Oct. 11&&$\gtrsim$18:00&-108&MR&N/A&N/A&&07:58&96&1.60$\pm$0.08&07:39$\pm$00:04&0.977&&08:29&9&2.18$\pm$0.14&08:01$\pm$00:08&0.958&&07:15\\
31&2013 Oct. 25&&13:49&-149&5.15$\pm$0.18&09:54$\pm$00:22&0.693&&\ldots&\ldots&\ldots&\ldots&\ldots&&09:41&-27&4.86$\pm$0.24&07:51$\pm$00:10&0.981&&07:59\\
32&2013 Oct. 28&&18:49&-116&N/A&N/A&N/A&&20:05&121&N/A&N/A&N/A&&17:01&36&4.77$\pm$0.22&15:36$\pm$00:10&0.984&&15:11\\
33&2013 Nov. 02&&No data!&N/A&N/A&N/A&N/A&&05:02&-73&1.33$\pm$0.05&04:40$\pm$00:03&0.988&&07:24&-145&2.63$\pm$0.31&06:56$\pm$00:13&0.900&&N/A\\
34&2013 Nov. 07&&No data!&N/A&N/A&N/A&N/A&&10:54&15&2.03$\pm$0.13&10:24$\pm$00:05&0.969&&11:11&-55&1.90$\pm$0.42&10:52$\pm$00:18&0.748&&N/A\\
35&2013 Dec. 26&&\ldots&\ldots&\ldots&\ldots&\ldots&&06:04&-12&N/A&N/A&N/A&&07:59&-85&N/A&N/A&N/A&&N/A\\
36&2013 Dec. 28&&18:32&52&2.06$\pm$0.01&17:55$\pm$00:01&0.995&&06:18\textsuperscript{+}&-89&N/A&N/A&N/A&&01:31\textsuperscript{+}&-141&12.60$\pm$4.45&19:55$\pm$02:32&0.728&&N/A\\
37&2014 Jan. 06&&N/A&22&N/A&N/A&N/A&&12:42&-133&4.93$\pm$3.02&11:45$\pm$01:52&0.471&&00:58\textsuperscript{+}&161&17.27$\pm$4.68&19:30$\pm$03:03&0.774&&07:43\\
38&2014 Jan. 07&&19:25&-58&3.71$\pm$0.02&18:10$\pm$00:03&0.990&&00:41\textsuperscript{+}&151&16.21$\pm$1.67&19:07$\pm$01:02&0.941&&01:31\textsuperscript{+}&68&8.53$\pm$1.97&22:43$\pm$01:18&0.826&&18:18\\
39&2014 Feb. 25&&03:05&-138&4.23$\pm$0.10&01:55$\pm$00:11&0.872&&01:35&59&1.11$\pm$0.07&01:29$\pm$00:04&0.965&&01:36&0&1.63$\pm$0.11&01:14$\pm$00:05&0.959&&00:46\\
40&2014 Mar. 04&&\ldots&\ldots&\ldots&\ldots&\ldots&&19:24&-129&1.87$\pm$0.34&19:02$\pm$00:12&0.820&&10:35\textsuperscript{+}&163&N/A&N/A&N/A&&18:22\\
41&2014 Mar. 05&&\ldots&\ldots&\ldots&\ldots&\ldots&&14:48&-42&4.38$\pm$0.58&13:35$\pm$00:21&0.880&&21:49&-89&N/A&N/A&N/A&&N/A\\
42&2014 Sep. 01&&20:41&169&9.20$\pm$0.27&17:39$\pm$00:31&0.786&&N/A&6&N/A&N/A&N/A&&11:49&-27&2.42$\pm$0.39&11:30$\pm$00:17&0.847&&N/A\\
43&2014 Sep. 10&&19:28&-71&1.90$\pm$0.16&19:44$\pm$00:14&0.918&&BG&N/A&N/A&N/A&N/A&&21:39&80&7.78$\pm$1.82&19:47$\pm$01:13&0.755&&17:33\\
44&2014 Sep. 25&&00:05&127&6.05$\pm$0.25&22:25$\pm$00:36\textsuperscript{*}&0.336&&N/A&-50&N/A&N/A&N/A&&22:47\textsuperscript{*}&-81&4.11$\pm$0.26&21:38$\pm$00:14\textsuperscript{*}&0.956&&N/A\\
45&2014 Dec. 13&&N/A&95&N/A&N/A&N/A&&N/A&N/A&N/A&N/A&N/A&&No data!&N/A&N/A&N/A&N/A&&N/A\\
46&2015 Jul. 01&&15:32&N/A&1.41$\pm$0.05&15:02$\pm$00:04&0.980&&N/A&N/A&N/A&N/A&N/A&&No data!&N/A&N/A&N/A&N/A&&N/A\\
\end{longtable}
\raggedright
\noindent *: The time given refers to the previous calendar day.\\
\noindent +: The time given refers to the following calendar day.\\
\noindent ++: The time given refers to the second calendar day after the date given.\\
Items marked with "N/A" or "No data!" could not be determined due to insufficient or completely missing data, respectively. The ellipsis (\ldots) indicates that the event in question was not detected. "BG" signifies that an event may have been present but is masked by elevated particle intensity after previous event(s); "MR" signifies that the event is marginal but probably present. The onset time preceded by $\gtrsim$ is a visually estimated lower limit. All times are rounded to the nearest minute.
\normalsize
\newpage

% Proton events with TSA params:

\scriptsize
\setlength{\tabcolsep}{2pt}
\begin{longtable}{llp{0.5pt}lrrlp{0.5pt}lrrlp{0.5pt}lrrlp{0.5pt}l}
\caption{\label{sep_protons_3} Proton event onset times, connection angles, and \textbf{TSA} parameters with the maximum of the soft X-ray flux time derivative of the associated solar flare for $>$55 MeV multi-spacecraft proton events in 2009--2016.}\\
\hline\hline
\noalign{\smallskip}
ID & Date &&\multicolumn{4}{c}{SOHO/ERNE 55--80 MeV proton flux}&&\multicolumn{4}{c}{STEREO-A/HET 40--100 MeV proton flux}&&\multicolumn{4}{c}{STEREO-B/HET 40--100 MeV proton flux}&&\multirow{2}{1.2cm}{X-ray flare max. d$I$/d$t$}\\
\cline{4-7}
\cline{9-12}
\cline{14-17}
\noalign{\smallskip}
&& & Onset & $\phi_{\rm C}$ [\textdegree]&\multirow{2}{0.8cm}{TSA $L$ [AU]}&\multirow{2}{1cm}{TSA $t_{\rm rel}$ + 500 s}&& Onset & $\phi_{\rm C}$ [\textdegree]&\multirow{2}{0.8cm}{TSA $L$ [AU]}&\multirow{2}{1cm}{TSA $t_{\rm rel}$ + 500 s}&&Onset & $\phi_{\rm C}$ [\textdegree]&\multirow{2}{0.8cm}{TSA $L$ [AU]}&\multirow{2}{1cm}{TSA $t_{\rm rel}$ + 500 s}&&\\
&&&[UT]&&&&&[UT]&&&&&[UT]&&&&&\\
&&&&&&[UT]&&&&&[UT]&&&&&[UT]&&[UT]\\
\hline
\endfirsthead
\caption{continued.}\\
\hline\hline
\noalign{\smallskip}
ID & Date &&\multicolumn{4}{c}{SOHO/ERNE 55--80 MeV proton flux}&&\multicolumn{4}{c}{STEREO-A/HET 40--100 MeV proton flux}&&\multicolumn{4}{c}{STEREO-B/HET 40--100 MeV proton flux}&&\multirow{2}{1.2cm}{X-ray flare max. d$I$/d$t$}\\
\cline{4-7}
\cline{9-12}
\cline{14-17}
\noalign{\smallskip}
&& & Onset & $\phi_{\rm C}$ [\textdegree]&\multirow{2}{0.8cm}{TSA $L$ [AU]}&\multirow{2}{1cm}{TSA $t_{\rm rel}$ + 500 s}&& Onset & $\phi_{\rm C}$ [\textdegree]&\multirow{2}{0.8cm}{TSA $L$ [AU]}&\multirow{2}{1cm}{TSA $t_{\rm rel}$ + 500 s}&&Onset & $\phi_{\rm C}$ [\textdegree]&\multirow{2}{0.8cm}{TSA $L$ [AU]}&\multirow{2}{1cm}{TSA $t_{\rm rel}$ + 500 s}&&\\
&&&[UT]&&&&&[UT]&&&&&[UT]&&&&&\\
&&&&&&[UT]&&&&&[UT]&&&&&[UT]&&[UT]\\
\hline%\hline
\noalign{\smallskip}
\endhead
\noalign{\smallskip}
\hline
\endfoot
\noalign{\smallskip}
1&2011 Jan. 28&&01:56&5&1.24&01:36$\pm$00:10&&02:37&-46&1.06&02:20$\pm$00:13&&\ldots&\ldots&\ldots&\ldots&&00:58\\
2&2011 Feb. 15&&03:39&-47&1.11&03:22$\pm$00:37&&\ldots&\ldots&\ldots&\ldots&&03:18&22&1.31&02:55$\pm$00:10&&01:52\\
3&2011 Mar. 07&&21:06&-16&1.16&20:48$\pm$00:12&&19:14\textsuperscript{+}&-84&1.05&18:57$\pm$00:57\textsuperscript{+}&&21:46&81&1.19&21:26$\pm$00:10&&19:56\\
4&2011 Mar. 21&&03:27&59&1.19&03:08$\pm$00:10&&02:56&-16&1.10&02:38$\pm$00:10&&\ldots&\ldots&\ldots&\ldots&&N/A\\
5&2011 Aug. 04&&04:40&-36&1.22&04:21$\pm$00:12&&06:49\textsuperscript{+}&-122&1.11&06:31$\pm$02:26\textsuperscript{+}&&20:26&66&1.23&20:05$\pm$01:11&&03:53\\
6&2011 Sep. 06&&23:27&-34&1.16&23:09$\pm$00:10&&\ldots&\ldots&\ldots&\ldots&&04:59\textsuperscript{+}&73&1.19&04:39$\pm$00:26&&22:18\\
7&2011 Sep. 22&&N/A&-53&1.17&N/A&&15:35&99&1.16&15:16$\pm$00:16&&11:18&-50&1.24&10:57$\pm$00:10&&10:44\\
8&2011 Oct. 04&&\ldots&\ldots&\ldots&\ldots&&18:35&50&1.10&18:17$\pm$00:55&&15:06&-95&1.17&14:47$\pm$00:11&&N/A\\
9&2011 Nov. 03&&23:39&135&1.20&23:20$\pm$00:12&&23:10&12&1.28&22:48$\pm$00:10&&23:51&-139&1.44&23:25$\pm$00:10&&N/A\\
10&2012 Jan. 23&&N/A&-33&1.12&N/A&&14:28&-137&1.09&14:10$\pm$00:26&&06:48&60&1.33&06:25$\pm$00:12&&03:49\\
11&2012 Jan. 27&&18:48&23&1.07&18:32$\pm$00:11&&21:08&-99&1.12&20:50$\pm$00:12&&08:26\textsuperscript{++}&120&1.25&08:05$\pm$05:28\textsuperscript{++}&&18:26\\
12&2012 Mar. 05&&09:57&-124&1.18&09:39$\pm$03:00&&\ldots&\ldots&\ldots&\ldots&&06:27&-7&1.25&06:06$\pm$00:14&&03:46\\
13&2012 Mar. 07&&02:11&-91&1.15&01:53$\pm$00:10&&05:57&146&1.20&05:37$\pm$00:26&&01:21&17&1.27&00:59$\pm$00:10&&00:18\\
14&2012 Mar. 24&&\ldots&\ldots&\ldots&\ldots&&00:56&24&1.11&00:38$\pm$00:10&&06:08&-117&1.23&05:47$\pm$00:28&&N/A\\
15&2012 May 17&&01:54&6&1.21&01:35$\pm$00:10&&12:28&-78&1.03&12:12$\pm$00:21&&\ldots&\ldots&\ldots&\ldots&&01:36\\
16&2012 Jul. 08&&N/A&22&1.17&N/A&&18:38&-102&1.14&18:19$\pm$00:35&&\ldots&\ldots&\ldots&\ldots&&16:29\\
17&2012 Jul. 12&&N/A&-45&1.16&N/A&&\ldots&\ldots&\ldots&\ldots&&17:35&57&1.24&17:14$\pm$00:10&&16:32\\
18&2012 Jul. 23&&05:57&79&1.15&05:39$\pm$00:10&&03:28&-38&1.09&03:10$\pm$00:10&&23:34&179&1.25&23:13$\pm$00:14&&N/A\\
19&2012 Sep. 20&&\ldots&\ldots&\ldots&\ldots&&16:42&16&1.12&16:24$\pm$00:24&&20:19&-119&1.33&19:56$\pm$00:34&&N/A\\
20&2012 Sep. 28&&N/A&-41&1.21&N/A&&07:29&-160&1.16&07:10$\pm$00:10&&02:38&84&1.27&02:16$\pm$00:14&&23:43\textsuperscript{*}\\
21&2012 Nov. 08&&N/A&105&1.13&N/A&&11:22&-27&1.14&11:03$\pm$00:10&&\ldots&\ldots&\ldots&\ldots&&N/A\\
22&2013 Mar. 05&&\ldots&\ldots&\ldots&\ldots&&03:57&14&1.16&03:38$\pm$00:10&&07:10&-103&1.44&06:44$\pm$00:10&&N/A\\
23&2013 Apr. 11&&08:10&-74&1.16&07:52$\pm$00:10&&\ldots&\ldots&\ldots&\ldots&&07:52&54&1.27&07:30$\pm$00:10&&07:10\\
24&2013 Apr. 24&&22:53&106&1.16&22:35$\pm$00:10&&23:16&-20&1.08&22:59$\pm$00:10&&\ldots&\ldots&\ldots&\ldots&&N/A\\
25&2013 May 15&&10:19&-130&1.19&10:00$\pm$00:10&&\ldots&\ldots&\ldots&\ldots&&BG&N/A&N/A&N/A&&01:42\\
26&2013 May 22&&13:47&-43&1.15&13:29$\pm$00:10&&02:55\textsuperscript{+}&168&1.16&02:36$\pm$00:55\textsuperscript{+}&&\ldots&\ldots&\ldots&\ldots&&13:17\\
27&2013 Aug. 20&&\ldots&\ldots&\ldots&\ldots&&01:41&-29&1.14&01:22$\pm$00:10&&08:12&-120&1.27&21:13$\pm$00:32&&N/A\\
28&2013 Sep. 30&&N/A&-65&1.36&N/A&&\ldots&\ldots&\ldots&\ldots&&09:56&75&1.44&09:30$\pm$00:30&&21:48\textsuperscript{*}\\
29&2013 Oct. 05&&\ldots&\ldots&\ldots&\ldots&&08:24&18&1.19&08:04$\pm$00:10&&21:46&-65&1.36&21:22$\pm$00:35&&N/A\\
30&2013 Oct. 11&&$\gtrsim$18:00&-108&1.16&N/A&&07:58&96&1.18&07:38$\pm$00:10&&08:29&9&1.39&08:04$\pm$00:10&&07:15\\
31&2013 Oct. 25&&13:49&-149&1.22&13:30$\pm$02:36&&\ldots&\ldots&\ldots&\ldots&&09:41&-27&1.46&09:15$\pm$00:10&&07:59\\
32&2013 Oct. 28&&18:49&-116&1.29&18:28$\pm$00:10&&20:05&121&1.13&19:46$\pm$00:35&&17:01&36&1.34&16:37$\pm$00:10&&15:11\\
33&2013 Nov. 02&&No data!&N/A&1.21&N/A&&05:02&-73&1.11&04:44$\pm$00:10&&07:24&-145&1.25&07:03$\pm$00:13&&N/A\\
34&2013 Nov. 07&&No data!&N/A&1.18&N/A&&10:54&15&1.06&10:37$\pm$00:10&&11:11&-55&1.19&10:51$\pm$00:10&&N/A\\
35&2013 Dec. 26&&\ldots&\ldots&\ldots&\ldots&&06:04&-12&1.10&05:46$\pm$00:14&&07:59&-85&1.33&07:36$\pm$00:12&&N/A\\
36&2013 Dec. 28&&18:32&52&1.19&18:13$\pm$00:10&&06:18\textsuperscript{+}&-89&1.13&05:59$\pm$01:27\textsuperscript{+}&&01:31\textsuperscript{+}&-141&1.25&01:10$\pm$00:10\textsuperscript{+}&&N/A\\
37&2014 Jan. 06&&N/A&22&1.16&N/A&&12:42&-133&1.17&12:23$\pm$01:27&&00:58\textsuperscript{+}&161&1.38&00:33$\pm$04:30\textsuperscript{+}&&07:43\\
38&2014 Jan. 07&&19:25&-58&1.15&19:07$\pm$00:12&&00:41\textsuperscript{+}&151&1.14&00:22$\pm$01:31\textsuperscript{+}&&01:31\textsuperscript{+}&68&1.46&01:05$\pm$00:43\textsuperscript{+}&&18:18\\
39&2014 Feb. 25&&03:05&-138&1.12&02:48$\pm$00:45&&01:35&59&1.14&01:16$\pm$00:10&&01:36&0&1.36&01:12$\pm$00:10&&00:46\\
40&2014 Mar. 04&&\ldots&\ldots&\ldots&\ldots&&19:24&-129&1.13&19:06$\pm$00:10&&10:35\textsuperscript{+}&163&1.38&10:10$\pm$09:56&&18:22\\
41&2014 Mar. 05&&\ldots&\ldots&\ldots&\ldots&&14:48&-42&1.16&14:29$\pm$00:13&&21:49&-89&1.29&21:27$\pm$01:03&&N/A\\
42&2014 Sep. 01&&20:41&169&1.17&20:23$\pm$00:19&&N/A&6&N/A&N/A&&11:49&-27&1.18&11:29$\pm$00:10&&N/A\\
43&2014 Sep. 10&&19:28&-71&1.20&19:09$\pm$00:28&&BG&N/A&N/A&N/A&&21:39&80&1.29&21:17$\pm$00:53&&17:33\\
44&2014 Sep. 25&&00:05&127&1.14&23:48$\pm$01:09\textsuperscript{*}&&N/A&N/A&N/A&N/A&&22:47\textsuperscript{*}&-81&1.23&22:26$\pm$00:10\textsuperscript{*}&&N/A\\
45&2014 Dec. 13&&N/A&95&1.07&N/A&&N/A&N/A&N/A&N/A&&No data!&N/A&N/A&N/A&&N/A\\
46&2015 Jul. 01&&15:32&N/A&1.18&15:14$\pm$00:10&&N/A&N/A&N/A&N/A&&No data!&N/A&N/A&N/A&&N/A\\
\end{longtable}
\raggedright
\noindent *: The time given refers to the previous calendar day.\\
\noindent +: The time given refers to the following calendar day.\\
\noindent ++: The time given refers to the second calendar day after the date given.\\
Items marked with "N/A" or "No data!" could not be determined due to insufficient or completely missing data, respectively. The ellipsis (\ldots) indicates that the event in question was not detected. "BG" signifies that an event may have been present but is masked by elevated particle intensity after previous event(s); "MR" signifies that the event is marginal but probably present. The onset time preceded by $\gtrsim$ is a visually estimated lower limit. All times are rounded to the nearest minute.
\normalsize
\newpage

% Electron events with TSA params:
\addtolength{\textheight}{2.5cm}
\scriptsize
\setlength{\tabcolsep}{1.2pt}
\begin{longtable}{llp{0.5pt}lrrllllp{0.5pt}lrrlllp{0.5pt}lrrlll}
\caption{\label{sep_electrons_1} Electron fluxes, flare-to-observer longitudinal distances, event rise times and durations, and TSA parameters for 0.18--0.31 MeV multi-spacecraft electron events in 2009--2016.}\\
\hline\hline
\noalign{\smallskip}
ID & Date &&\multicolumn{7}{c}{ACE/EPAM 0.18--0.31 MeV electron flux}&&\multicolumn{6}{c}{STEREO-A/SEPT 0.165--0.335 MeV electron flux}&&\multicolumn{6}{c}{STEREO-B/SEPT 0.165--0.335 MeV electron flux}\\
\cline{4-10}
\cline{12-17}
\cline{19-24}
\noalign{\smallskip}
&&& Onset & $\Delta \phi$& \multirow{3}{1cm}{$I_{\rm max}$ [pfu MeV$^\textrm{-1}$]} & \multirow{3}{0.8cm}{Rise time [h]} & \multirow{2}{0.7cm}{TSA $L$ [AU]} & \multirow{2}{0.9cm}{TSA $t_{\rm rel}$ + 500 s}&\multirow{2}{0.7cm}{Data type}&& Onset & $\Delta \phi$& \multirow{3}{1cm}{$I_{\rm max}$ [pfu MeV$^\textrm{-1}$]} & \multirow{3}{0.8cm}{Rise time [h]} & \multirow{2}{0.7cm}{TSA $L$ [AU]} & \multirow{2}{0.9cm}{TSA $t_{\rm rel}$ + 500 s}&& Onset & $\Delta \phi$& \multirow{3}{1cm}{$I_{\rm max}$ [pfu MeV$^\textrm{-1}$]} & \multirow{3}{0.8cm}{Rise time [h]} &\multirow{2}{0.7cm}{TSA $L$ [AU]} & \multirow{2}{0.9cm}{TSA $t_{\rm rel}$ + 500 s}\\
&&&[UT]&[\textdegree]&&&&&&&[UT]&[\textdegree]&&&&&&[UT]&[\textdegree]&&\\
&&&&&&&&&&&&&&&&&&&&&\\
&&&&&&&&[UT]&&&&&&&&[UT]&&&&&&&[UT]\\
\hline
\endfirsthead
\caption{continued.}\\
\hline\hline
\noalign{\smallskip}
ID & Date &&\multicolumn{7}{c}{ACE/EPAM 0.18--0.31 MeV electron flux}&&\multicolumn{6}{c}{STEREO-A/SEPT 0.165--0.335 MeV electron flux}&&\multicolumn{6}{c}{STEREO-B/SEPT 0.165--0.335 MeV electron flux}\\
\cline{4-10}
\cline{12-17}
\cline{19-24}
\noalign{\smallskip}
&&& Onset & $\Delta \phi$& \multirow{3}{1cm}{$I_{\rm max}$ [pfu MeV$^\textrm{-1}$]} & \multirow{3}{0.8cm}{Rise time [h]} & \multirow{2}{0.7cm}{TSA $L$ [AU]} & \multirow{2}{0.9cm}{TSA $t_{\rm rel}$ + 500 s}&\multirow{2}{0.7cm}{Data type}&& Onset & $\Delta \phi$& \multirow{3}{1cm}{$I_{\rm max}$ [pfu MeV$^\textrm{-1}$]} & \multirow{3}{0.8cm}{Rise time [h]} & \multirow{2}{0.7cm}{TSA $L$ [AU]} & \multirow{2}{0.9cm}{TSA $t_{\rm rel}$ + 500 s}&& Onset & $\Delta \phi$& \multirow{3}{1cm}{$I_{\rm max}$ [pfu MeV$^\textrm{-1}$]} & \multirow{3}{0.8cm}{Rise time [h]} &\multirow{2}{0.7cm}{TSA $L$ [AU]} & \multirow{2}{0.9cm}{TSA $t_{\rm rel}$ + 500 s}\\
&&&[UT]&[\textdegree]&&&&&&&[UT]&[\textdegree]&&&&&&[UT]&[\textdegree]&&\\
&&&&&&&&&&&&&&&&&&&&&\\
&&&&&&&&[UT]&&&&&&&&[UT]&&&&&&&[UT]\\
\hline%\hline
\noalign{\smallskip}
\endhead
\noalign{\smallskip}
\hline
\endfoot
\noalign{\smallskip}
1&2011 Jan. 28&&01:34&88&8.7$\times 10^{1}$&1.5$\pm$0.1&1.24&01:28$\pm$00:05&LF&&01:27&1&7.5$\times 10^{1}$&1.3$\pm$0.1&1.06&01:23$\pm$00:05&&\ldots&\ldots&\ldots&\ldots&\ldots&\ldots\\
2&2011 Feb. 15&&02:22&10&6.7$\times 10^{1}$&7.9$\pm$0.1&1.11&02:18$\pm$00:08&LF&&\ldots&\ldots&\ldots&\ldots&\ldots&\ldots&&02:40&104&2.4$\times 10^{2}$&1.8$\pm$0.1&1.31&02:33$\pm$00:05\\
3&2011 Mar. 07&&20:58&48&1.8$\times 10^{3}$&15.1$\pm$0.1&1.16&20:53$\pm$00:05&DE&&22:02&-40&IC&30.3$\pm$0.1&1.05&21:58$\pm$00:08&&20:19&143&1.3$\times 10^{3}$&0.1$\pm$0.1&1.19&20:14$\pm$00:05\\
4&2011 Mar. 21&&03:23&130&2.3$\times 10^{2}$&4.6$\pm$0.2&1.19&03:18$\pm$00:13&DE&&02:32&42&7.3$\times 10^{3}$&1.1$\pm$0.1&1.10&02:28$\pm$00:05&&01:38&-135&6.5$\times 10^{1}$&6.9$\pm$2.3&1.18&01:33$\pm$02:20\\
5&2011 Aug. 04&&04:36&36&2.4$\times 10^{3}$&3.1$\pm$0.1&1.22&04:30$\pm$00:05&DE&&\ldots&\ldots&\ldots&\ldots&\ldots&\ldots&&\ldots&\ldots&\ldots&\ldots&\ldots&\ldots\\
6&2011 Sep. 06&&23:32&28&2.1$\times 10^{2}$&2.4$\pm$0.1&1.16&23:27$\pm$00:05&DE&&\ldots&\ldots&\ldots&\ldots&\ldots&\ldots&&01:04\textsuperscript{+}&123&2.6$\times 10^{1}$&4.6$\pm$0.1&1.19&00:59$\pm$00:08\textsuperscript{+}\\
7&2011 Sep. 22&&14:23&-89&N/A&N/A&1.17&14:18$\pm$00:06&DE&&IC&167&IC&N/A&1.16&N/A&&10:57&8&2.2$\times 10^{4}$&18.5$\pm$0.1&1.24&10:51$\pm$00:05\\
8&2011 Oct. 04&&\ldots&\ldots&\ldots&\ldots&\ldots&\ldots&\ldots&&14:36&105&7.4$\times 10^{2}$&13.4$\pm$0.1&1.10&14:32$\pm$00:05&&IC&-53&IC&N/A&1.17&N/A\\
9&2011 Nov. 03&&23:20&-152&1.7$\times 10^{2}$&6.7$\pm$0.1&1.20&23:15$\pm$00:08&DE&&22:39&103&5.6$\times 10^{3}$&2.6$\pm$0.1&1.28&22:33$\pm$00:05&&23:32&-50&1.6$\times 10^{2}$&5.7$\pm$0.1&1.44&23:24$\pm$00:08\\
10&2012 Jan. 23&&04:02&25&1.8$\times 10^{4}$&4.0$\pm$0.3&1.12&03:58$\pm$00:17&LF&&06:17&-83&5.8$\times 10^{3}$&54.3$\pm$0.9&1.09&06:13$\pm$00:54&&IC&139&2.0$\times 10^{3}$&N/A&1.33&N/A\\
11&2012 Jan. 27&&18:49&71&7.9$\times 10^{3}$&19.0$\pm$0.1&1.07&18:45$\pm$00:05&LF&&20:55&-37&IC&N/A&1.12&20:51$\pm$00:08&&03:10\textsuperscript{+}&-175&2.7$\times 10^{2}$&56.0$\pm$0.2&1.25&03:04$\pm$00:15\textsuperscript{+}\\
12&2012 Mar. 05&&\ldots&\ldots&\ldots&\ldots&\ldots&\ldots&\ldots&&05:45&-163&5.2$\times 10^{1}$&12.0$\pm$0.2&1.11&05:41$\pm$00:09&&05:23&64&6.5$\times 10^{3}$&3.3$\pm$0.3&1.25&05:17$\pm$00:19\\
13&2012 Mar. 07&&00:32&-27&7.0$\times 10^{4}$&13.9$\pm$0.1&1.15&00:27$\pm$00:05&LF&&01:30&-136&6.7$\times 10^{2}$&30.8$\pm$0.1&1.20&01:25$\pm$00:06&&00:48&91&2.3$\times 10^{4}$&14.0$\pm$0.2&1.27&00:42$\pm$00:11\\
14&2012 Mar. 24&&\ldots&\ldots&\ldots&\ldots&\ldots&\ldots&\ldots&&00:35&84&4.5$\times 10^{3}$&1.2$\pm$0.1&1.11&00:31$\pm$00:05&&03:15&-48&2.6$\times 10^{2}$&16.2$\pm$0.2&1.23&03:09$\pm$00:13\\
15&2012 May 17&&01:48&76&4.5$\times 10^{3}$&0.9$\pm$0.1&1.21&01:43$\pm$00:05&LF&&06:20&-39&IC&N/A&1.03&06:17$\pm$00:09&&03:57&-166&1.3$\times 10^{1}$&11.9$\pm$0.2&1.17&03:52$\pm$00:11\\
16&2012 Jul. 08&&17:42&83&1.4$\times 10^{3}$&6.0$\pm$0.1&1.17&17:37$\pm$00:05&DE&&BG&N/A&N/A&N/A&1.14&N/A&&\ldots&\ldots&\ldots&\ldots&\ldots&\ldots\\
17&2012 Jul. 12&&16:59&15&1.7$\times 10^{3}$&9.9$\pm$0.1&1.16&16:54$\pm$00:05&LF&&BG&N/A&N/A&N/A&N/A&N/A&&17:05&130&1.2$\times 10^{3}$&4.1$\pm$0.1&1.24&16:59$\pm$00:05\\
18&2012 Jul. 23&&05:14&137&1.9$\times 10^{3}$&31.0$\pm$0.2&1.15&05:09$\pm$00:13&DE&&IC&16&1.3$\times 10^{6}$&N/A&1.09&N/A&&13:11&-108&4.3$\times 10^{3}$&31.3$\pm$0.2&1.25&13:05$\pm$00:13\\
19&2012 Sep. 20&&BG&-159&N/A&N/A&N/A&N/A&N/A&&15:51&76&2.0$\times 10^{4}$&12.5$\pm$0.1&1.12&15:47$\pm$00:05&&15:06&-42&3.9$\times 10^{4}$&65.6$\pm$0.1&1.33&14:59$\pm$00:06\\
20&2012 Sep. 28&&00:17&32&2.2$\times 10^{3}$&2.0$\pm$0.1&1.21&00:12$\pm$00:05&LF&&BG&N/A&2.0$\times 10^{4}$&N/A&1.16&N/A&&00:26&150&9.4$\times 10^{2}$&5.2$\pm$0.1&1.27&00:20$\pm$00:08\\
21&2012 Nov. 08&&11:54&164&6.3$\times 10^{1}$&11.1$\pm$0.2&1.13&11:49$\pm$00:09&LF&&10:59&37&1.0$\times 10^{3}$&1.9$\pm$0.1&1.14&10:54$\pm$00:05&&BG&N/A&N/A&N/A&N/A&N/A\\
22&2013 Mar. 05&&BG&-145&N/A&N/A&N/A&N/A&N/A&&03:43&84&2.0$\times 10^{4}$&6.5$\pm$0.1&1.16&03:38$\pm$00:05&&05:07&-5&3.8$\times 10^{3}$&37.3$\pm$0.1&1.44&04:59$\pm$00:08\\
23&2013 Apr. 11&&07:51&-12&1.6$\times 10^{3}$&4.8$\pm$0.1&1.16&07:46$\pm$00:05&LF&&17:27&-145&1.0$\times 10^{1}$&63.2$\pm$0.7&1.08&17:23$\pm$00:42&&07:26&130&1.2$\times 10^{4}$&1.5$\pm$0.1&1.27&07:20$\pm$00:05\\
24&2013 Apr. 24&&22:29&167&1.1$\times 10^{2}$&3.8$\pm$0.1&1.16&22:24$\pm$00:05&DE&&00:02\textsuperscript{+}&32&5.4$\times 10^{2}$&3.5$\pm$0.3&1.08&23:58$\pm$00:21&&\ldots&\ldots&\ldots&\ldots&\ldots&\ldots\\
25&2013 May 15&&10:02&-64&7.3$\times 10^{3}$&12.7$\pm$0.1&1.19&09:57$\pm$00:05&LF&&BG&N/A&N/A&N/A&N/A&N/A&&IC&78&IC&N/A&1.13&N/A\\
26&2013 May 22&&13:45&15&1.8$\times 10^{4}$&13.4$\pm$0.1&1.15&13:40$\pm$00:05&LF&&20:28&-122&1.4$\times 10^{3}$&30.6$\pm$0.5&1.16&20:23$\pm$00:28&&21:54&157&1.7$\times 10^{2}$&3.1$\pm$0.6&1.18&21:49$\pm$00:34\\
27&2013 Aug. 20&&BG&178&N/A&N/A&N/A&N/A&N/A&&00:42&34&9.7$\times 10^{3}$&33.6$\pm$0.2&1.14&00:37$\pm$00:09&&00:27&-44&2.0$\times 10^{3}$&34.4$\pm$0.1&1.27&00:21$\pm$00:08\\
28&2013 Sep. 30&&N/A&33&3.1$\times 10^{3}$&N/A&1.36&N/A&LF&&10:04&-114&4.3$\times 10^{1}$&28.5$\pm$0.7&1.16&09:59$\pm$00:44&&01:31&172&8.9$\times 10^{1}$&23.0$\pm$0.1&1.44&01:23$\pm$00:08\\
29&2013 Oct. 05&&\ldots&\ldots&\ldots&\ldots&\ldots&\ldots&\ldots&&06:43&92&2.6$\times 10^{3}$&6.9$\pm$0.4&1.19&06:38$\pm$00:25&&10:37&19&IC&N/A&1.36&10:30$\pm$00:25\\
30&2013 Oct. 11&&BG&-44&1.8$\times 10^{1}$&N/A&1.16&N/A&LF&&07:32&169&6.2$\times 10^{3}$&1.4$\pm$0.1&1.18&07:27$\pm$00:05&&07:48&96&1.0$\times 10^{3}$&5.1$\pm$0.1&1.39&07:41$\pm$00:05\\
31&2013 Oct. 25&&BG&-73&N/A&N/A&1.22&N/A&N/A&&09:26&139&1.3$\times 10^{1}$&14.3$\pm$0.1&1.09&09:22$\pm$00:08&&08:39&69&2.2$\times 10^{3}$&10.3$\pm$0.1&1.46&08:31$\pm$00:05\\
32&2013 Oct. 28&&BG&-28&N/A&N/A&1.29&N/A&N/A&&17:23&-176&4.0$\times 10^{1}$&5.2$\pm$0.3&1.13&17:18$\pm$00:19&&16:19&114&2.9$\times 10^{2}$&4.7$\pm$0.1&1.34&16:12$\pm$00:05\\
33&2013 Nov. 02&&07:05&133&2.1$\times 10^{2}$&9.9$\pm$0.4&1.21&07:00$\pm$00:23&DE&&04:40&-15&2.0$\times 10^{3}$&1.6$\pm$0.1&1.11&04:36$\pm$00:05&&05:49&-84&7.2$\times 10^{1}$&3.9$\pm$0.1&1.25&05:43$\pm$00:05\\
34&2013 Nov. 07&&BG&-151&N/A&N/A&1.18&N/A&N/A&&10:35&60&9.3$\times 10^{3}$&0.4$\pm$0.1&1.06&10:31$\pm$00:05&&10:37&-7&1.0$\times 10^{4}$&8.3$\pm$0.1&1.19&10:32$\pm$00:05\\
35&2013 Dec. 26&&05:07&-164&8.8$\times 10^{1}$&17.5$\pm$0.2&1.32&05:00$\pm$00:09&LF&&03:49&45&1.5$\times 10^{3}$&9.8$\pm$0.1&1.10&03:45$\pm$00:05&&04:10&-13&1.0$\times 10^{3}$&9.4$\pm$0.1&1.33&04:03$\pm$00:06\\
36&2013 Dec. 28&&18:33&125&3.3$\times 10^{2}$&2.0$\pm$0.1&1.19&18:28$\pm$00:08&DE&&IC&-26&IC&N/A&1.13&N/A&&BG&N/A&N/A&N/A&N/A&N/A\\
37&2014 Jan. 06&&08:09&89&1.9$\times 10^{3}$&2.3$\pm$0.1&1.16&08:04$\pm$00:05&DE&&09:31&-62&1.5$\times 10^{2}$&26.9$\pm$0.1&1.17&09:26$\pm$00:05&&17:08&-118&2.1$\times 10^{1}$&18.8$\pm$0.3&1.38&17:01$\pm$00:19\\
38&2014 Jan. 07&&19:10&8&3.9$\times 10^{4}$&15.4$\pm$0.1&1.15&19:05$\pm$00:08&DE&&17:48&-143&1.8$\times 10^{3}$&23.2$\pm$0.1&1.14&17:43$\pm$00:05&&21:03&161&3.2$\times 10^{2}$&31.7$\pm$0.2&1.46&20:55$\pm$00:11\\
39&2014 Feb. 25&&02:38&-82&2.1$\times 10^{3}$&59.3$\pm$0.1&1.12&02:34$\pm$00:06&LF&&01:11&125&3.9$\times 10^{3}$&2.8$\pm$0.1&1.14&01:06$\pm$00:05&&01:17&78&9.1$\times 10^{3}$&12.7$\pm$0.1&1.36&01:10$\pm$00:05\\
40&2014 Mar. 04&&\ldots&\ldots&\ldots&\ldots&\ldots&\ldots&\ldots&&18:50&-65&6.0$\times 10^{2}$&2.4$\pm$0.1&1.13&18:46$\pm$00:05&&20:40&-111&1.3$\times 10^{2}$&16.5$\pm$0.2&1.38&20:33$\pm$00:11\\
41&2014 Mar. 05&&\ldots&\ldots&\ldots&\ldots&\ldots&\ldots&\ldots&&15:02&28&5.6$\times 10^{2}$&12.0$\pm$0.9&1.16&14:57$\pm$00:56&&BG&N/A&N/A&N/A&N/A&N/A\\
42&2014 Sep. 01&&18:07&-128&1.4$\times 10^{2}$&40.5$\pm$0.6&1.17&18:02$\pm$00:34&LF&&No data!&N/A&N/A&N/A&N/A&N/A&&11:32&33&1.8$\times 10^{5}$&4.1$\pm$0.1&1.18&11:27$\pm$00:05\\
43&2014 Sep. 10&&21:30&-2&2.1$\times 10^{3}$&8.0$\pm$0.1&1.20&21:25$\pm$00:05&LF&&No data!&N/A&N/A&N/A&N/A&N/A&&19:04&159&2.2$\times 10^{2}$&13.8$\pm$0.2&1.29&18:58$\pm$00:15\\
44&2014 Sep. 25&&00:07&-175&2.4$\times 10^{1}$&10.8$\pm$0.2&1.14&00:02$\pm$00:11&LF&&No data!&N/A&N/A&N/A&N/A&N/A&&21:49\textsuperscript{*}&-14&2.1$\times 10^{3}$&8.9$\pm$0.1&1.23&21:43$\pm$00:08\textsuperscript{*}\\
45&2014 Dec. 13&&15:19&141&7.6$\times 10^{1}$&19.8$\pm$0.1&1.07&15:15$\pm$00:06&LF&&No data!&N/A&N/A&N/A&N/A&N/A&&No data!&N/A&N/A&N/A&N/A&N/A\\
46&2015 Jul. 01&&15:13&N/A&1.6$\times 10^{2}$&1.4$\pm$0.1&1.18&15:08$\pm$00:05&DE&&No data!&N/A&N/A&N/A&N/A&N/A&&No data!&N/A&N/A&N/A&N/A&N/A\\
\end{longtable}
\raggedright
\noindent *: The time given refers to the previous calendar day.\\
\noindent +: The time given refers to the following calendar day.\\
Items marked with "N/A" or "No data!" could not be determined due to insufficient or completely missing data, respectively. The ellipsis (\ldots) indicates that the event in question was not detected. "BG" signifies that an event may have been present but is masked by elevated particle intensity after previous event(s); "IC" signifies that ion contamination occurred in the detector during the time of event onset or maximum. All times are rounded to the nearest minute.
\normalsize
\end{landscape}
\newpage
% The appendix:

\begin{appendix}
\section{Notes on Particle Data}
\label{App1}
\subsection*{Proton Intensity Intercalibration}
\label{App1_proton_intercal}
Although a detailed attempt to intercalibrate the measured proton intensities accurately between different instruments, so as to account for differences in instrument efficiency, was felt to be outside the scope of this article, an intensity comparison and a simple tentative intercalibration was performed to evaluate the need for correction factors. To assure that the spacecraft observations were made of nearly identical SEP populations, the decay phase of the high-energy proton event of 14 December 2006 was chosen for this purpose. The time range of interest was taken to span from the noon of 15 December to 20:00 UT on 16 December, eliminating the rise phase and ending approximately when the STEREO-A/HET measured 40\textendash 100 MeV proton intensity fell below $\approx$2.0$\times 10^{-3}$ pfu MeV$^{-1}$. To compensate for the imperfect match between the energy bins of HET and ERNE, the intensities recorded by the former were separately examined in two energy ranges, 40\textendash 100 MeV and 60\textendash 100 MeV. The background intensities were visually estimated as $\approx$5.0$\times 10^{-4}$ pfu MeV$^{-1}$ for ERNE, $\approx$1.0$\times 10^{-4}$ pfu MeV$^{-1}$ for the HET 40\textendash 60 MeV channels, $\approx$4.0$\times 10^{-4}$ pfu MeV$^{-1}$ for the HET 60\textendash 100 MeV channels, and $\approx$3.0$\times 10^{-4}$ pfu MeV$^{-1}$ for the combined HET 40\textendash 100 MeV channels of both STEREO-A and -B; these were subtracted from the measured intensities before the intercalibration analysis. The time resolution was one hour for all data sets. In the following, (the intensities of) the energy channels mentioned above are referred to as $I_{\rm ERNE}$, $I_{\rm HET 40\text{--}100 MeV}$, $I_{\rm HET 40\text{--}60 MeV}$, and $I_{\rm HET 60\text{--}100 MeV}$; the respective quiet-time background intensities have been subtracted.

\begin{figure*}
\centering
\includegraphics[width=0.7\columnwidth]{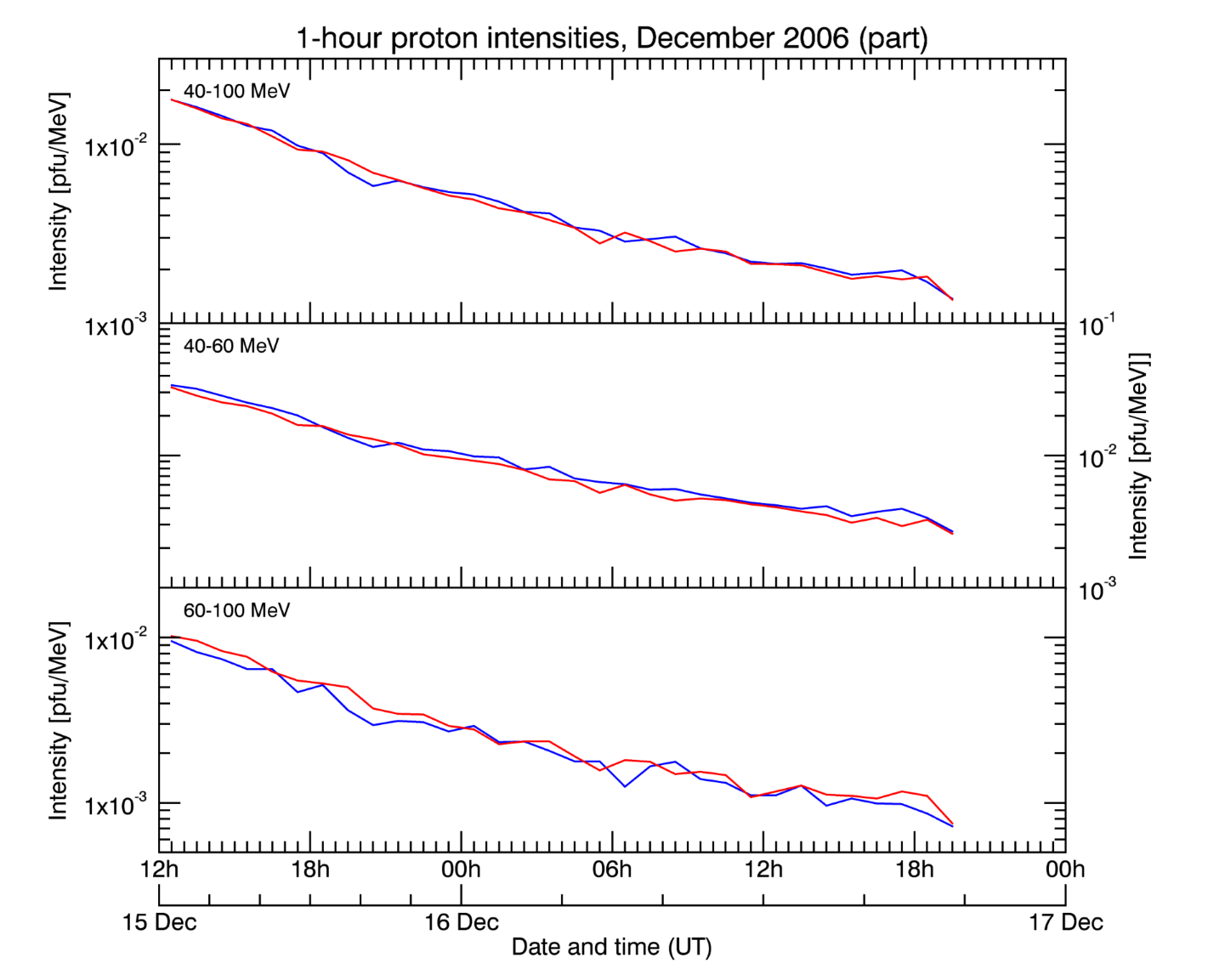}
\caption{\small HET-A (blue) and HET-B (red) proton intensities, with quiet-time background intensities subtracted, during the comparison period. From the upper panel down: $I_{\rm HET 40\text{--}100 MeV}$, $I_{\rm HET 40\text{--}60 MeV}$, and $I_{\rm HET 60\text{--}100 MeV}$. The time resolution is one hour.}
\label{HET-A-B-intercal}
\end{figure*}

The average and the standard deviation of the intensity ratio HET-A/HET-B during the period of interest are 1.09 and 0.09 for the 40\textendash 60 MeV channels, 0.92 and 0.11 for the 60\textendash 100 MeV channel, and 1.02 and 0.08 for the 40\textendash 100 MeV channel; these intensities, with the background intensities given above subtracted, are shown in Figure \ref{HET-A-B-intercal}. The results suggest that the proton intensities measured in this energy range by both HETs usually differ by not more than some 20\%, possibly less still in the 40\textendash 100 MeV combined channel, which enjoys better count statistics than its constituent channels during times of elevated particle flux. For this reason, an intercalibration between HET-A and HET-B was not considered necessary. In the following, the HET-A proton intensities are taken as representative of both HET-A and HET-B as regards any comparison with ERNE.

\begin{figure*}
\centering
\includegraphics[width=0.7\columnwidth]{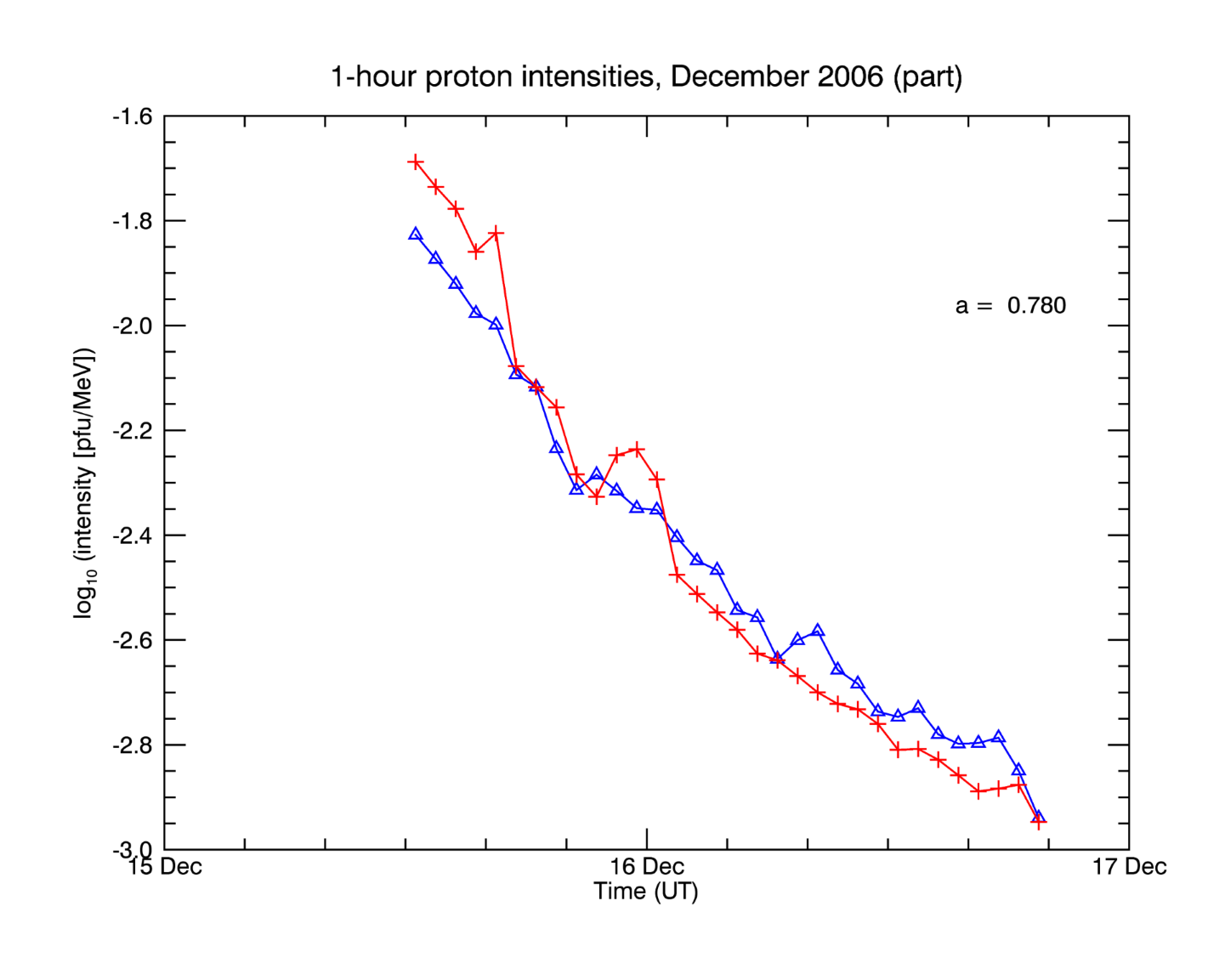}
\caption{\small  $I_{\rm  ERNE}$ (red line and crosses) and $I_{\rm HET,lin}$ (blue line and triangles), when $a$ = 0.78 (see text). Time resolution is one hour.}
\label{HET-ERNE-intercal}
\end{figure*}

For the purpose of the intercalibration, a linear combination of $I_{\rm HET 40\text{--}60 MeV}$ and $I_{\rm HET 60\text{--}100 MeV}$ was defined as follows:

\begin{equation}
I_{\rm HET,lin} = aI_{\rm HET 40\text{--}60 MeV} + (1-a)I_{\rm HET 60\text{--}100 MeV},
\end{equation}
where $a$ is a parameter within the range [0,1]. Choosing $a$ so that $I_{\rm HET,lin}$ coincides as closely as possible with $I_{\rm ERNE}$ and then comparing the uncorrected maximum intensities in $I_{\rm HET 40\text{--}100 MeV}$ with $I_{\rm HET,lin}$ gives a coarse estimate of the measured intensity difference between ERNE and HET and, therefore, of the intercalibration factor. Minimizing the sum of the squares of the quantity $\log_{10}(I_{\rm ERNE}) - \log_{10}(I_{\rm HET,lin})$ over the time range of interest suggests that $I_{\rm HET,lin}$ and $I_{\rm ERNE}$ coincide best when $a$ $\approx$ 0.78. However, it must be emphasized that this result is strictly applicable only to periods of moderate, decreasing proton intensities. The intensities $I_{\rm HET,lin}$ and $I_{\rm ERNE}$, with $a$ = 0.78, are shown in Figure \ref{HET-ERNE-intercal} for the comparison period. 

To study the effects of this correction briefly, the peak intensities of the events considered in this work were determined using both $I_{\rm HET,lin}$ and $I_{\rm HET 40\text{--}100 MeV}$, with $a$ set to 0.78. The maximum intensity values derived from $I_{\rm HET,lin}$ are between 1.3 and 2.2 times greater than those derived from $I_{\rm HET 40\text{--}100 MeV}$, which implies that the intercalibration factor for the ERNE and HET high-energy proton channels of interest would also lie approximately in this range.

A visual inspection of ERNE and HET data recorded in December 2006 confirms that $I_{\rm ERNE}$ and $I_{\rm HET 40\text{--}100 MeV}$ typically agree to within a factor of $\approx$2 during periods of clearly enhanced proton flux and good data coverage. The only notable exceptions to this are the short, very high intensity spike in HET data on 12 December 2006 and a period of about 24 hours during the decay phase of the 13 December 2006 event. ERNE did not detect anything corresponding to the first feature, but in contrast appears to have resolved some structure not readily visible in the 1-hour-averaged HET data in the latter case. Such discrepancies are likely to be due to local structures in the proton flux.

Considering the fact that when widely separated, SOHO and the STEREOs encounter different particle populations, a precise proton intensity intercalibration between ERNE and the HETs would be, in general, not possible. For this reason, it was not pursued here any further and the measured proton intensity values are reported and analysed as such for each observing spacecraft, without applying any correction. In the light of the results presented above, it nevertheless does not seem unreasonable to expect that the measured intensities are likely accurate and comparable to each other within a factor of $\approx$2 in most cases.

\subsection*{Electron Data Intercalibration}
\label{App1_electron_intercal}

\begin{figure*}
\centering
\includegraphics[width=0.7\columnwidth]{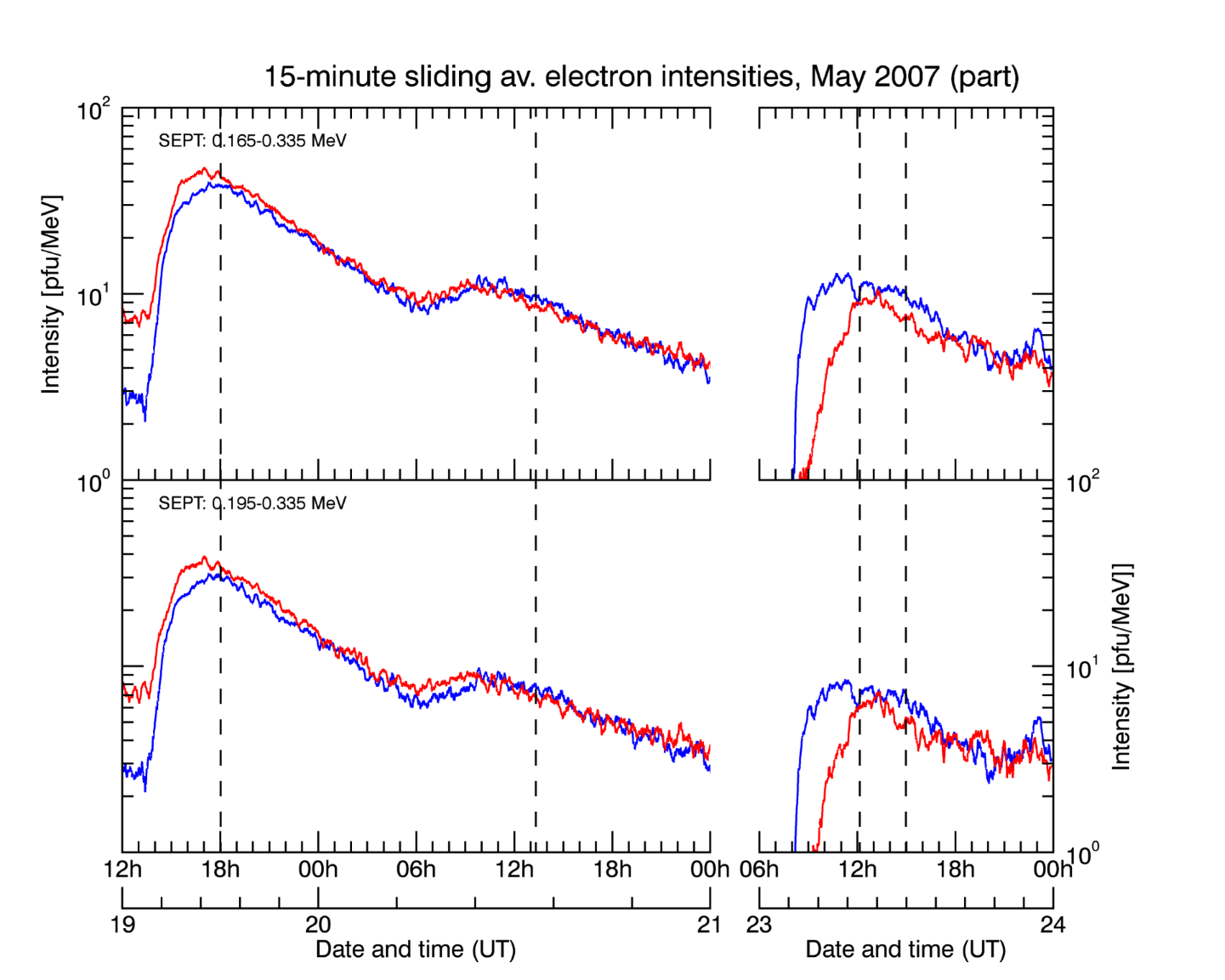}
\caption{\small SEPT-A (blue) and SEPT-B (red) electron intensity data with quiet-time background intensity of 2.0 pfu MeV$^{-1}$ subtracted for the 0.165\textendash 0.335 MeV (upper panel) and the 0.195\textendash 0.335 MeV (lower panel) combined energy channels. All data are smoothed with a 15-minute sliding average. The dashed vertical lines mark the limits of the comparison time selection.}
\label{sept_epam_intercal1}
\end{figure*}

Approximate intercalibration factors were determined for STEREO/SEPT and ACE/EPAM omnidirectional electron intensities. The decay phases of two small electron events in late May of 2007\textemdash during which STEREO-A and -B were separated from the Earth by $\approx$6\textdegree \,and $\approx$3\textdegree, respectively\textemdash were chosen for comparison and the data points selected from the approximate time of intensity maximum onward until the 0.165\textendash 0.335 MeV electron intensity recorded by SEPT-A fell below 10 pfu MeV$^{-1}$. Since electron intensities are usually more isotropic during the event decay than the onset and peak phases (see \textit{e.g.} \citealp{Dresing2014}), only the decay phases were investigated to minimise the effect of differing instrument responses due to particle flux anisotropies. SEPT intensities were separately examined in two energy ranges, 0.165\textendash 0.335 MeV and 0.195\textendash 0.335 MeV, to compensate for the imperfect match between the energy bins of SEPT and EPAM. After visually estimated quiet-time background intensities were subtracted (2.0 pfu MeV$^{-1}$ for SEPT-A and -B in both energy ranges, 8.0 pfu MeV$^{-1}$ for EPAM LEFS60, 20.0 pfu MeV$^{-1}$ for EPAM DE30), the data were smoothed with 15-minute sliding average.\footnote{This somewhat simpler method was preferred over the one used in proton intensity intercalibration between SOHO/ERNE and STEREO/HET due to the electron observations having much better particle count statistics and the fact that the ACE/EPAM and STEREO/SEPT electron energy channels of interest match each other better than the SOHO/ERNE and STEREO/HET proton energy channels.}

\begin{figure*}
\centering
\includegraphics[width=0.7\columnwidth]{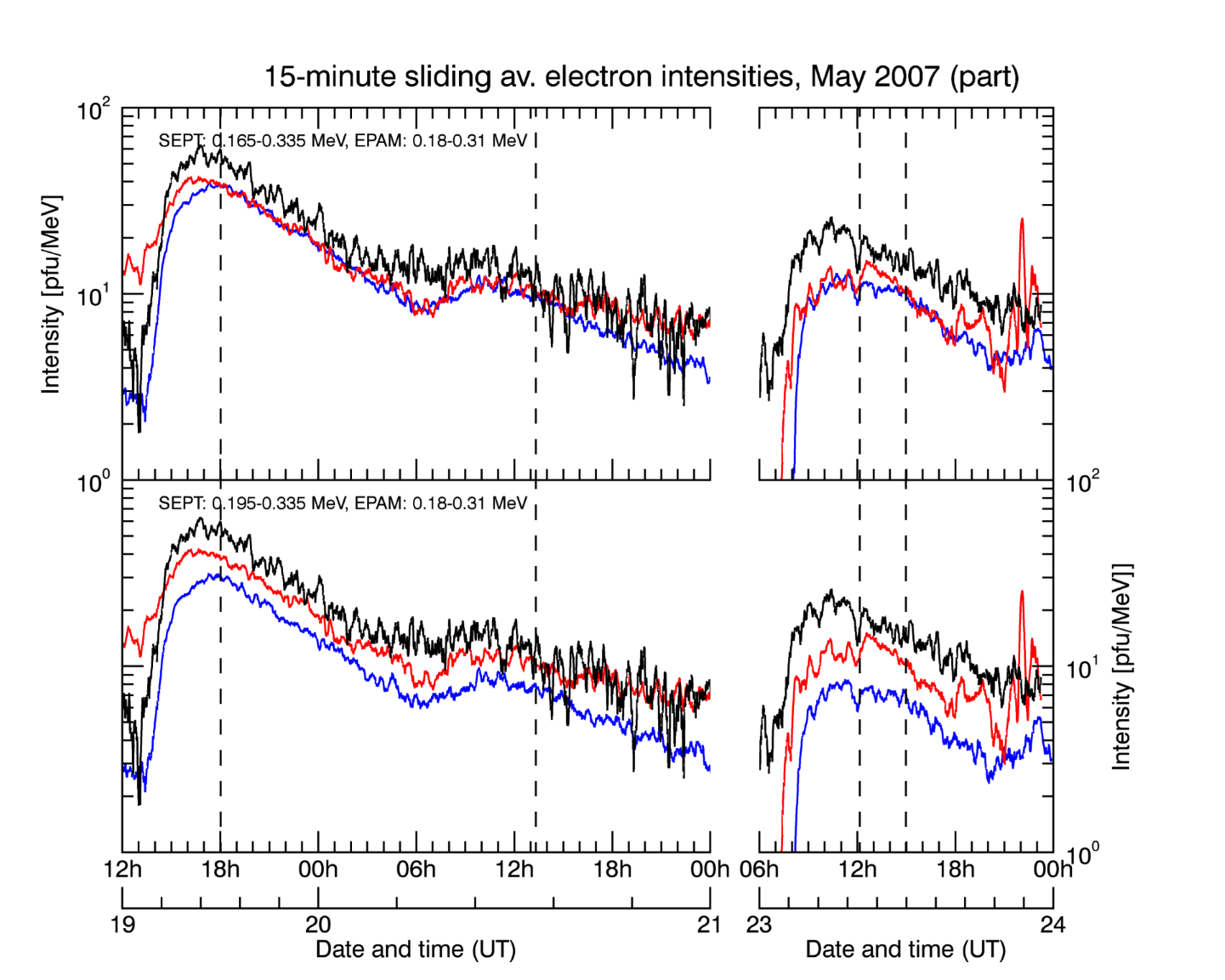}
\caption{\small SEPT-A electron intensity data, with quiet-time background intensity of 2.0 pfu MeV$^{-1}$ subtracted, for the 0.165\textendash 0.335 MeV (upper panel) and the 0.195\textendash 0.335 MeV (lower panel) combined energy channels (blue), each compared with the 0.18\textendash 0.31 MeV EPAM LEFS60 (red) and DE30 (black) electron intensity data, with quiet-time background intensities of 8.0 pfu MeV$^{-1}$ and 20.0 pfu MeV$^{-1}$, respectively, subtracted. All data are smoothed with a 15-minute sliding average. The dashed vertical lines mark the limits of the comparison time selection.}
\label{sept_epam_intercal2}
\end{figure*}

\begin{figure*}
\centering
\includegraphics[width=0.7\columnwidth]{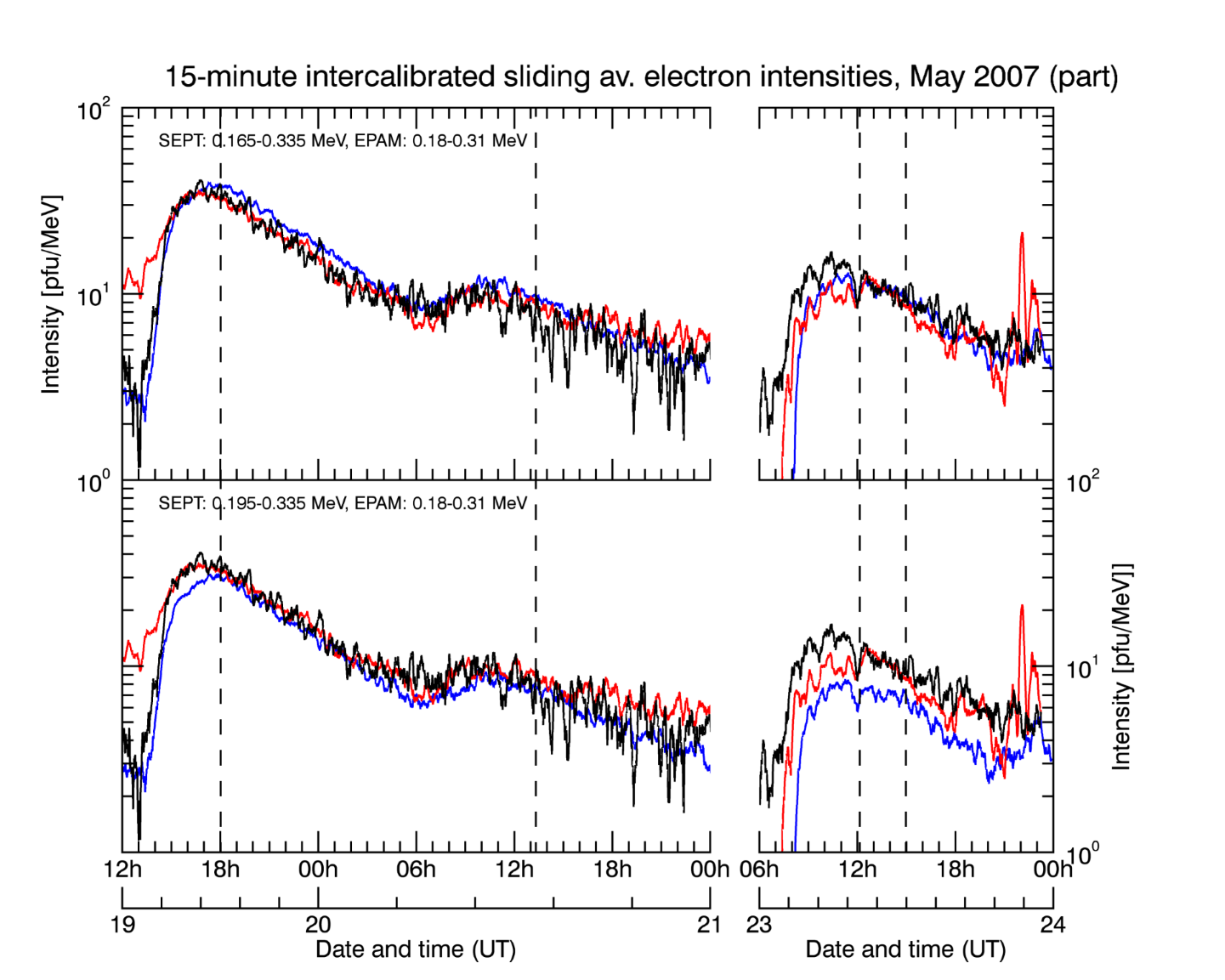}
\caption{\small Same as Fig. \ref{sept_epam_intercal2}, but with correction factors of 0.84 and 0.65 applied to EPAM LEFS60 and DE30 data, respectively. The dashed vertical lines mark the limits of the comparison time selection.}
\label{sept_epam_intercal3}
\end{figure*}

SEPT-A and SEPT-B were considered first. Since the average difference of their intensities was less than 2\% in both energy ranges during the period of interest, no intercalibration between them was regarded as necessary. However, the averages and standard deviations of the SEPT-A/EPAM (LEFS60) and the SEPT-A/EPAM (DE30) intensity ratios were next determined, respectively, as 0.96 $\pm$ 0.10 and 0.74 $\pm$ 0.12 for the SEPT 0.165\textendash 0.335 MeV energy range and as 0.72 $\pm$ 0.10 and 0.56 $\pm$ 0.10 for the SEPT 0.195\textendash 0.335 MeV energy range. Based on these results, the intercalibration factor was selected as 0.84 for EPAM LEFS60 and 0.65 for EPAM DE30, these being the mean values of the averages given above for each EPAM data type. It must be noted that as in the case of protons, explained in Section \ref{Sec2.1}, these factors are fully applicable only when moderate and decreasing electron fluxes are being observed.

Figures \ref{sept_epam_intercal1}, \ref{sept_epam_intercal2}, and \ref{sept_epam_intercal3} demonstrate the intensity intercalibration process for electron data. Figure \ref{sept_epam_intercal1} shows the data with background intensity subtracted for SEPT-A and SEPT-B in the energy ranges 0.165\textendash 0.335 MeV (upper panel) and 0.195\textendash 0.335 MeV (lower panel); Figure \ref{sept_epam_intercal2} shows a comparison between SEPT-A 0.165\textendash 0.335 MeV intensity with EPAM LEFS60 and DE30 intensity (upper panel) and SEPT-A 0.195\textendash 0.335 MeV intensity with the same EPAM data (lower panel), all with background intensity subtracted; and finally, Figure \ref{sept_epam_intercal3} displays the same data as Figure \ref{sept_epam_intercal2} but with the intercalibration factors applied to both EPAM data types.
\end{appendix}

\begin{acks}
The research described in this paper was supported by ESA contract 4000120480/17/NL/LF/hh. M.P. and R.V. acknowledge the funding from Academy of Finland (decisions 267186 and 297395). N.D. and B.H. acknowledge the funding from Deutscher Akademischer Austauschdienst (DAAD 57247608) and the STEREO/SEPT, Chandra/EPHIN and SOHO/EPHIN project which is supported under grant 50OC1702 by the Federal Ministry of Economics and Technology on the basis of a decision by the German Bundestag. We would like to acknowledge and express our gratitude to the organizations and teams responsible for maintaining the data sources used in this article (SEPServer, CDAW SOHO LASCO CME Catalog, SolarSoft Latest Events Archive, NOAA/Solar-Terrestrial Physics at the National Centers for Environmental Information, Coordinated Data Analysis Web, and the ACE Science Center). CDAW SOHO LASCO CME Catalog: this CME catalog is generated and maintained at the CDAW Data Center by NASA and The Catholic University of America in cooperation with the Naval Research Laboratory. SOHO is a project of international cooperation between ESA and NASA. The Radio Monitoring website: this survey is generated and maintained at the Observatoire de Paris by the LESIA UMR CNRS 8109 in cooperation with the Artemis team, Universities of Athens and Ioanina and the Naval Research Laboratory. The STEREO/SECCHI/COR2 CME catalog (the Dual-Viewpoint CME Catalog from the SECCHI/COR Telescopes): this catalogue is generated and maintained at JHU/APL, in collaboration with the NRL and GSFC, and is supported by NASA.
\end{acks}

\begin{acks}[Disclosure of Potential Conflicts of Interest]
The authors declare that they have no conflicts of interest.
\end{acks}

\bibliographystyle{spr-mp-sola}
\bibliography{multisc_cat_bib}

\end{article}
\end{document}